\documentclass[final,10pt,5p,times,twocolumn]{elsarticle}

\usepackage{subcaption}
\usepackage{afterpage}
\usepackage{amsmath}
\usepackage{dblfloatfix}
\usepackage{lscape}
\usepackage{rotating} 

\usepackage[modulo]{lineno}

\usepackage{amssymb}
\usepackage{array,multirow}
\newcolumntype{C}[1]{>{\centering\arraybackslash}p{#1}}
\usepackage[utf8]{inputenc}

\usepackage[%
    colorlinks=true,
    pdfborder={0 0 0},
    linkcolor=red,
    breaklinks
]{hyperref}
\usepackage{url}

\usepackage{breakurl}

\usepackage{url}

\usepackage[capitalize]{cleveref}

\journal{NIM-A}

\begin{document}

\begin{frontmatter}


 \title{Detailed simulation for the ClearMind prototype detection module and event reconstruction using artificial intelligence}
 \author[1]{C.-H. Sung\corref{cor1}}
 \ead{e830430@gmail.com}
 \author[3]{L. Cappellugola}
 \author[1]{M. Follin\fnref{fn1}}
 \author[3]{S. Curtoni}
 \author[3]{M. Dupont}
 \author[3]{C. Morel}
 \author[1]{A. Galindo-Tellez}
 \author[1]{R. Chyzh}
 \author[4]{D. Breton}
 \author[4]{J. Maalmi}
 \author[1,2]{D. Yvon}
 \author[1,2]{V. Sharyy}
 \cortext[cor1]{Corresponding author}
 \address[1]{Université Paris-Saclay, CEA, IRFU, Département de Physique des Particules, Gif-sur-Yvette, France}
 \address[2]{Université Paris-Saclay, CEA, CNRS, Inserm, Laboratoire d’Imagerie Biomédicale Multimodale Paris Saclay, Orsay, France}
 \address[3]{Aix-Marseille Université, CNRS/IN2P3, CPPM, Marseille, France}
 \address[4]{Université Paris-Saclay, CNRS, IJCLab, Orsay, France}
 \fntext[fn1]{Currently at EOS Imaging, Paris, France}




\begin{abstract}
  The ClearMind project aims to develop the TOF-PET position sensitive detection module optimized for the time resolution, spatial resolution, and detection efficiency. For this, the ClearMind project uses a large (59 $\times$ 59 mm$^2$) monolithic PbWO$_4$ (PWO) scintillating crystal with a bialkali photo-electric layer deposited directly on the crystal. Scintillation and Cherenkov photons result together from the 511 keV gamma-ray interation into the PWO crystal. A micro-channel plate photomultiplier tube (MCP-PMT) encapsulating the PWO crystal amplifies photoelectrons generated at the photocathode, and the corresponding anode signals are collected through the transmission lines read out at both ends and digitized by a SAMPIC module. In this work, we present a realistic Geant4 simulation of the ClearMind prototype detector, including the propagation of the visible photons in the crystal, the modelling of a realistic response of the photocathode and of the PMT, and the propagation of the electrical signals over the transmission lines. The reconstruction of the gamma conversion in the detector volume is performed from the signals registered at both ends of the transmission lines. We compare the reconstruction precision of a statistical algorithm against machine learning algorithms developed using the TMVA package. We expect to reach a spatial resolution down to a few mm$^3$ (FWHM). Finally, we will discuss prospects for the ClearMind detector.
\end{abstract}




\end{frontmatter}



\section{Introduction}

Positron emission tomography (PET) is a medical imaging technique widely used in oncology \cite{Gallamini2014}. It can be helpful for the diagnosis of diseases and cancers. In PET, the radioactive tracer decays by emitting positrons, which then annihilate with atomic electrons, resulting in the emission of two back-to-back 511 keV gamma-rays. The activity in each organ can be reconstructed from the detection in coincidence of pairs of 511 keV gamma-rays. PET image quality, i.e., SNR  (Signal-to-Noise Ratio), strongly depends on the statistics of coincidences detected by the PET scanner. The time-of-flight (TOF) technique, which measures the difference between the detection times of the two annihilation photons, allows either to improve SNR in the reconstructed image or to reduce tracers doses delivered to the patients for equivalent scan time and image SNR, or to reduce scan time for equivalent patient dose and image SNR \cite{Karp2008}. Image SNR improvement is inversely proportional to the square root of the coincidence time resolution (CTR). The best CTR ($\sim$200 ps FWHM) of commercial PET camera is for now achieved by Siemens \cite{VanSluis2019}. Recent developments of the new types of fast photo-detectors allow the improvement of CTR by detecting Cherenkov photons, emitted by electrons resulting from the photoelectric interaction of the 511 keV gamma-rays in the scintillating crystal. In Refs. \cite{Kwon2016, Gundacker2019, Ota2019}, the possibility to obtain CTRs in the range of 30 ps to 100 ps has been demonstrated on a single pair of photodetectors with a limited detection efficiency. The ClearMind project aims to develop TOF-PET detection modules with a 3D spatial resolution down to a few mm$^3$ FWHM, a CTR $\leq$ 100 ps FWHM, and high detection efficiency. Good performances are expected due to detecting both the Cherenkov and the scintillating photons generated inside a PbWO$_4$ (PWO) crystal. 


In this article we present the principle of ClearMind (CM) detector (Section \ref{section:intro}) and provide a realistic simulation of detection module in Section \ref{section:CM}. This includes the simulation of the photon generation and propagation in the crystal in Sections \ref{subsection:interaction} and \ref{subsection:crystal}, an innovative approach to photocathode simulation in Section \ref{subsection:photocathode}, simulation of the PMT response and signal processing by electronics in Section \ref{subsection:MCPPMT}. In Section \ref{section:CMresults}, we discuss the performances of the CM detection module evaluated in the simulation. In Section \ref{section:Reco}, we demonstrate the feasibility of the machine learning-based position reconstruction of 511 keV gamma-ray interaction in the detector. Finally, in Section \ref{section:conclusion}, we discuss the main factors that affect the time resolution of the module.

\section{ClearMind detector objective and principle}\label{section:intro}
\begin{figure}[!]
  \centering
  \includegraphics[width=0.7\linewidth] {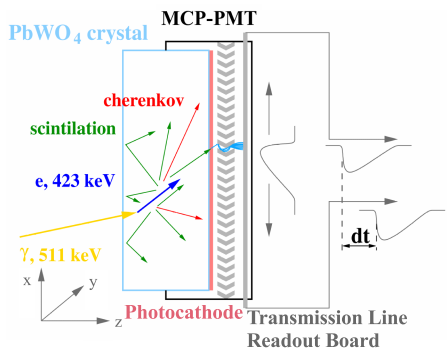}
  \caption{Schematic diagram of the CM detection module. A 511 keV gamma-ray interaction in the crystal produces scintillation and Cherenkov photons that are converted by the photocathode to photoelectrons. These photoelectrons are then multiplied by the MCP-PMT and induce signals on the transmission lines(TLs). Signals from the left and right ends of each TL are amplified by 40 dB amplifiers and digitized by a SAMPIC module.}
  \label{CMD}
\end{figure}

\begin{figure}[!]
  \centering
  \includegraphics[width=0.7\linewidth] {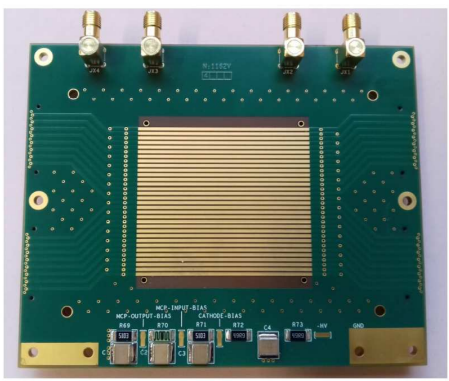}
  \caption{Transmission lines readout board.}
  \label{TL}
\end{figure}

The CM detection module (Fig. \ref{CMD}) is composed of a micro-channel plate photomultiplier tube (MCP-PMT) sealed by a monolithic PWO crystal with a photoelectric layer of high quantum efficiency directly deposited on its inner face. The direct deposition of a photocathode with a refractive index superior to the refractive index of the PWO crystal allows to avoid total reflection at the crystal/photocathode interface, thus maximizing the photon collection efficiency of the module \cite{Yvon2020}. The use of this "scintronic" crystal as an entrance window of an MCP-PMT makes it possible to optimize the time resolution thanks to the excellent electron transit time to the detection an anodes provided by this type of photodetector. The PWO crystal, which is homogeneously doped and has a 59 $\times$ 59 mm$^2$ surface, is provided by CRYTUR \cite{CRYTUR}. A photocathode is directly deposited on the crystal face by PHOTEK \cite{Photek}, which then uses it to form the optical window of the MCP-PMT. We developed a signal readout system for this device using 32 transmission lines (TL) as shown in Fig. \ref{TL} \cite{Sharyy2021}. The signals are read out at both ends of the TLs and allow reconstructing the 511 keV gamma-ray interaction positions as explained in section \ref{section:Reco}.

\section{Simulation of the ClearMind detector}\label{section:CM}

In this study, a complete simulation, done by Geant4 version 10.7 \cite{G4_1, G4_2, G4_3}, of the CM detection module prototype is performed. For the first prototype, the detector has a thin crystal, 5 mm thick, with a photocathode deposited on one side. We simulate the complete signal formation starting from the gamma-ray interaction in the crystal and the production of Cherenkov and scintillation photons. The realistic simulation of the photocathode considers the reflection of visible photons from the photocathode, absorption of photons by the photocathode, and extraction of generated photoelectrons as a function of the photon wavelength. We simulate the propagation and the multiplication of individual photoelectrons generated by the photocathode in the MCP-PMT. Finally, we simulate the signal readout with realistic signal shapes.

\subsection{Interaction in the PWO crystal}\label{subsection:interaction}
511 keV gamma-rays interact within the PWO crystal mainly through photoelectric interaction, Compton scattering, or Rayleigh scattering. The relative probabilities of each interaction are 42.7\%, 51.0\%, and 6.3\%, respectively \cite{nistxcom}.

\begin{figure}[hbt!]
\centering
\includegraphics[width=0.7\linewidth]{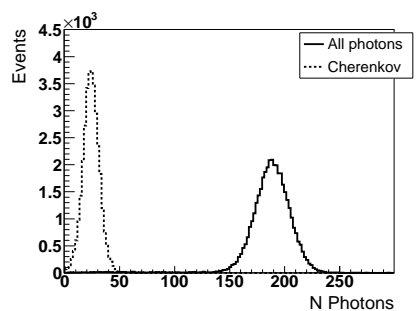}
\caption{Simulated photon production in PWO for one 511 keV gamma-ray conversion with dashed curve the number of Cherenkov photons per event and solid curve the total number of visible photons that are generated.} \label{P}
\end{figure}

Our simulation is performed with the G4EmPenelopePhysics model since it is used for gamma-rays, electrons, and positrons below 1 GeV with a good accuracy \cite{Jodal2012}. The electron emits visible photons through the two following processes: Cherenkov radiation and scintillation mechanism. We applied the default Geant4 electron range cut \cite{Sung}. Fig. \ref{P} shows the simulated results of the photon production in PWO resulting from the photoelectric interaction of 511 keV gamma-rays. The dashed curve shows that an average of $\sim$22 Cherenkov photons are generated from the gamma interaction and the solid curve shows that there are $\sim$187 photons produced in total, i.e., the scintillation light yield amounts $\sim$165 photons, as specified in the simulation\footnote{In the presented plots, we consider only the photons that travel longer than 100 $\mu$m}. The simulation details will be discussed in section \ref{subsubsection:sci}.

\subsection{PWO crystal simulation}\label{subsection:crystal}
\begin{figure}[hbt!]
\centering
\begin{minipage}[t]{0.235\textwidth}
\includegraphics[width=\linewidth]{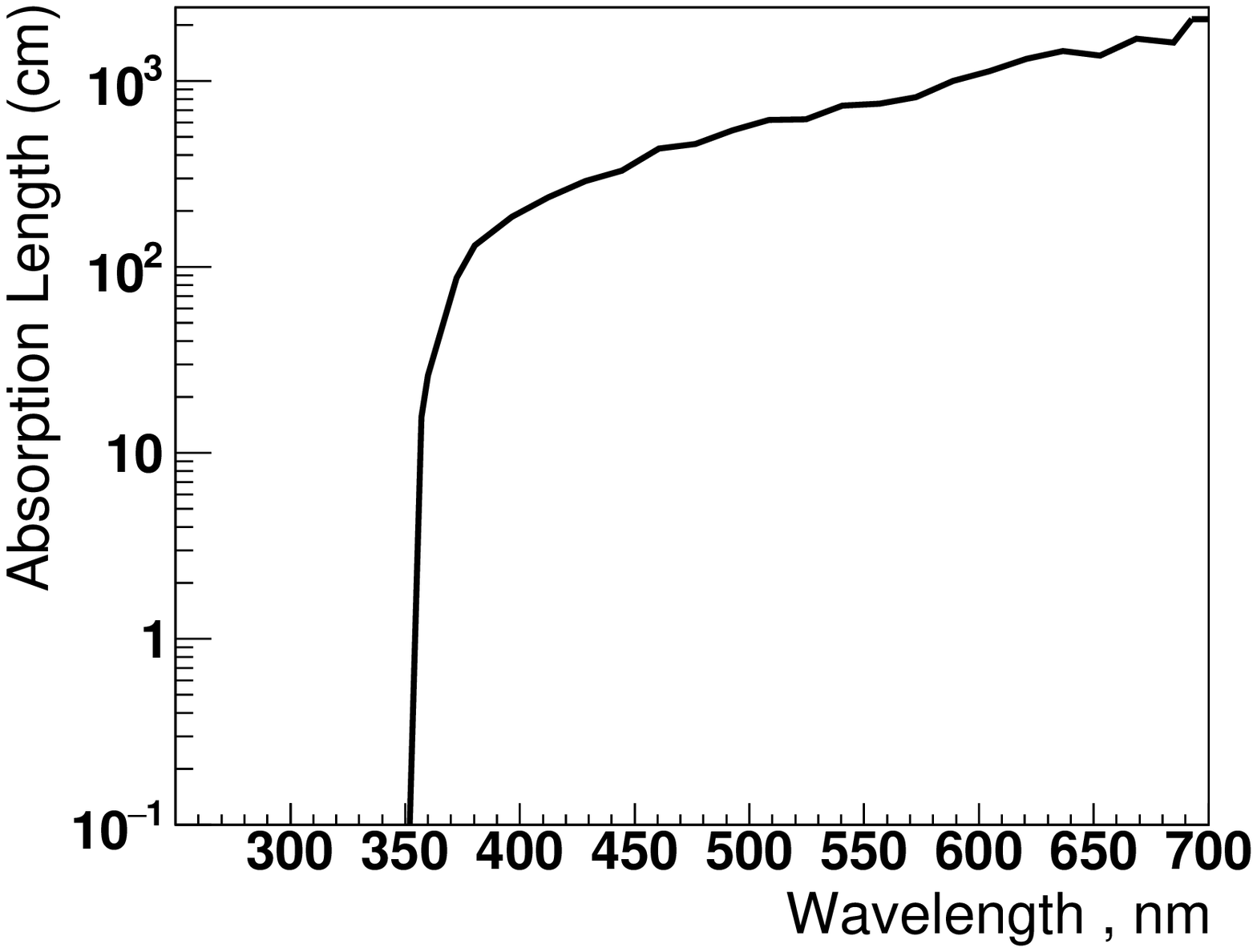}
\caption{Absorption length of PWO \cite{ANNENKOV2002}.} \label{AL_PWO}
\end{minipage}
\hspace*{\fill}
\begin{minipage}[t]{0.235\textwidth}
\includegraphics[width=\linewidth]{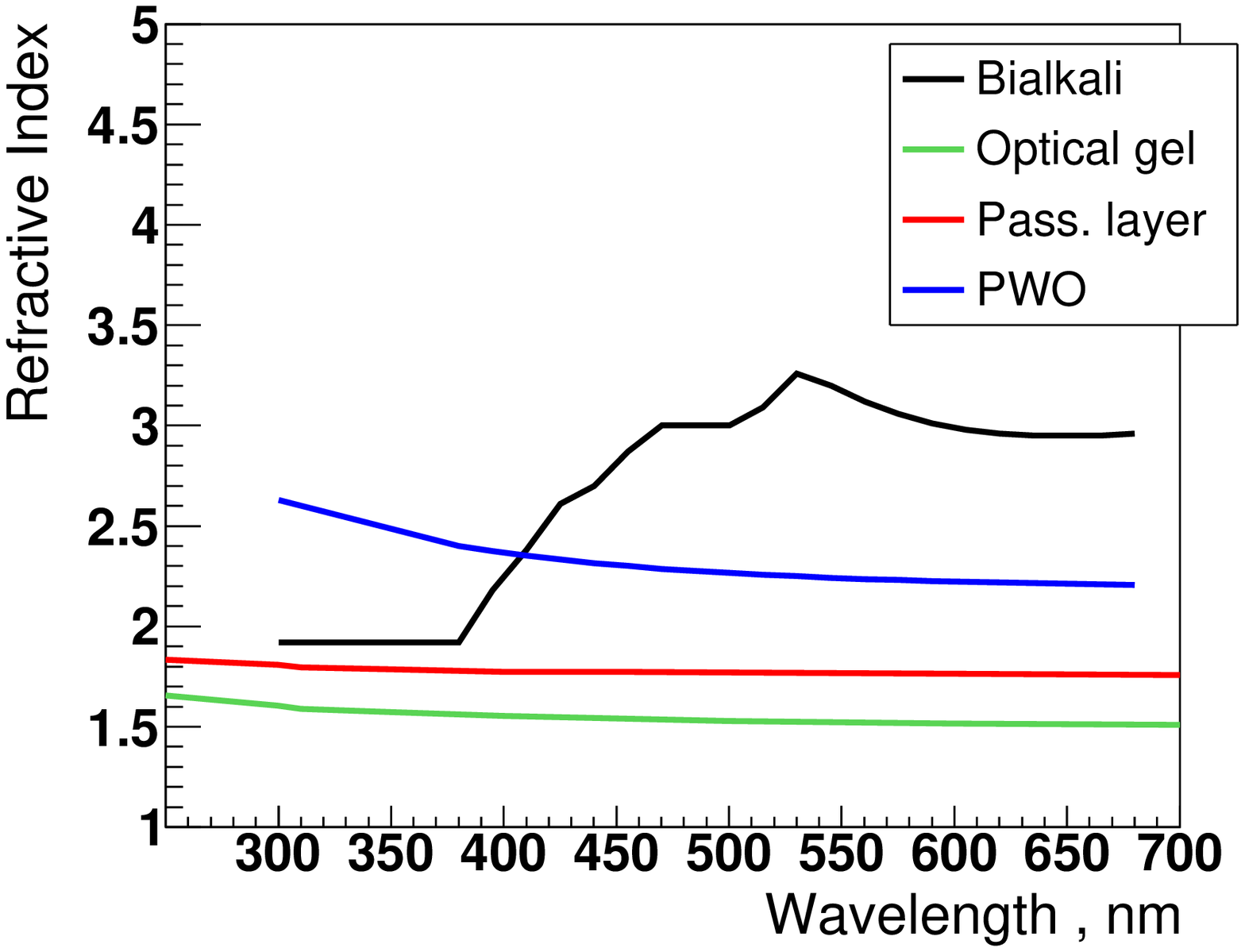}
\caption{Refractive indexes.}\label{RI_PWO}
\end{minipage}
\end{figure}

CRYTUR provides the PWO crystal for the CM detection module using the technology developed for the Panda-II experiment \cite{Borisevich2016}. It has a high density of 8.28 g/cm$^3$. Four main processes can happen to the visible photons in the PWO crystal: absorption in the crystal, reflection on the crystal faces, escape from the crystal into the air, and transmission from the crystal to the photocathode. The reflection is simulated with the unified model \cite{Levin1996}. The absorption length and the refractive indexes used in Geant4 are shown in Figs. \labelcref{AL_PWO,RI_PWO}. The absorption length versus wavelength is computed from the transmission measurement of a PWO crystal published by Ref. \cite{ANNENKOV2002}. The refractive index of PWO is interpolated according to the measurements of Ref. \cite{HUANG2007}. Since Geant4 does not allow users to simulate the birefringence, we chose to specify the average refractive index $n$ of the crystal ordinary and extraordinary refractive indexes by the following polynomial approximation:
\begin{equation} \label{}
n  = 0.0567 \; E_{ph}^2 - 0.1546 \; E_{ph} + 2.3006 \;,
\end{equation}
where $E_{ph}$ is the photon energy in eV. 

\subsubsection{Cherenkov and scintillation photons simulation details}\label{subsubsection:sci}
\begin{figure}[hbt!]
\centering
\includegraphics[width=0.7\linewidth]{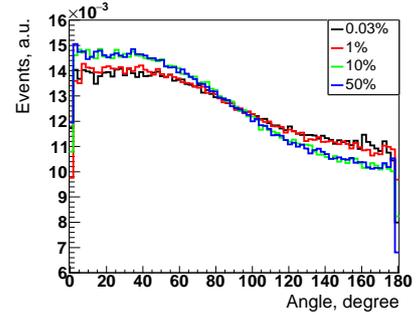}
\caption{Angle distribution of the Cherenkov photons relatively to the direction of the impinging gamma-ray.} \label{Angle_che}
\end{figure}

The Cherenkov photon simulation is a crucial element to assess the timing performance. The Cherenkov photons are emitted along a generatrix of the Cherenkov cone, whose angle depends on the energy of the charged particle. Ref. \cite{Trigila2022} demonstrates a strong dependence between the electron mean step length and the Cherenkov photon production, including the number of photons and their angular distribution. One can control the step size by limiting the electron velocity change per step and specifying the maximum number of Cherenkov photons created in the step. Fig. \ref{Angle_che} shows the angular distribution between the Cherenkov photons and the impinging gamma-ray in the crystal simulated with the default electron velocity change 10\% and with different values ranging from 0.03\% to 50\%. We noticed only a slight difference of $\sim$2\% in the angles of the Cherenkov photons relative to the direction of the impinging gamma-ray. This study uses the default electron velocity change 10\% resulting in $\sim$55\% of the Cherenkov photons going forward ($<$ 90$^\circ$), i.e., noticeably toward the photocathode.

\begin{figure}[bht!]
\centering
\includegraphics[width=0.7\linewidth]{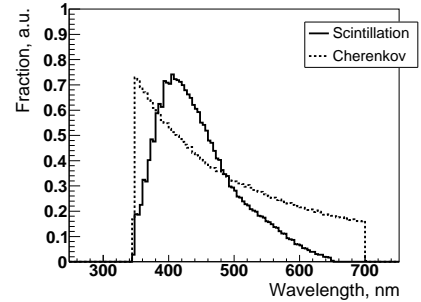}
\caption{Simulated energy spectra of Cherenkov and scintillation photons that have traveled at least 0.1 mm in PWO \cite{PWOsci}.} \label{sci}
\end{figure}

Scintillation properties of several PWO crystals were measured at different temperatures by Ref. \cite{Follin2021}. We applied the properties measured at 20 $^\circ$C for the simulation. For the CRYTUR crystal, the scintillation is described with two components: fast and slow, whose spectra are defined identically. Fig. \ref{sci} shows the simulated energy spectrum of the Cherenkov and scintillation photon, which travel at least 0.1 mm in PWO.
The details of the scintillation photons, such as the time constants, and light yield of both components, are shown in Table \ref{Table:sci}.
\begin{table*}[hbt!]
\centering
\caption{Scintillation production parameters. \cite{Follin2021}} \label{Table:sci}
\begin{tabular}{lcc}
Parameter & Fast component & Slow component\\
\hline
Total light yield (photons/MeV) & \multicolumn{2}{c}{330}\\
Fraction (\%)               & 58.6 & 41.4 \\
Time constant (ns)          & 1.79 & 6.41\\
\hline
\end{tabular}\\  
\end{table*}
Approximately 165 scintillation photons result from the photoelectric interaction of a 511 keV gamma-ray (as shown in Fig. \ref{P}).

\subsubsection{Crystal surface treatment}
The first CM detection module prototype uses a PWO crystal of 59 mm $\times$ 59 mm $\times$ 5 mm. The side for the photocathode deposition and the opposite side are polished to the optical quality. The four other sides are used for sealing. Therefore, we consider them to be black in the simulation.

\begin{figure}[hbt!]
\centering
\includegraphics[width=0.7\linewidth]{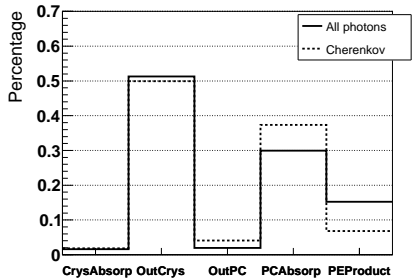}
\caption{Photon destiny for the Cherenkov and scintillation photons altogether (solid line) and the Cherenkov photons only (dashed line). "CrysAbsorp": the photon is absorbed within the crystal. "OutCrys": the photons escape from the crystal to the air or are absorbed by the black surfaces. "OutPC": the photons are transmitted through the photocathode into the air. "PCAbsorp": the photons are absorbed by the photocathode, but no photoelectron is generated. "PEProduct": the photons are absorbed by the photocathode and generate photoelectrons.} \label{term}
\end{figure}

Fig. \ref{term} shows that there are $\sim$2\% of photons absorbed in the crystal, $\sim$52\% are escaping outside of the crystal or absorbed by the black surfaces ("OutCrys" and "OutPC").

\subsection{Photocathode simulation}\label{subsection:photocathode}
The simplest approach to simulate a photocathode assumes that the photocathode perfectly absorbs the photons. The user then defines a quantum efficiency curve, i.e., the probability of converting the photon into a photoelectron. However, Refs. \cite{HARMER2000, Motta2005, HARMER2006, MATSUOKA2018} document that such a model is reductive. They observed experimentally that the quantum efficiencies of the photoelectric layers increase when the incidence angle decreases. They assumed that a photocathode behaves like an absorbing optical medium described by a complex refractive index depending on the wavelength while neglecting the thin layer effects. The propagation of visible photons, including the photon absorption within the photoelectric layer and the reflections at its interfaces follow Fresnel's laws. In order to reproduce the measured quantum efficiency curves, the authors have added to the model a photocathode absorption probability, which is a function of the photon wavelength, and a photoelectron extraction probability averaged over the photocathode thickness, which also depends on the photon wavelength. As the photoelectric layers show significant variations from the detector to detector, the authors averaged the results measured on several PMTs.

\begin{figure}[!]
\centering
\begin{subfigure}[t]{0.235\textwidth}
\includegraphics[width=\linewidth]{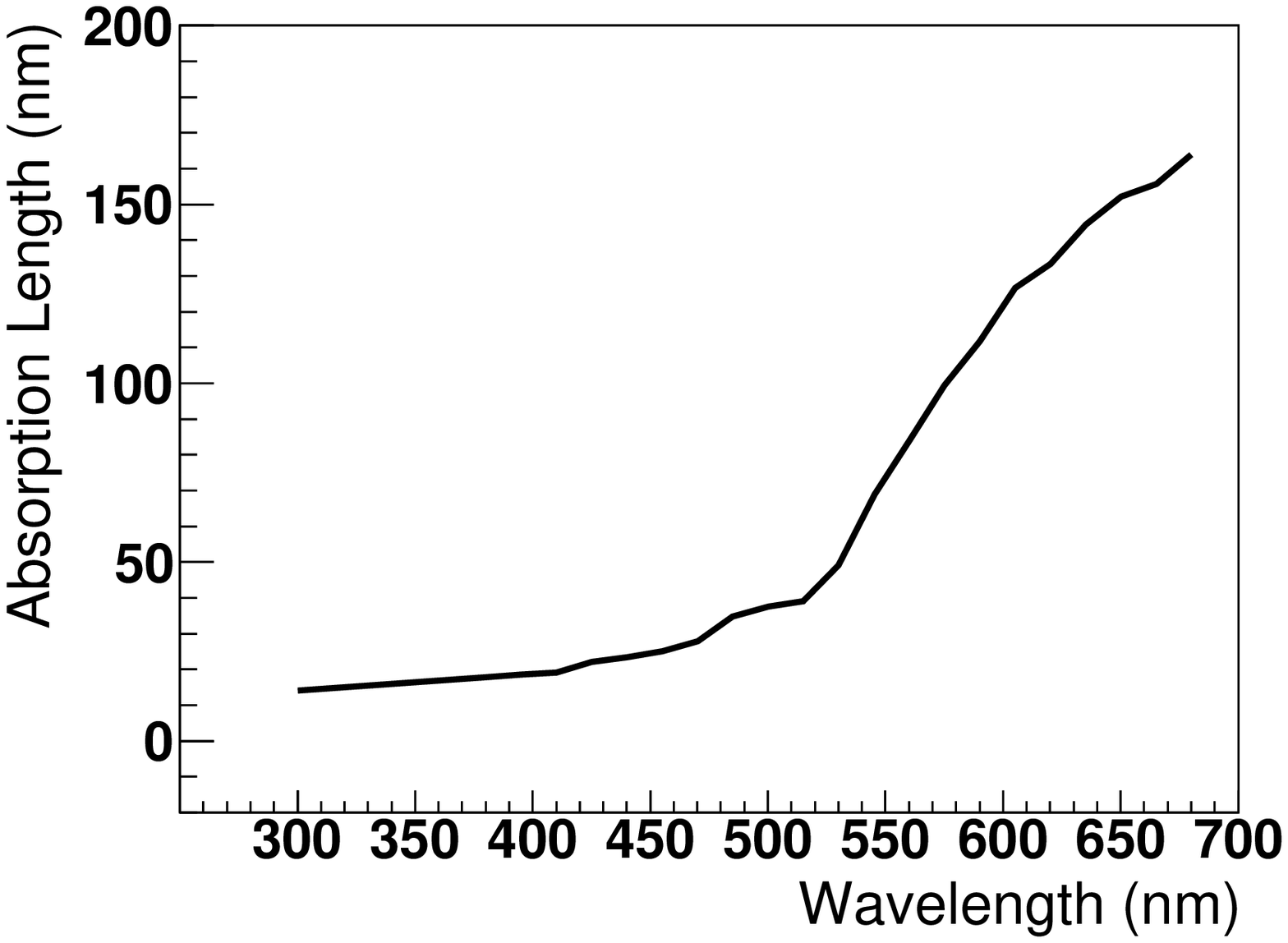}
\caption{Absorption length of a bialkali photocathode.} \label{AL_PC}
\end{subfigure}
\hspace*{\fill} 
\begin{subfigure}[t]{0.235\textwidth}
\includegraphics[width=\linewidth]{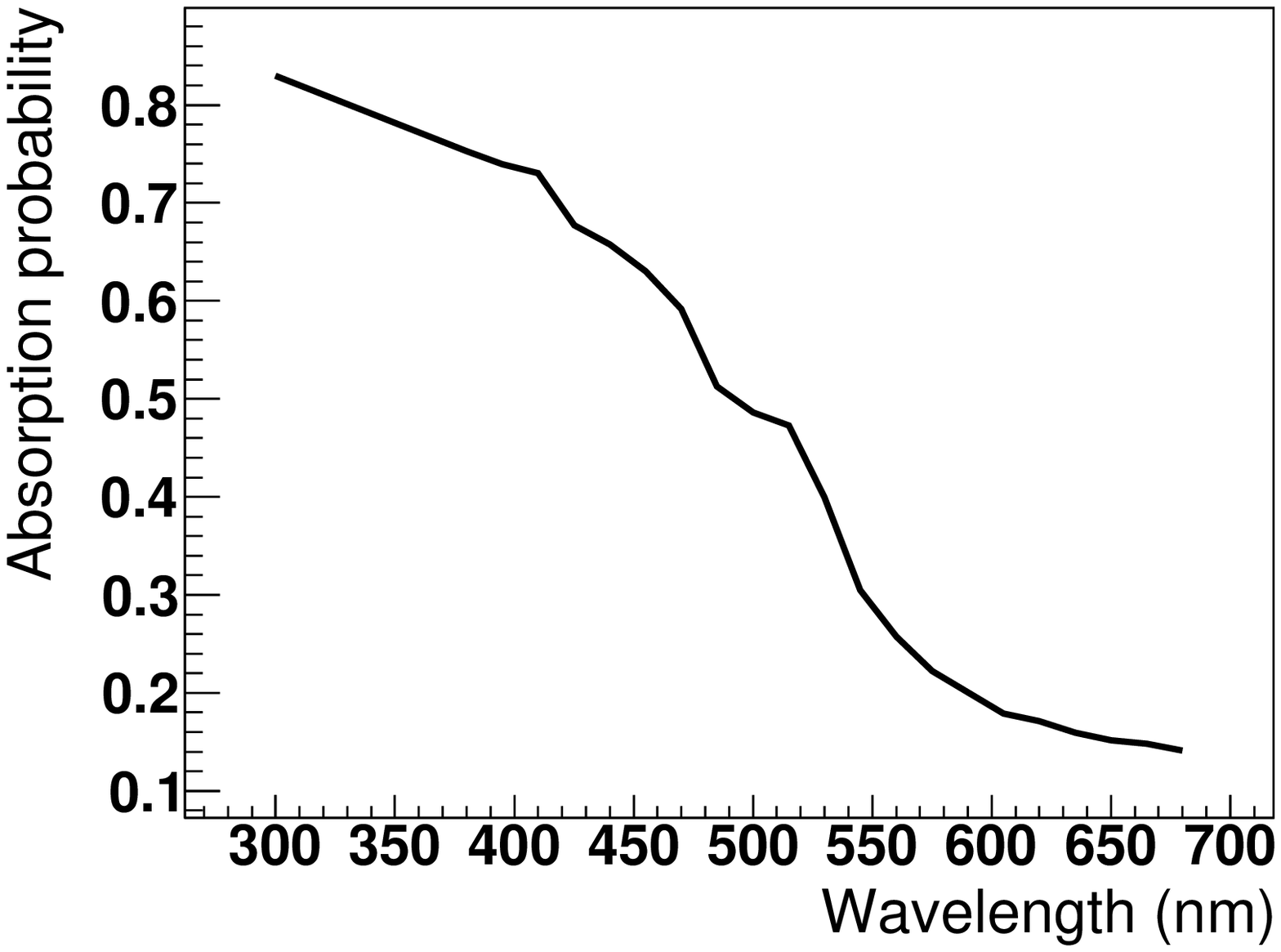}
\caption{Absorption probability of a bialkali photocathode.}\label{PCabs}
\end{subfigure}\\

\begin{subfigure}[t]{0.235\textwidth}
\includegraphics[width=\linewidth]{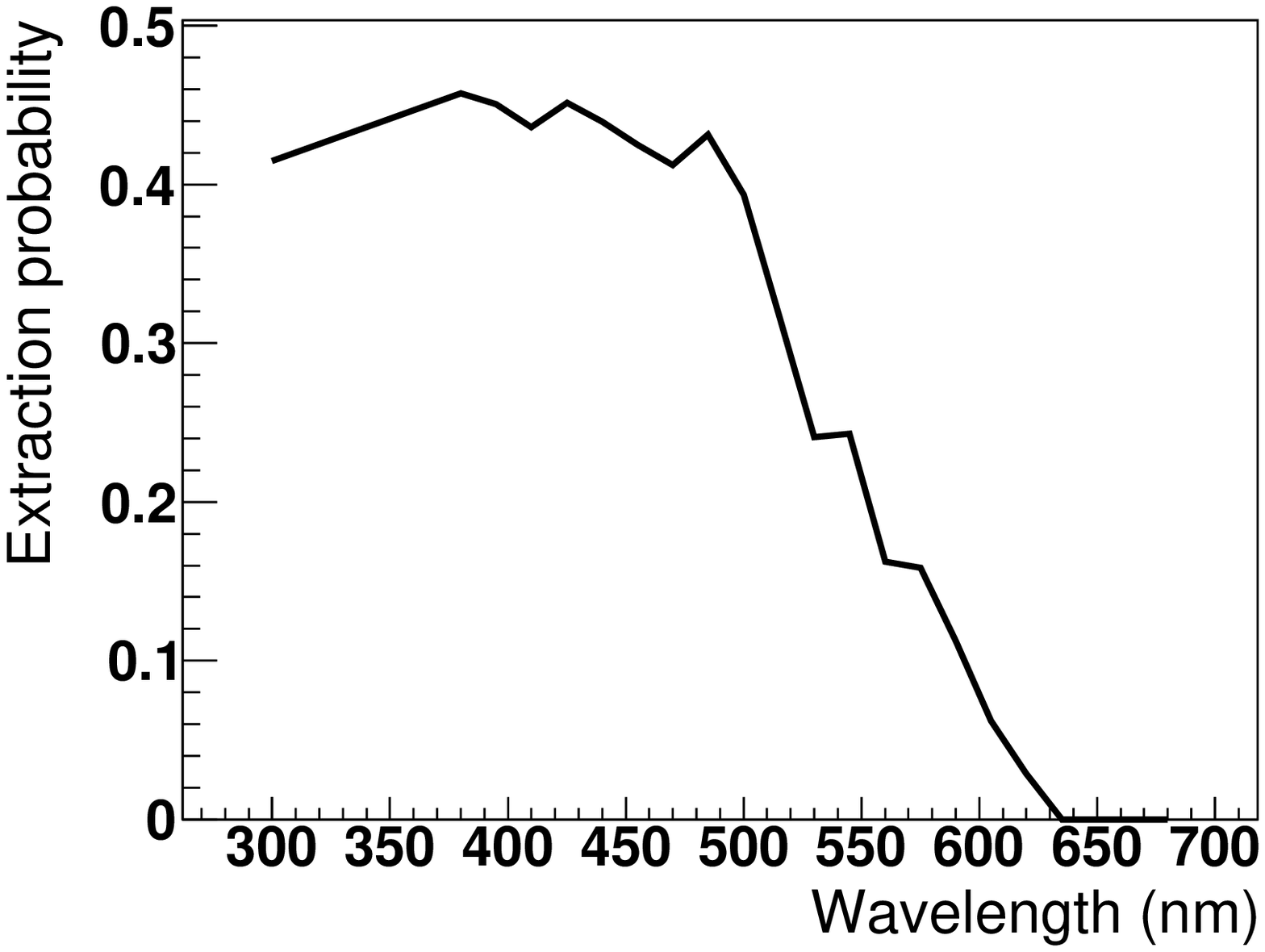}
\caption{Extraction probability of bialkali photocathode. Note that this is lower than 50\% since half of the photoelectrons generated in the photocathode are indeed backscattered.}\label{EX_PC}
\end{subfigure}
\hspace*{\fill}
\begin{subfigure}[t]{0.235\textwidth}
\includegraphics[width=\linewidth]{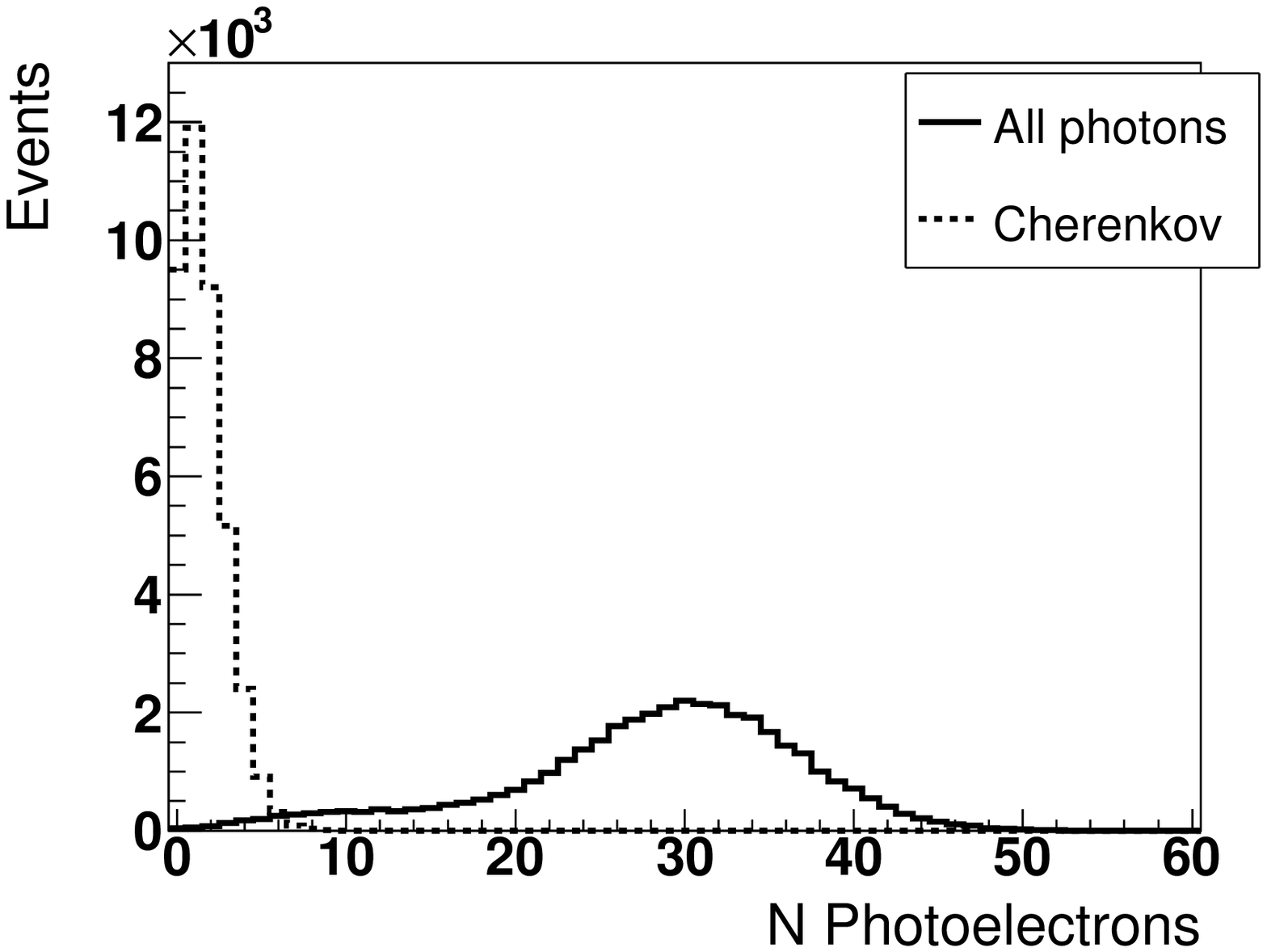}
\caption{Visible photons converted in photoelectrons (solid line) from Cherenkov and scintillation photons altogether and Cherenkov photons only (dashed line). About 30 photons on average are converted to photoelectrons by the photocathode.}\label{Pe}
\end{subfigure}
\caption{Photocathode specifications \cite{Motta2005}.} \label{pc}
\end{figure}

We chose the latter approach for our simulation. We have extracted from Ref. \cite{Motta2005} the complex refractive index of the blue-sensitive bialkali (KCsSb) and green-sensitive bialkali (RbCsSb) photocathodes as a function of wavelength. Here we present the simulation results for the blue-sensitive photocathode only since we expect that it corresponds better to the photocathode used for the CM detection module prototype. We calculated the refractive index (Fig. \ref{RI_PWO}) and the absorption length (Fig. \ref{AL_PC}) of visible photons in these media. To determine the extraction probability of photoelectrons, we used the quantum efficiency measurements of the PMTs ETL 9102 and 9902 presented in Ref. \cite{Motta2005}. The incident optical flux in the PMT borosilicate window is attenuated by reflections at the interface, assuming a refractive index $n=1.5$. We computed the absorption probability of incident visible photons in the photocathode for a typical bialkali photocathode thickness of 25 nm. The ratio between the measured quantum efficiency of the PMTs and the absorption probability gives the photoelectrons extraction probability as a function of the photon wavelength (Fig. \ref{EX_PC}). 

From a theoretical point of view, direct deposition of the photocathode layer on the rear face of the PWO crystal is expected to maximize the transmission of the visible photons by avoiding total reflection at the crystal/photocathode interface. In practice, it has been observed that such a process would induce chemical contamination of the photocathode, which can be oxidized at the contact with the PWO crystal. A passivation layer is therefore deposited between the crystal and the photocathode. Its thickness should be enough to preserve the photocathode chemical stability but must remain as small as possible to minimize its effect on the visible photon transport. Due to its small thickness compared to the photon wavelength, it can be considered as a thin film acting like an anti-reflective optical coating, i.e., which involves interference effects described by Fresnel's laws. Moreover, for wavelengths in the visible spectrum, the refractive index of the typical passivation layer is typically inferior to the refractive index of PWO. That introduces a potential total internal reflection at the diopter for incidence angles superior to the corresponding critical angle. However, since the passivation layer is a thin film and its refractive index is inferior to the photocathode one, frustrated total internal reflection (FTIR) occurs \cite{Zhu1986,Court1964}. This optical phenomenon, quite analogous to an optical tunneling effect, makes it possible to obtain a non-zero transmittance above the critical angle. Both processes (interferences and FTIR) affect the actual transmittance and have to be considered in the simulation. For this reason, we implemented a dedicated function managing these processes in our Geant4 model and applied it to the CM experiment \cite{Cappellugola2021}. The results presented in this study rely on this custom model. 

Our photocathode model allows one to more accurately simulate the positions and times of the generated photoelectrons on the large area of MCP-PMT. It should be noted that a significant part of the visible photons is not absorbed by the photocathode but can be reflected from the backside of the photocathode due to the large difference between the refractive indexes of the photocathode and vacuum. The photocathode thickness is small and comparable to the optical quality roughness of $\sim$$\lambda/10$. We modeled the rear face of the photocathode as grounded using the unified model with $\sigma_\alpha \sim30^{\circ}$ in order to have a continuous angular distribution of photon backscatter.

Fig. \ref{term} shows that the photocathode absorbs $\sim$45\% of the total number of visible photons, and one-third of these photons only will generate photoelectrons. There are $\sim$7\% of the Cherenkov photons that are absorbed and generate photoelectrons in the photocathode. The solid line in Fig. \ref{Pe} shows the number of photoelectrons resulting from the conversion of the visible photons generated in the PWO crystal. The dashed line shows the number of photoelectrons converted from Cherenkov photons only. With the given refractive index, absorption length, and extraction probability, the photocathode generates an average of 30 photoelectrons  from the 511 keV gamma-ray interaction in the crystal. According to the dashed line, 75\% of the events comprise at least one Cherenkov photon converted into a photoelectron. 

\subsection{Simulation of the MCP-PMT and signal formation}\label{subsection:MCPPMT}
MCP-PMTs provide the best time resolution up to now \cite{Photonis, Hamamatsu, MAPMT253}. Our MCP-PMT simulation uses the photoelectrons generated by the photocathode in Geant4. We tracked the position and time of every photoelectron. Each photoelectron either enters the 15 $\mu$m-pore-sized microchannels or absorbed by the microchannel plate and re-emit backscattered electrons \cite{Krizan2009}, or absorbed without producing any electrons. For the last situation, we assume a probability of 10\% in the simulation, corresponding to the electron collection efficiency of 90\% \cite{Lehmann2020}. This behavior affects the distribution of the time response of the MCP-PMT (Section \ref{subsubsection:tts}). We applied a gain value to the signal induced by each electron according to the distribution described in Section \ref{subsubsection:gain} so that the photoelectrons induce the charge on the anode pads individually. The distance between the MCP surface and the anode plane is 3 mm, so the drifted electrons induce a charge on several pads according to the distribution described in Section \ref{subsubsection:chargesharing}. A charge profile is applied to the anode pads to simulate the charge sharing effect. The signal readout considers the total charge from the pads associated with the same line is divided in two, and the signal propagation time to the left and right ends are calculated according to the measured signal propagation speed (Section \ref{subsubsection:RO}). Due to the signal pile-up effect on a TL, we simulate realistic signal shapes instead of implementing a simple model to represent the signals (Section \ref{subsubsection:signal}). At last, the signals are digitized, including the sampling period, electronics noise, and signal saturation with a 64-channel SAMPIC module (Section \ref{subsubsection:sampic}). Information such as amplitude, charge, and time response is extracted from the signals. Furthermore, we intend to use the simulated signals as the input of the event reconstruction algorithm.

\subsubsection{Time response}\label{subsubsection:tts}
\begin{figure}[!]
\centering
\includegraphics[width=0.7\linewidth]{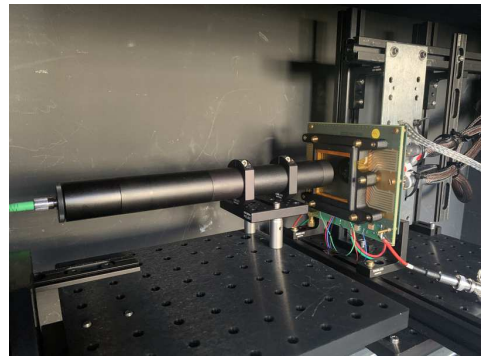}
\caption{Calibration setup using the MAPMT253 photodetector illuminated by a pulsed laser. The horizontal direction is defined as the X-axis and the vertical direction is defined as the Y-axis represented in Fig. \ref{CMD}.} \label{Laser}
\end{figure}

The MCP-PMT model is tuned to the measurement of the commercial MCP-PMT MAPMT253 \cite{MAPMT253,Milnes2020} since the CM detection module uses the same MCP-PMT but with a PWO optical window. Electrons induce charge on 64 by 64 anode pads. Each pair of pad raws is connected to one TL through the Shin-Etsu MT-type of Inter-Connector$^{\text{\textregistered}}$ \cite{ADF}. Induced signals are split into two equal parts and propagated to the both ends of TLs. The signals at both ends of the TLs propagate through a first stage amplification board, followed by 50 Ohm cables connecting to a second stage amplification board. All signals are digitized by a 64-channel SAMPIC module \cite{SAMPIC1,SAMPIC2,SAMPIC3,SAMPIC4}. To measure the time response of the MCP-PMT, we used the pulsed laser Pilas by ALS \cite{Sharyy2021,LASER} as a light source. The light beam from the laser fiber was collimated by a pin-hole of 100 $\mu$m diameter. The calibration setup is presented in Fig. \ref{Laser}. We chose distances and light intensity in such a way that the MCP-PMT was working in a single-photon regime with a fraction of detected photon of 2\%, corresponding to a ratio of two-photon/one-photon events of 1\%. We acquired MCP-PMT data in coincidence with the laser trigger and scanned the whole detector surface with the step of 3 mm along lines (X-axis) and 0.8 mm across lines (Y-axis).

\begin{figure*}[hbt!]
\centering
\begin{subfigure}[t]{0.31\textwidth}
\includegraphics[width=\linewidth]{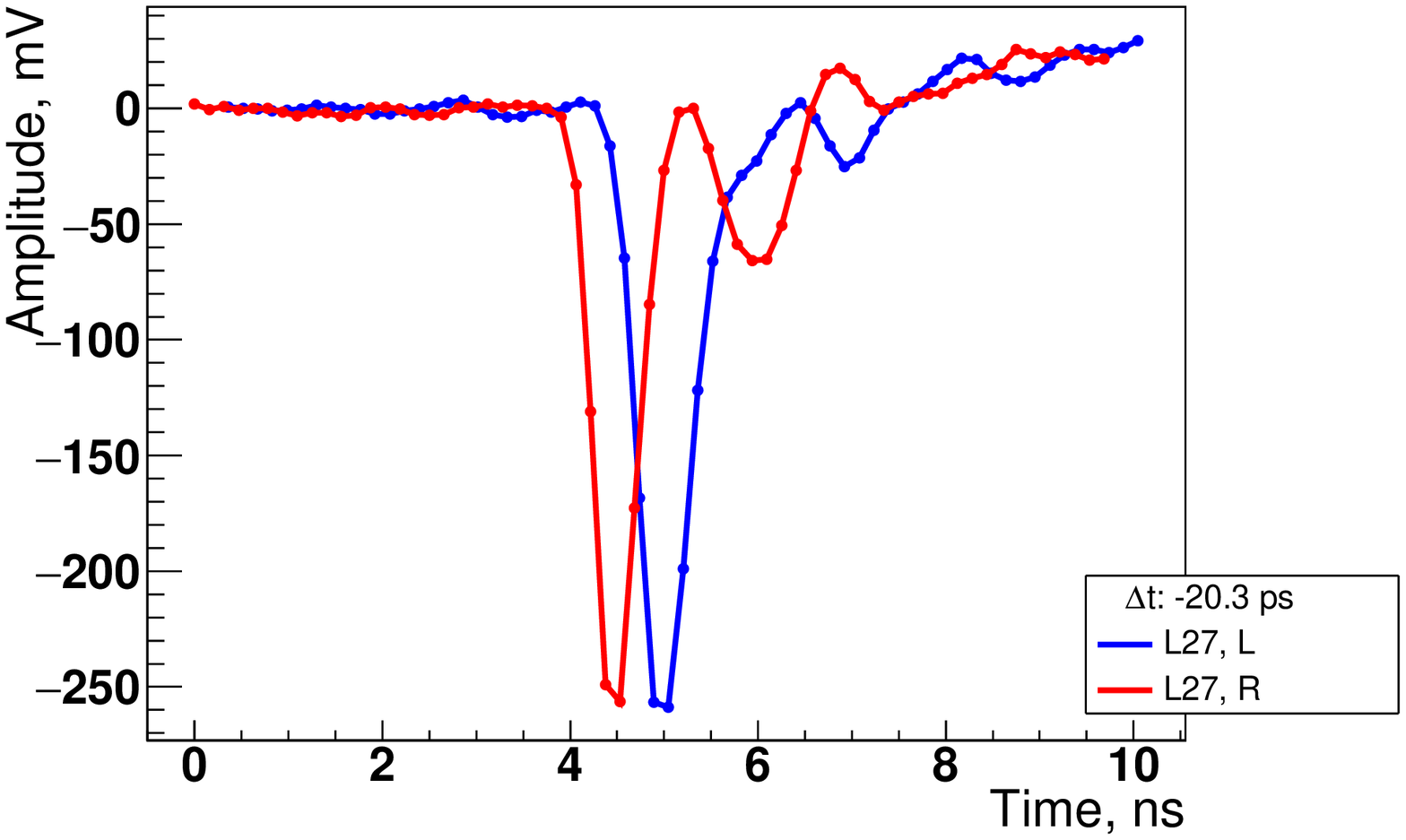}
\caption{Laser at (X,Y)=(28.0 mm, 63.2 mm).}\label{sig_l}
\end{subfigure}
\hspace*{\fill} 
\begin{subfigure}[t]{0.31\textwidth}
\includegraphics[width=\linewidth]{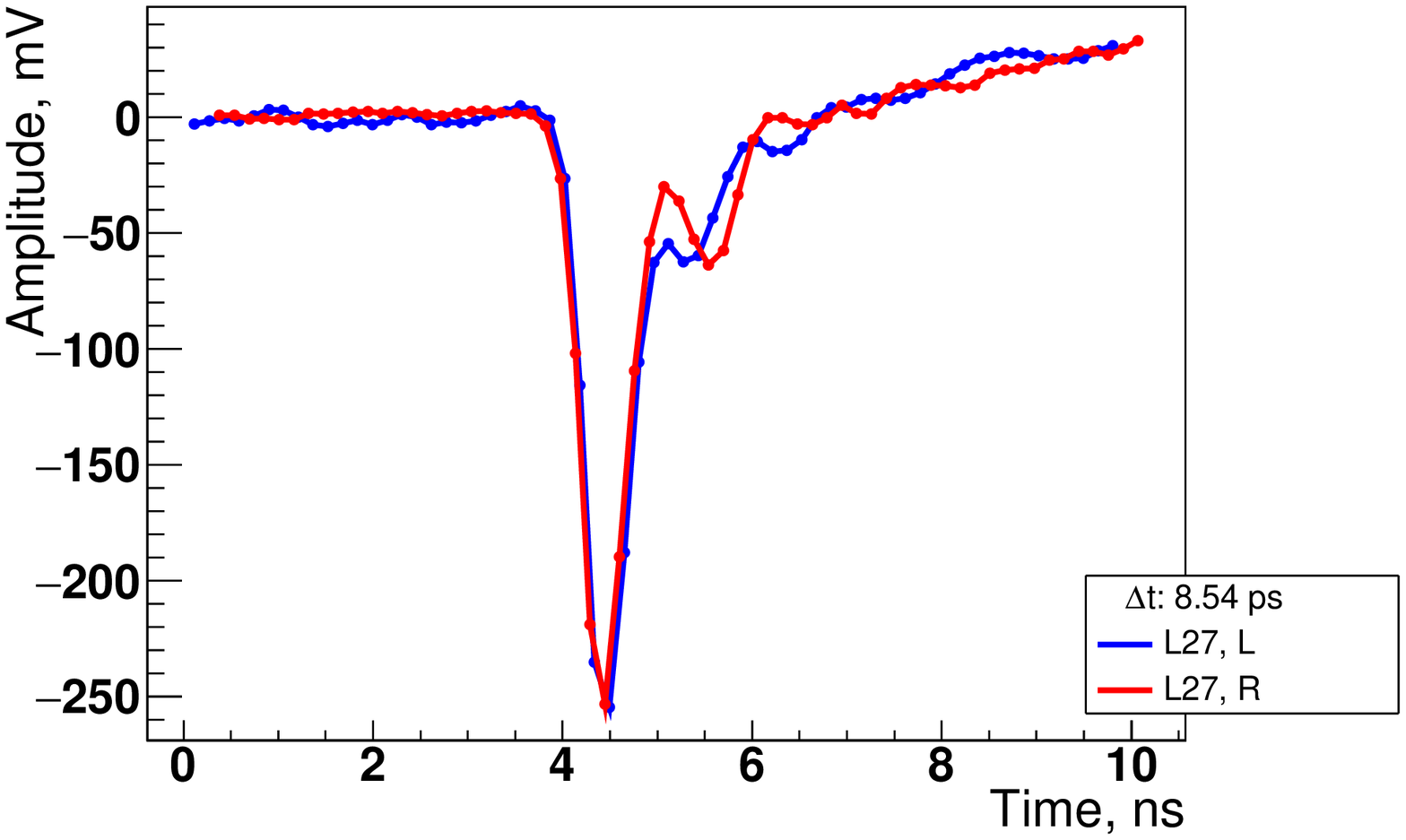}
\caption{Laser at (X,Y)=(49.0 mm, 63.2 mm).}\label{sig_c}
\end{subfigure}
\hspace*{\fill} 
\begin{subfigure}[t]{0.31\textwidth}
\includegraphics[width=\linewidth]{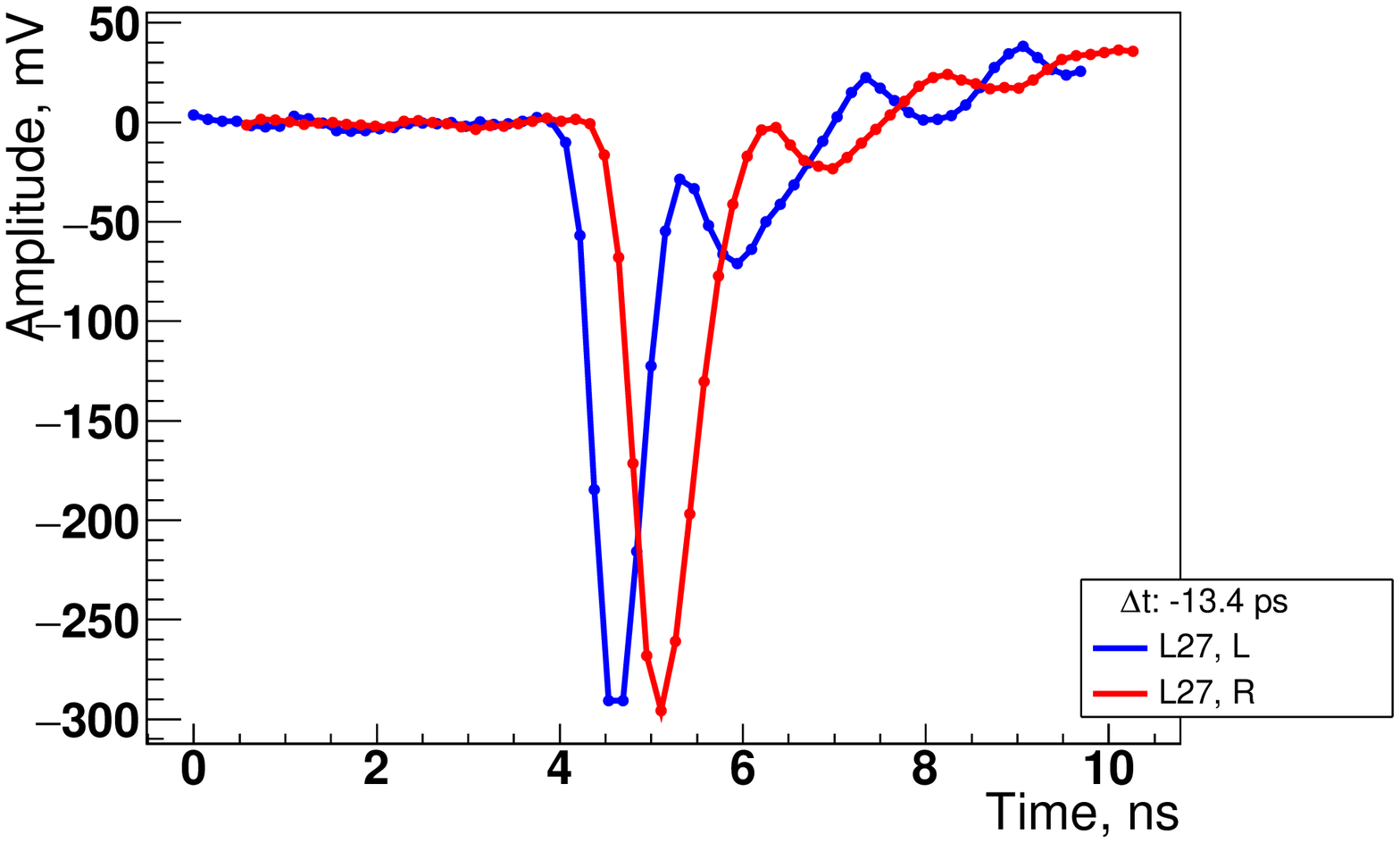}
\caption{Laser at (X,Y)=(70.0 mm, 63.2 mm).} \label{sig_r}
\end{subfigure}
\hspace*{\fill} 
\caption{Registered signals read out at the left (in blue) and right (in red) ends of TL \#27 for different positions of illumination along the line.} \label{sig}
\end{figure*}

\begin{figure}[hbt!]
\centering
\begin{subfigure}[t]{0.235\textwidth}
\includegraphics[width=\linewidth]{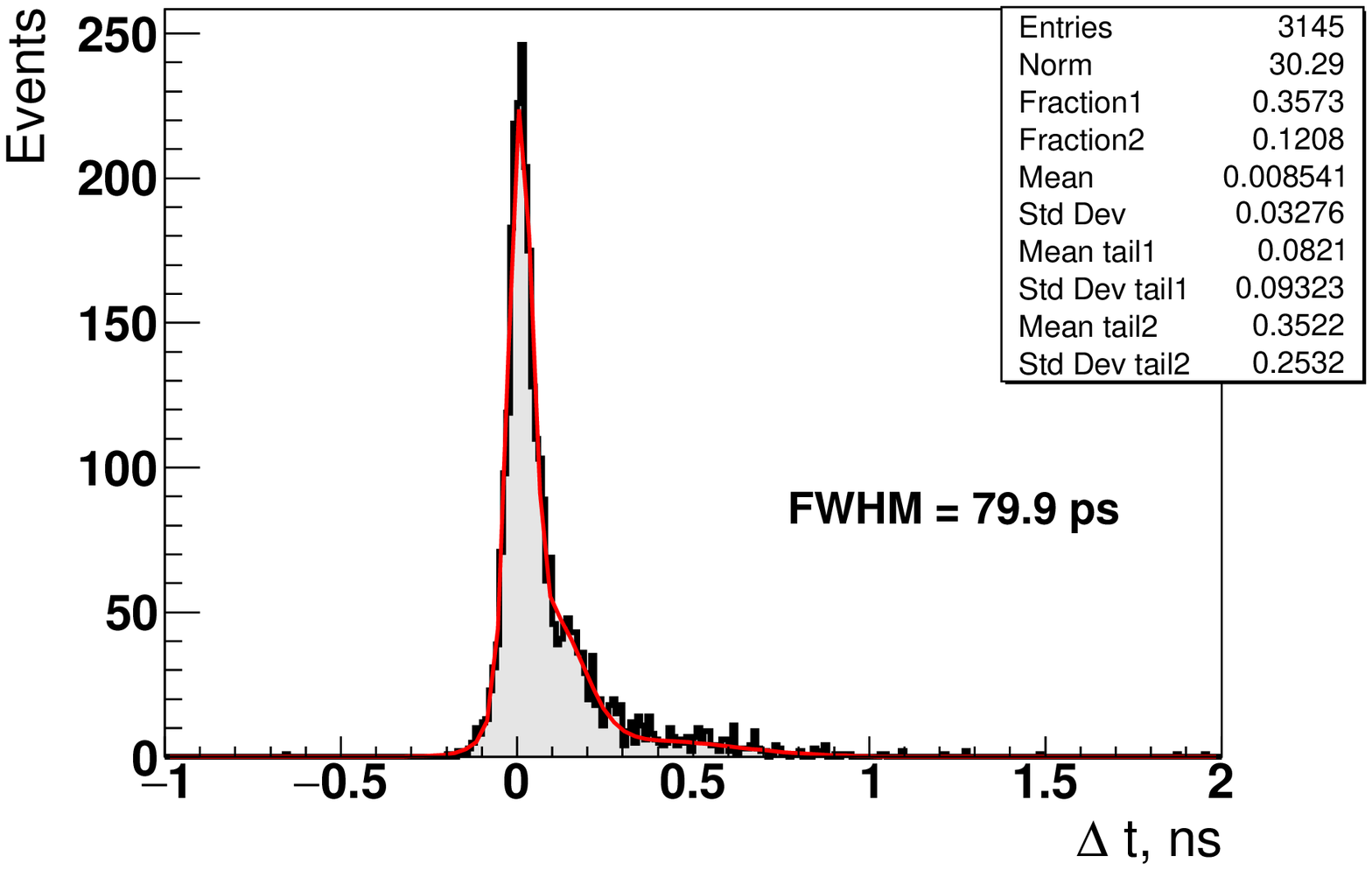}
\caption{Measurement (in black) and fit of Eq. \ref{eq:tts} (in red) of the time difference between the laser trigger time for a pulse located at coordinates (X,Y)=(49.0 mm, 63.2 mm) and the signal time.} \label{tts_photek}
\end{subfigure}
\hspace*{\fill} 
\begin{subfigure}[t]{0.235\textwidth}
\includegraphics[width=\linewidth]{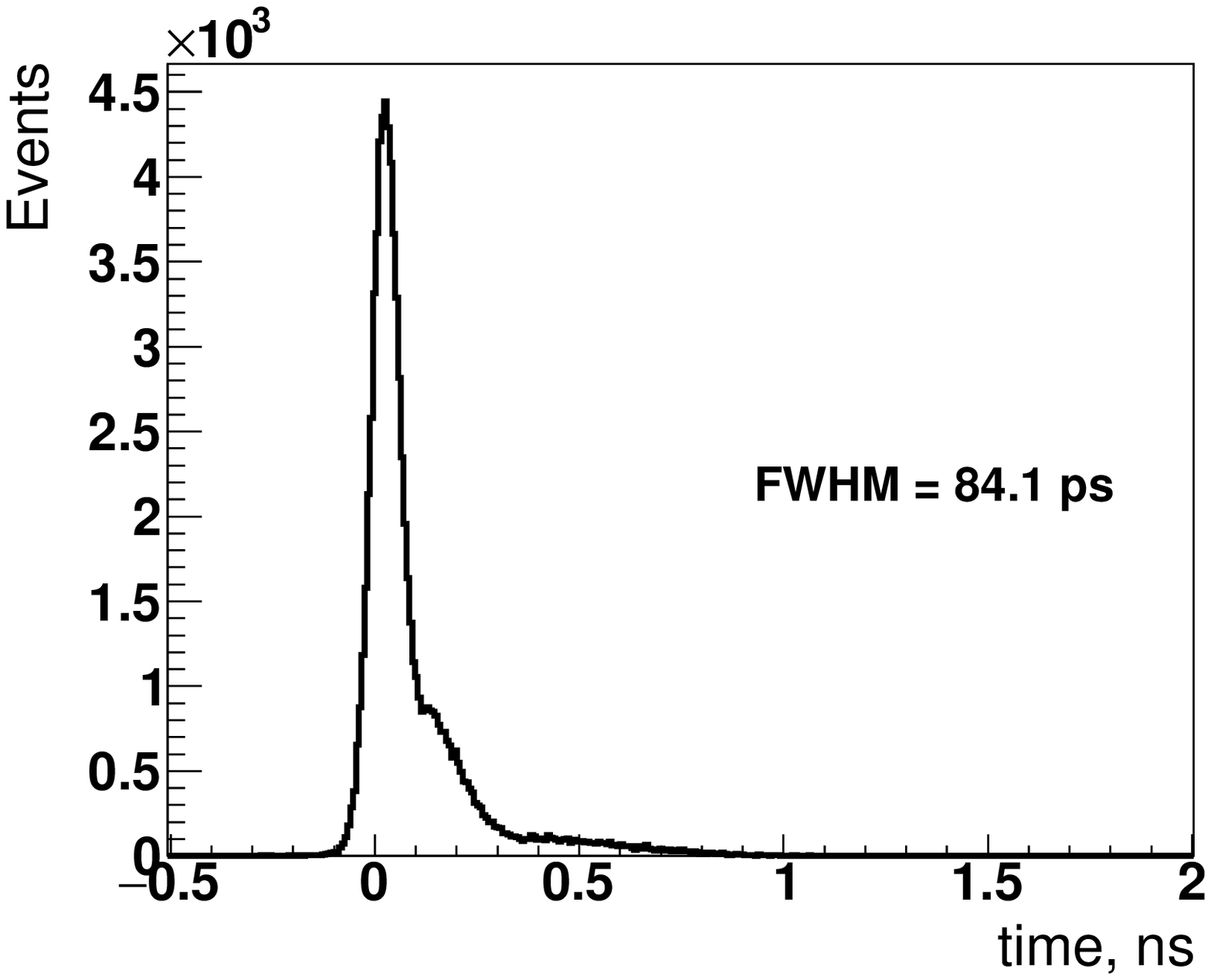}
\caption{Simulated time difference between photoelectron collection time and the signal time.} \label{tts_sim}
\end{subfigure}
\caption{Time response measured and simulated results.} \label{}
\end{figure}

In this study, we used a constant fraction discriminator (CFD) algorithm with a threshold of 50\% of amplitude to determine the time of a signal. Fig. \ref{sig} shows the typical signal shape read out at both ends of the TL \#27 for different positions of illumination along this line. 
To implement the transit time spread (TTS) of the MCP-PMT, we first measured its time response. Fig. \ref{tts_photek} is an example of the time difference measured between the laser trigger and the signal when the laser is at a fixed position. The distribution has a main peak with a tail corresponding to the backscattered electrons. A triple-Gaussian function, $f(t)$ was used to fit the distribution \cite{Sharyy2021}:
\begin{equation}\label{eq:tts}
\begin{aligned}
f(t) = & \frac{A}{\sqrt{2 \pi}}(\frac{1-f_1-f_2}{\sigma_1}e^{-\frac{1}{2}(\frac{t-t_1}{\sigma_1})^2}\\
& +\frac{f_1}{\sigma_2}e^{-\frac{1}{2}(\frac{t-t_1-t_2}{\sigma_2})^2}+\frac{f_2}{\sigma_3}e^{-\frac{1}{2}(\frac{t-t_1-t_3}{\sigma_3})^2}) \;,
\end{aligned}
\end{equation}
where $A$ is a normalization coefficient, $f_1$, $f_2$ are fractions of events in the second and third Gaussian terms, $t_1$ is the mean of the first term, $t_2$, $t_3$ are the additional delays for the second and third terms, and $\sigma_1$, $\sigma_2$, $\sigma_3$ are the corresponding standard deviations. We fitted the spectrum within [$-0.5$ ns, 2.5 ns] from the laser trigger time. We consider the all time response distributions from the different laser positions. Fig. \ref{tts_sim} shows the time difference between the simulated signal time and the photoelectron detection time using the parameters presented in Table \ref{FitResults}, which result from the fit of Eq. \ref{eq:tts}. The different shapes of the time difference shown in Fig. \ref{tts_photek} comparing to Fig. \ref{tts_sim} results from the fact that Fig. \ref{tts_photek} demonstrates a fixed position, whereas the entire detector surface was used for Fig. \ref{tts_sim}.
\begin{table}[!]
\centering
\caption{Results of the time response fitted with Eq. \ref{eq:tts}.} \label{FitResults}
\begin{tabular}{lccc}
{} & Probability & Mean & SD$^*$\\
\hline
First Gaussian & 62.5 \% & 0.00 ns & 0.033 ns\\
Second Gaussian & 27.6 \% & 0.11 ns & 0.078 ns\\
Third Gaussian & \hspace{1ex}9.9 \% & 0.38 ns & 0.22 ns\\
\hline
\multicolumn{4}{l}{$^*$SD: Standard deviation}
\end{tabular}\\  
\end{table}

\subsubsection{Gain and fluctuations}\label{subsubsection:gain}
\begin{figure}[hbt!]
\centering
\begin{subfigure}[t]{0.235\textwidth}
\includegraphics[width=\linewidth]{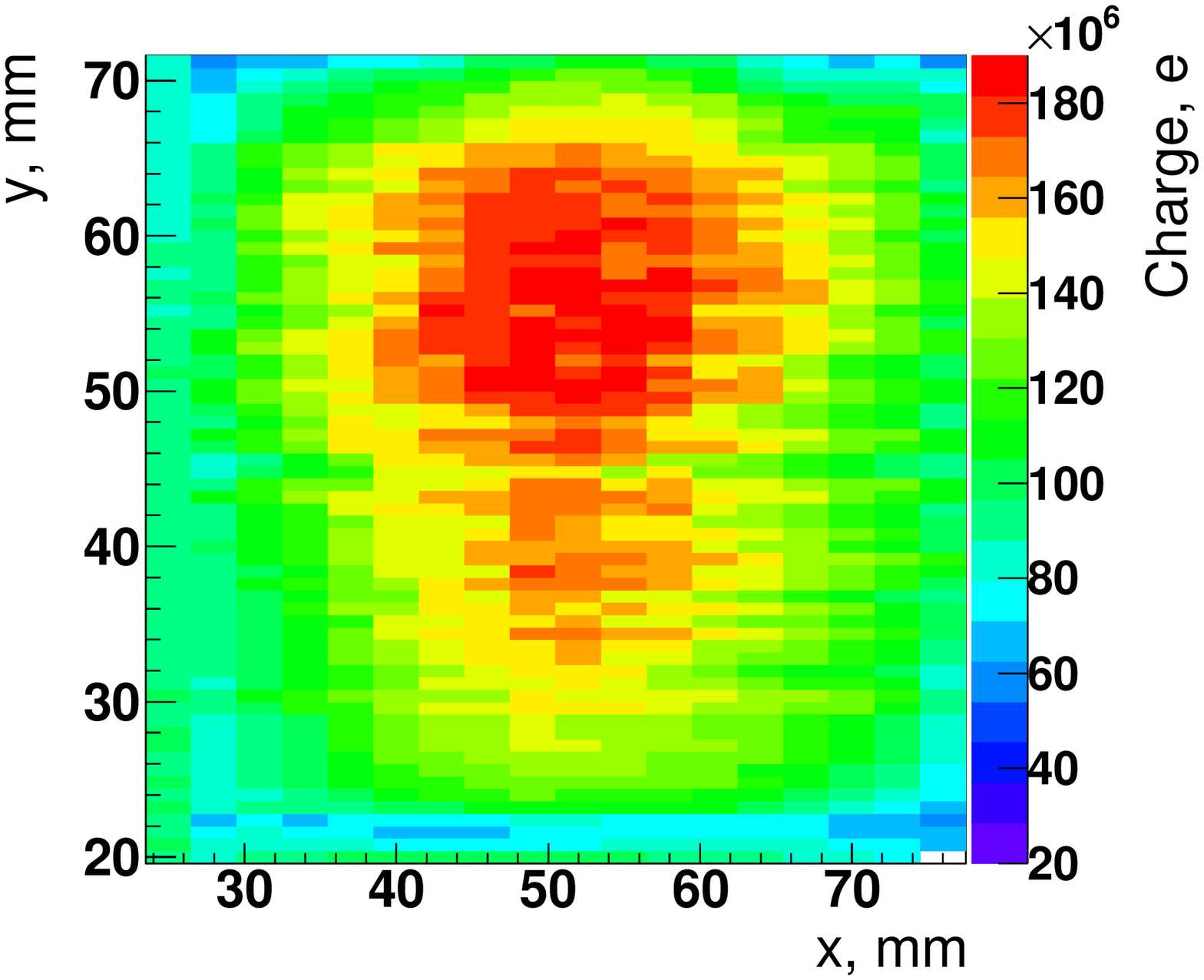}
\caption{} \label{C2d}
\end{subfigure}	
\hspace*{\fill}
\begin{subfigure}[t]{0.235\textwidth}
\includegraphics[width=\linewidth]{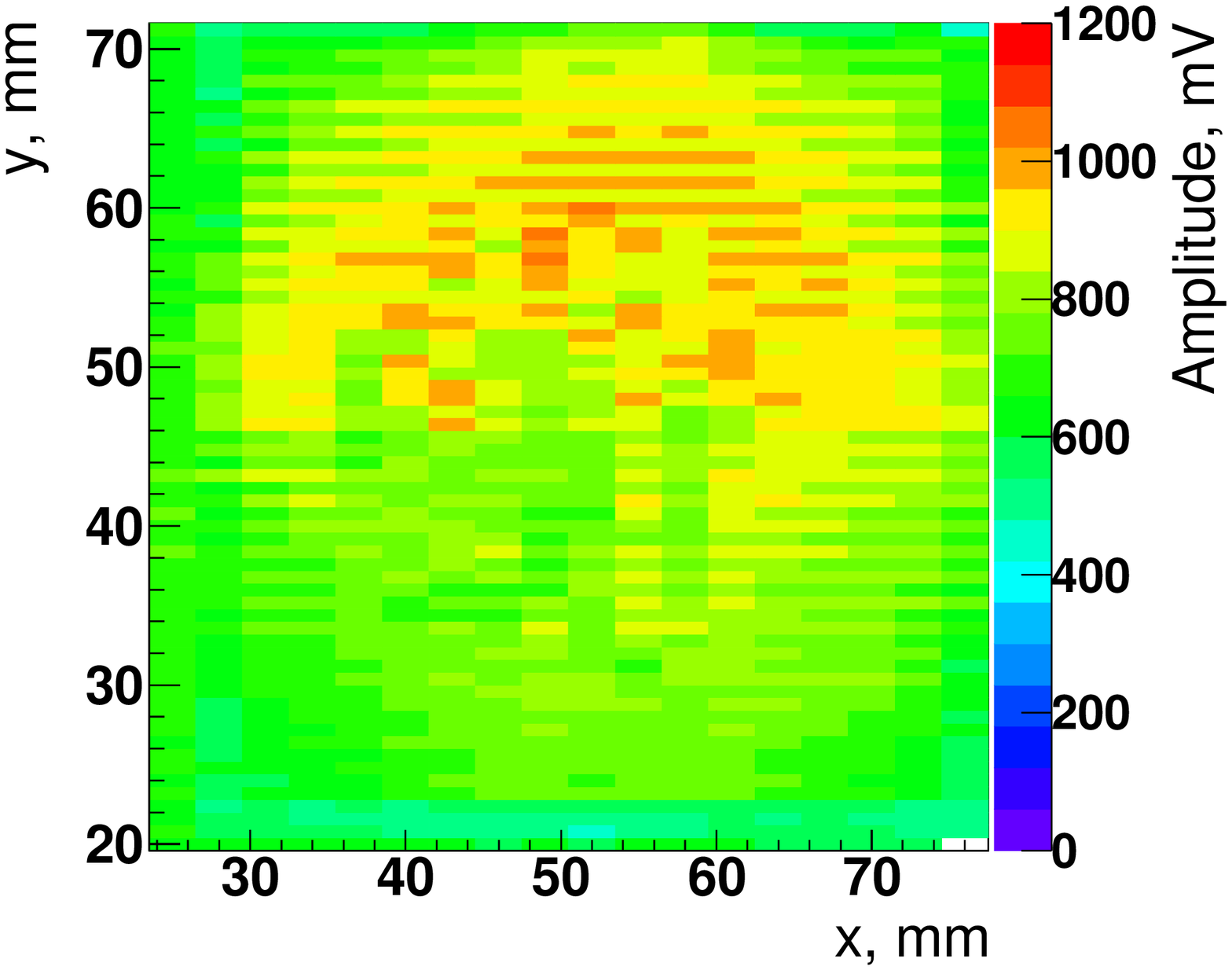}
\caption{} \label{Amp2d}
\end{subfigure}	
\caption{Charge (a) and amplitude of the TL with the maximum value (b) measured over the entire sensitive surface of the detector. The X and Y coordinates correspond to the position of the laser. The detector surface was scanned with steps of 3 mm in X and 0.8 mm in Y.} \label{CA2D}
\end{figure}

\begin{figure}[hbt!]
\centering
\begin{subfigure}[t]{0.235\textwidth}
\includegraphics[width=\linewidth]{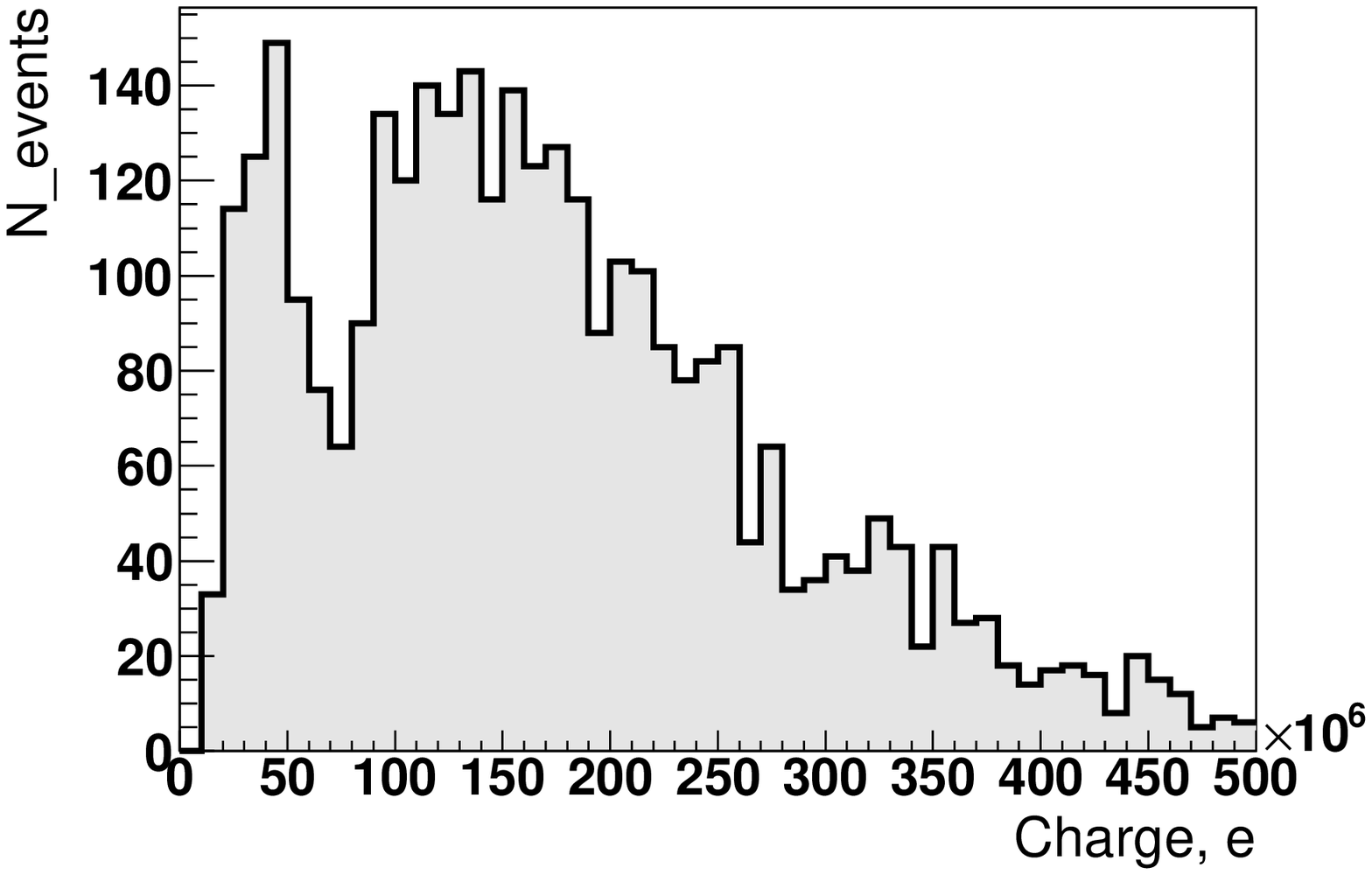}
\caption{Charge at (X,Y) = (49.0 mm, 63.2 mm).} \label{C1d575}
\end{subfigure}	
\hspace*{\fill} 
\begin{subfigure}[t]{0.235\textwidth}
\includegraphics[width=\linewidth]{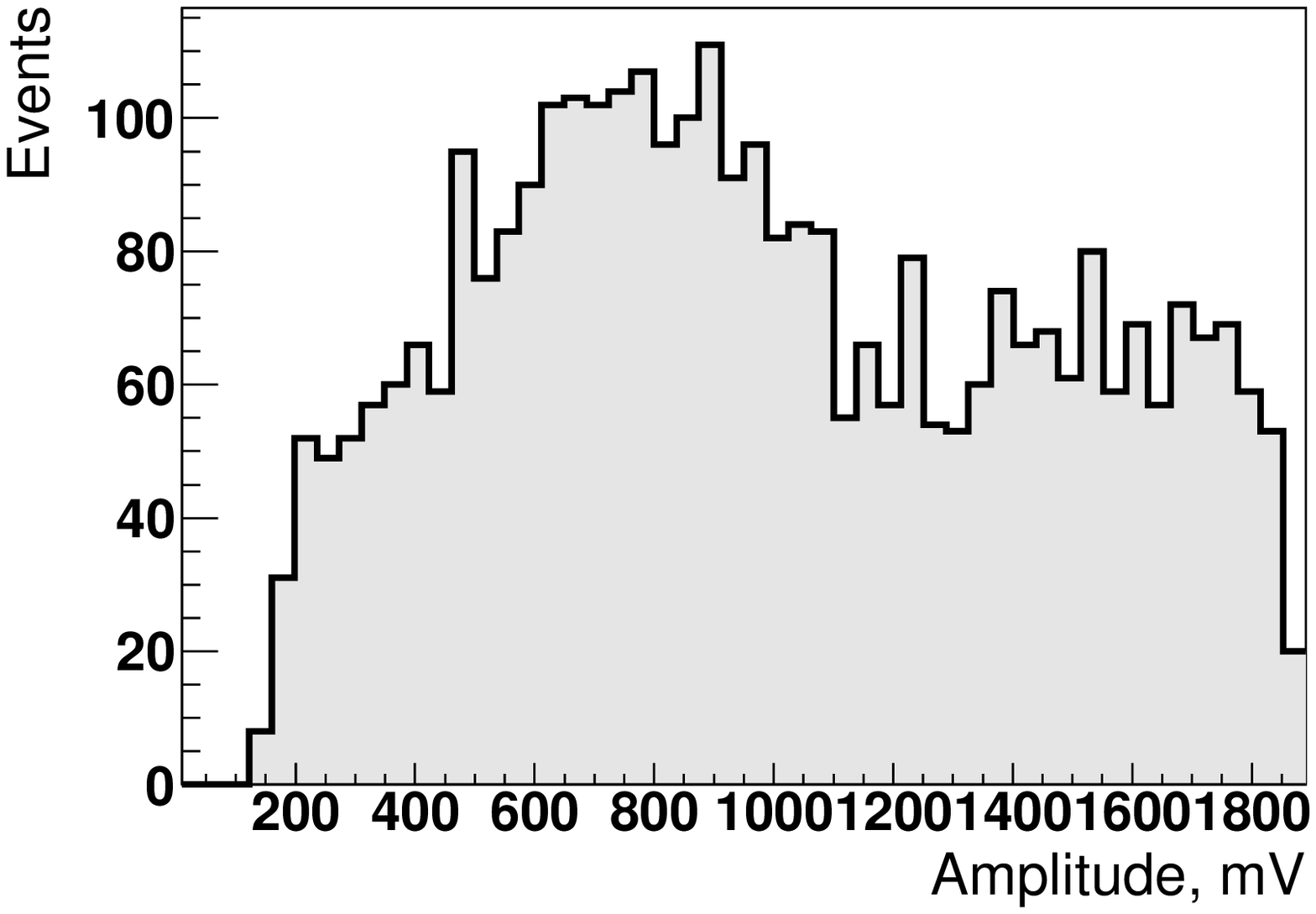}
\caption{Amplitude at (X,Y) = (49.0 mm, 63.2 mm).} \label{Amp1d575}
\end{subfigure}\vskip1ex
\begin{subfigure}[t]{0.235\textwidth}
\includegraphics[width=\linewidth]{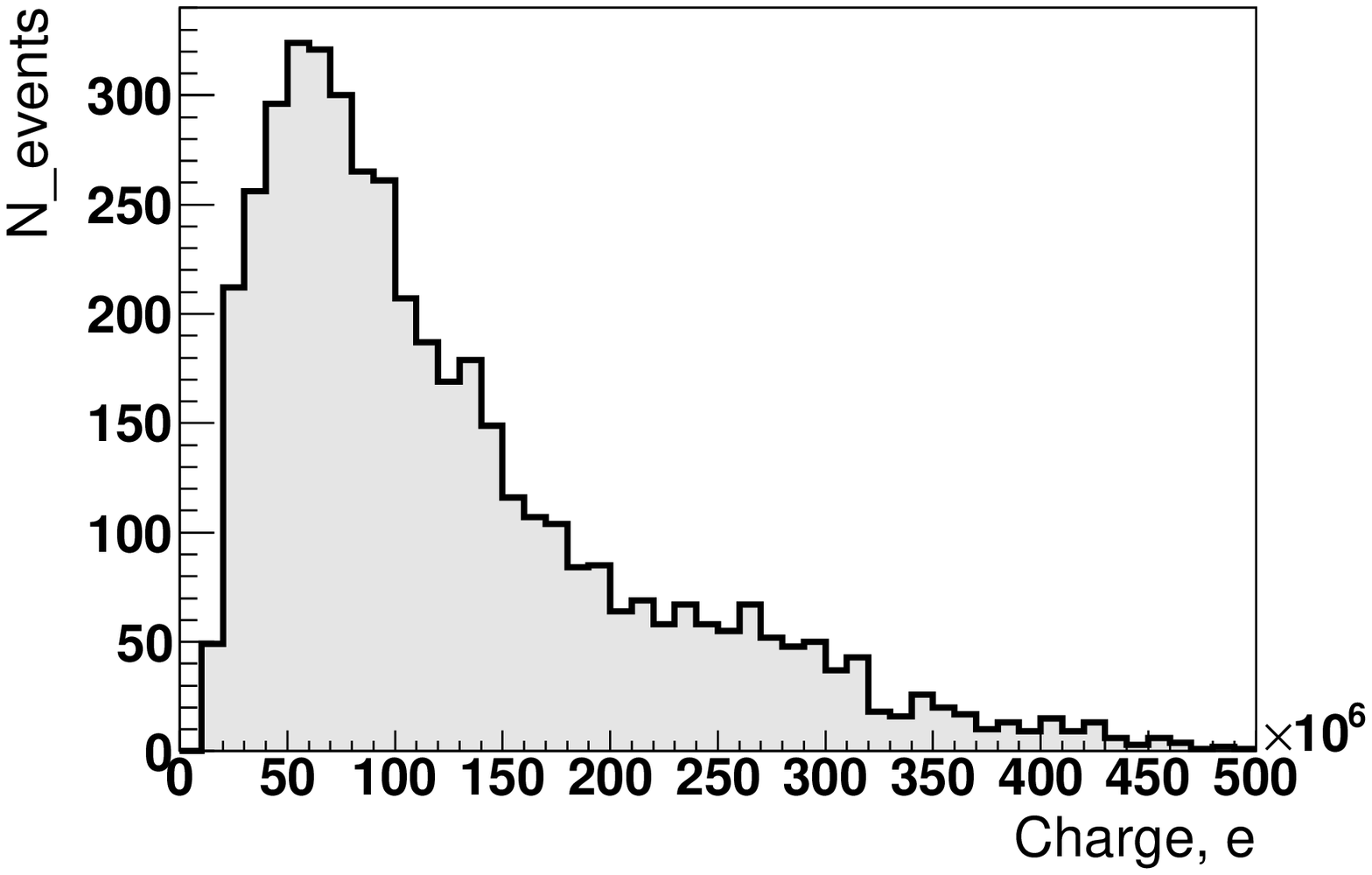}
\caption{Charge at (X,Y) = (70.0 mm, 63.2 mm).} \label{C1d986}
\end{subfigure}	
\hspace*{\fill} 
\begin{subfigure}[t]{0.235\textwidth}
\includegraphics[width=\linewidth]{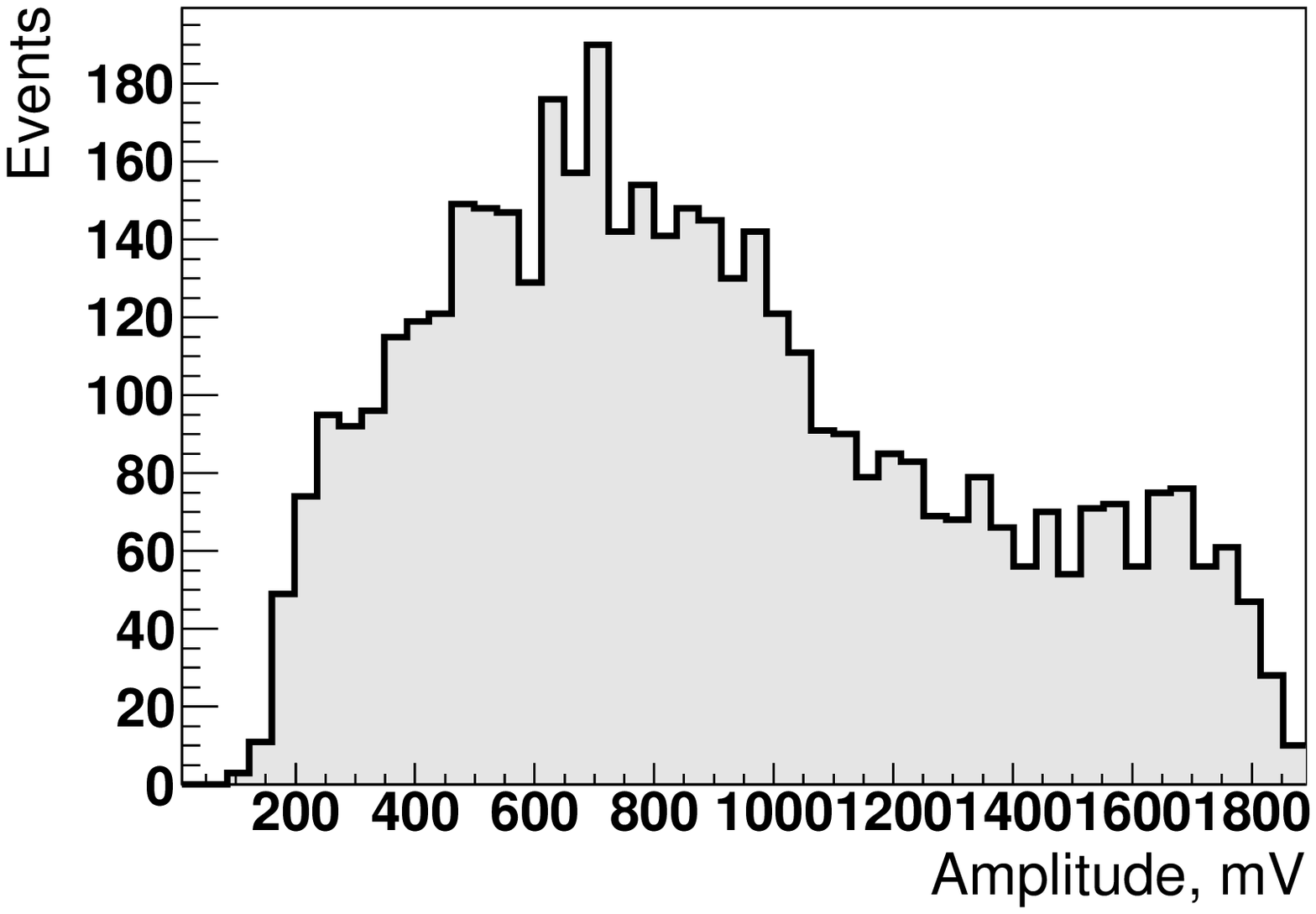}
\caption{Amplitude at (X,Y) = (70.0 mm, 63.2 mm).} \label{Amp1d986}
\end{subfigure}
\caption{Charge and amplitude measurement results.} \label{CAmea}
\end{figure}

To determine the PMT gain, we first observed the charge and amplitude collected from the measurement. Figs. \labelcref{CA2D,CAmea} show 2D and 1D distributions of the charge and amplitude measured over the entire sensitive surface of the detector. The charge is calculated as the integral of the negative part of both the left and the right signals summed for all the lines triggered in the event. The amplitude is determined by the peak value of the signal of the line with the maximum amplitude. The means of charge and amplitude amount $1.8\times 10 ^8$ electrons and $\sim$900 mV in the center of the MCP-PMT after the amplifiers. The nonuniformity was caused by the contact between each layer of the detector and the border effects. In current simulation, we did not consider the border effect to the gain value. Therefore, we took into account the average value from the different positions. For example, in Figs. \labelcref{C1d575,Amp1d575,C1d986,Amp1d986}, the amplitude and charge distributions for two different positions are shown. The charge and amplitude peak values depend on the MCP-PMT high voltage and can be adjusted in the measurement. Thus, we selected the gain which was reasonably close to the measurements and focus more on adjusting the fluctuations to fit the charge and amplitude distributions.

\begin{figure}[hbt!]
\centering
\begin{subfigure}[t]{0.235\textwidth}
\includegraphics[width=\linewidth]{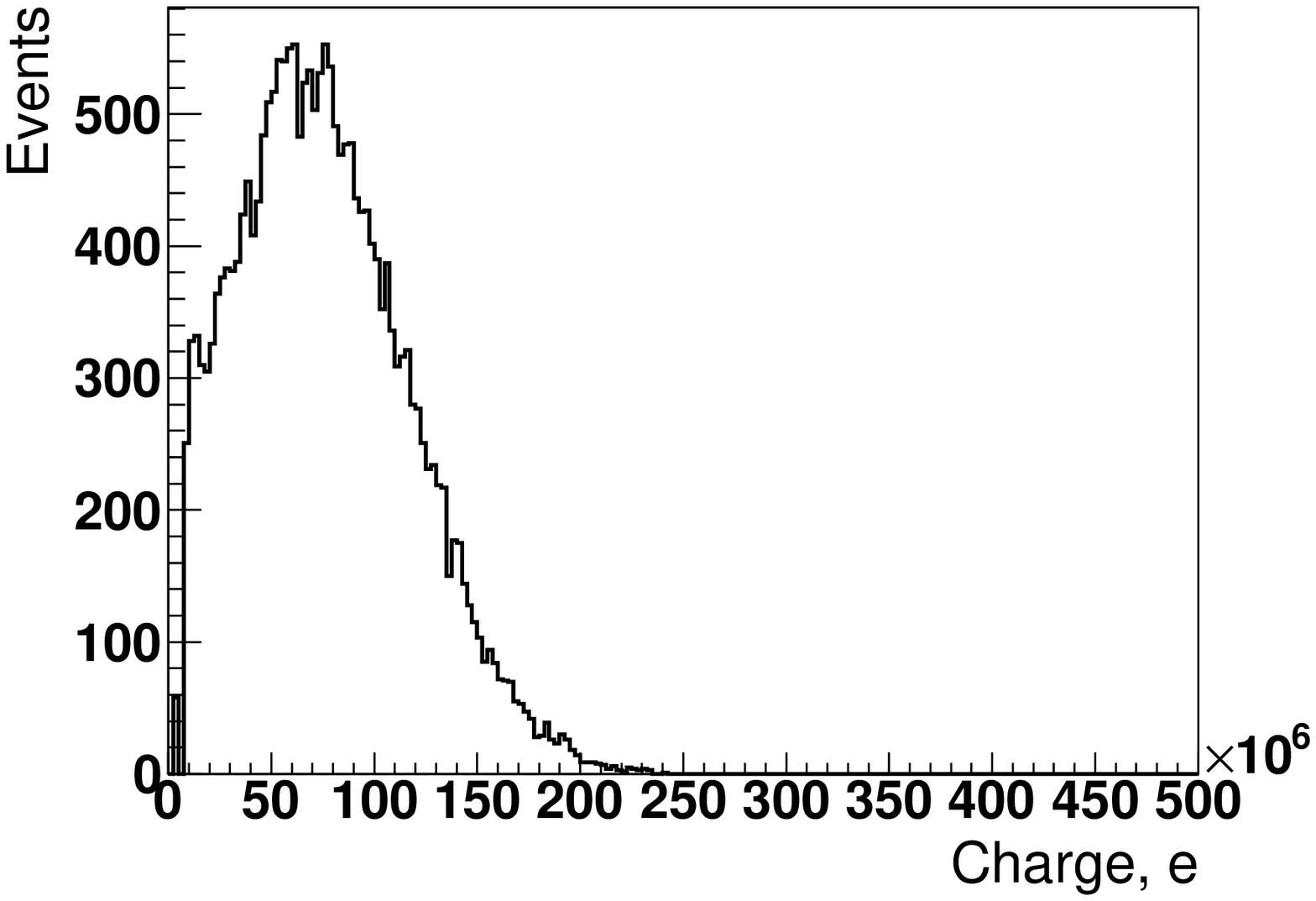}
\caption{Charge distribution.} \label{C1dsim}
\end{subfigure}	
\hspace*{\fill} 
\begin{subfigure}[t]{0.235\textwidth}
\includegraphics[width=\linewidth]{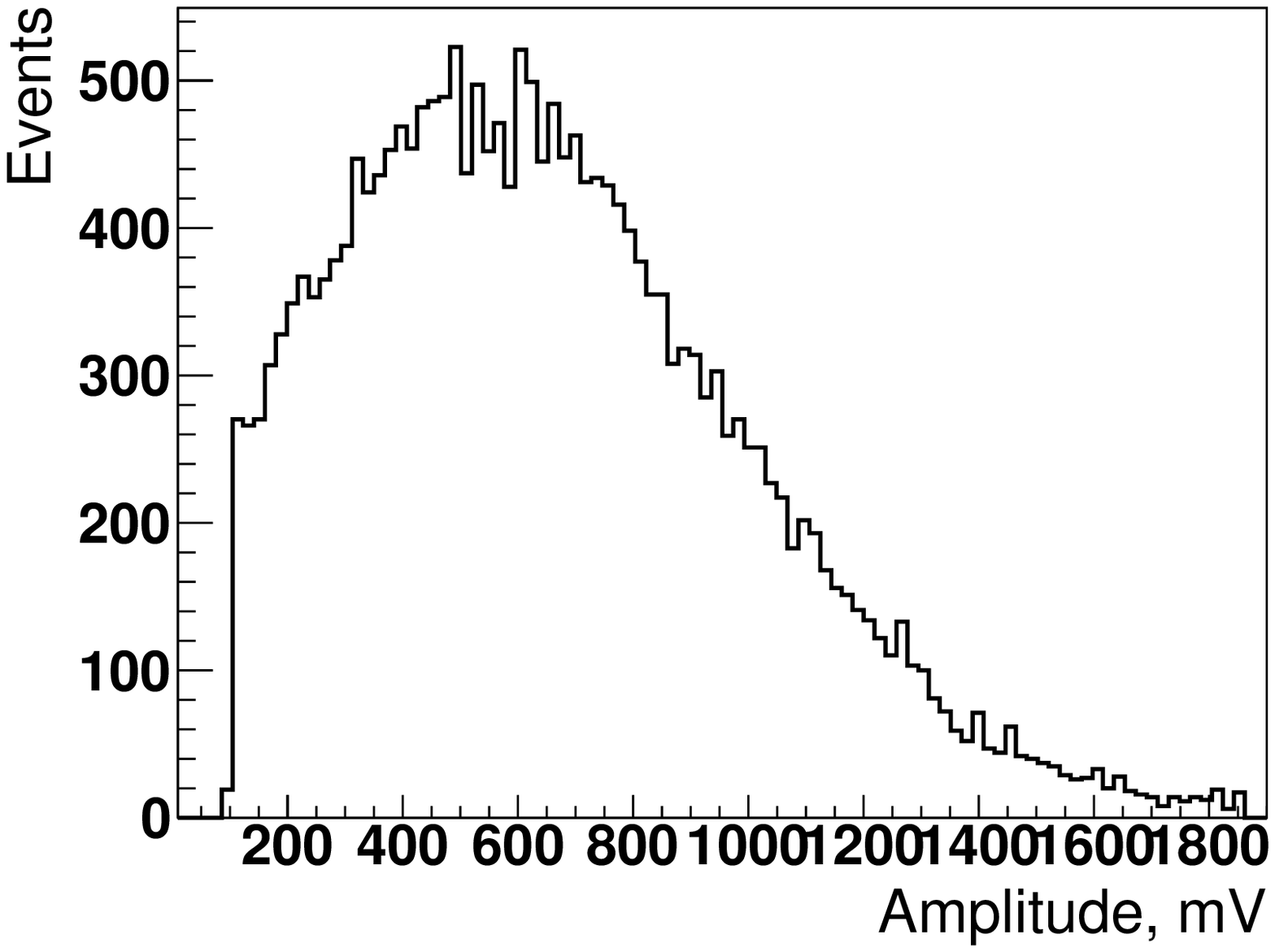}
\caption{Amplitude distribution.} \label{Amp1dsim} 
\end{subfigure}	
\caption{Charge and amplitude simulation results.} \label{CAsim}
\end{figure}

We tuned the gain using a Gaussian model with a mean of 1.125 $\times$ 10$^6$ and a standard deviation of 70\% times the mean value. Fig. \ref{CAsim} shows the one-photon simulation results with such an implementation. The charge and amplitude distributions have similar fluctuations as observed in the measured data.

\subsubsection{Charge sharing}\label{subsubsection:chargesharing}
\begin{figure*}[hbt!]
\centering
\begin{subfigure}[t]{0.31\textwidth}
\includegraphics[width=\linewidth]{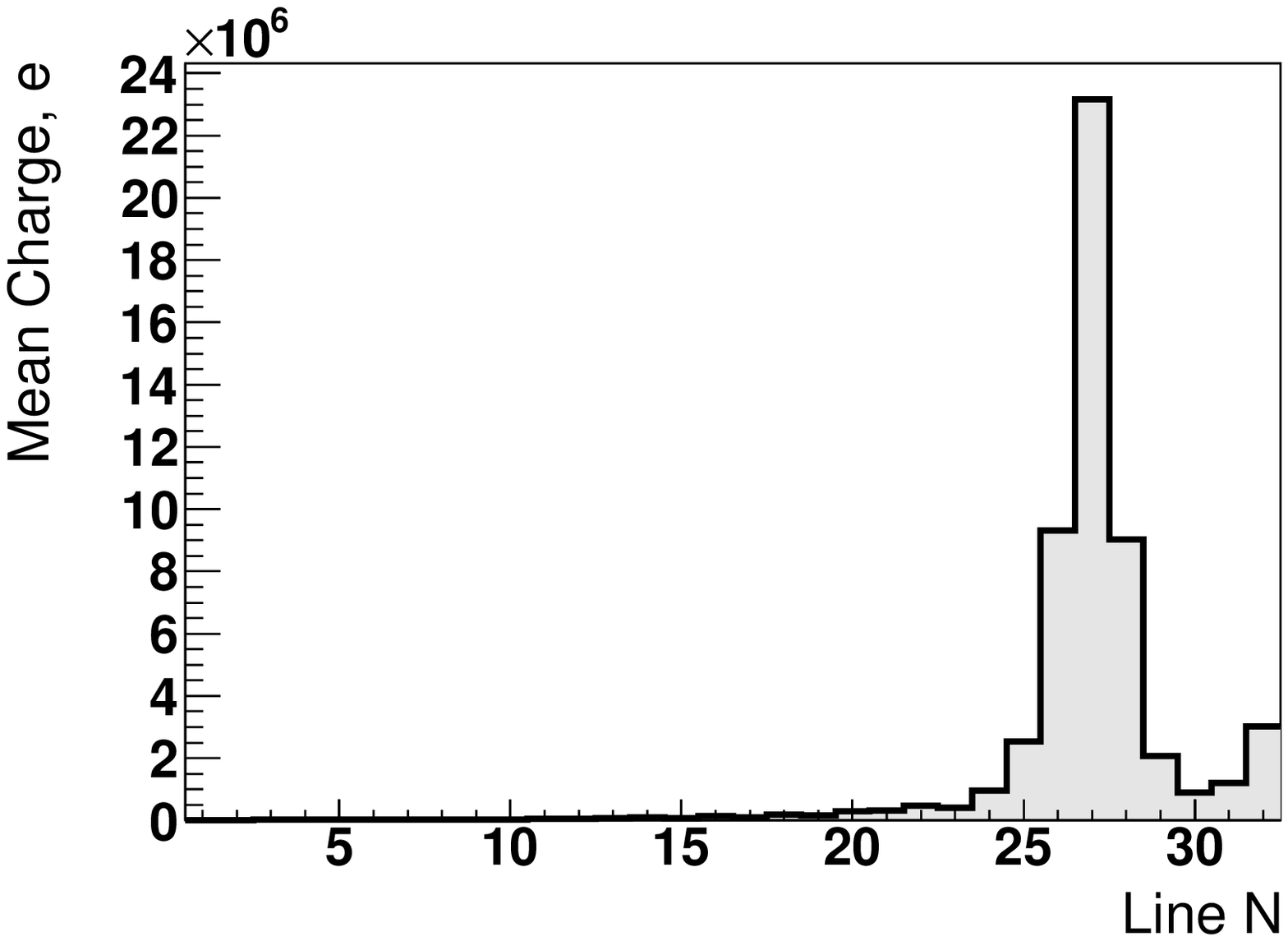}
\caption{Average measured charge on TLs for the laser at laser position (X,Y) = (49.0 mm, 63.2 mm).} \label{Coverline_mea575}
\end{subfigure}	
\hspace*{\fill}
\begin{subfigure}[t]{0.31\textwidth}
\includegraphics[width=\linewidth]{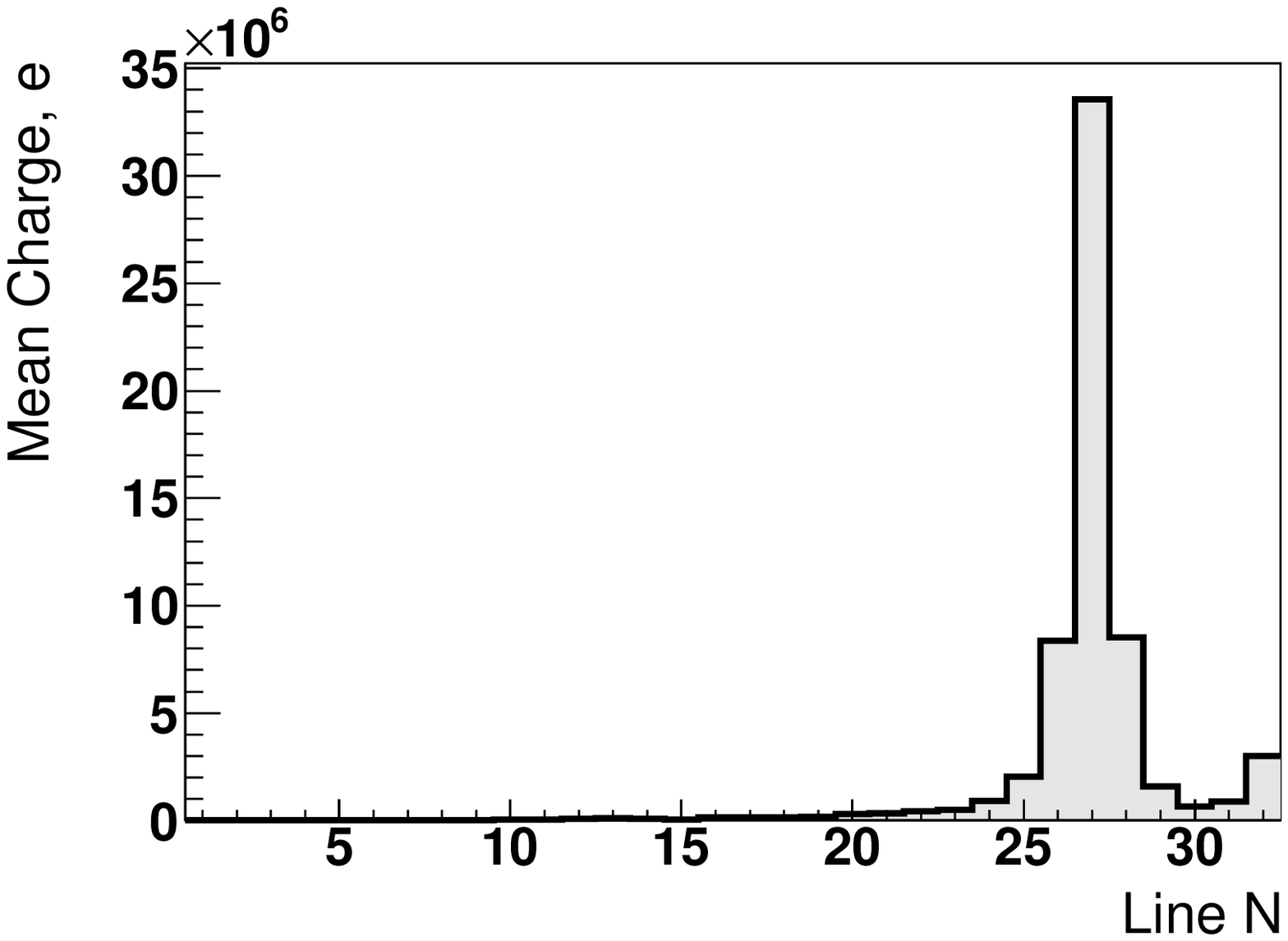}
\caption{Average measured charge on TLs for the laser at laser position (X,Y) = (70.0 mm, 63.2 mm).} \label{Coverline_mea986}
\end{subfigure}
\hspace*{\fill}
\begin{subfigure}[t]{0.31\textwidth}
\includegraphics[width=0.9\linewidth]{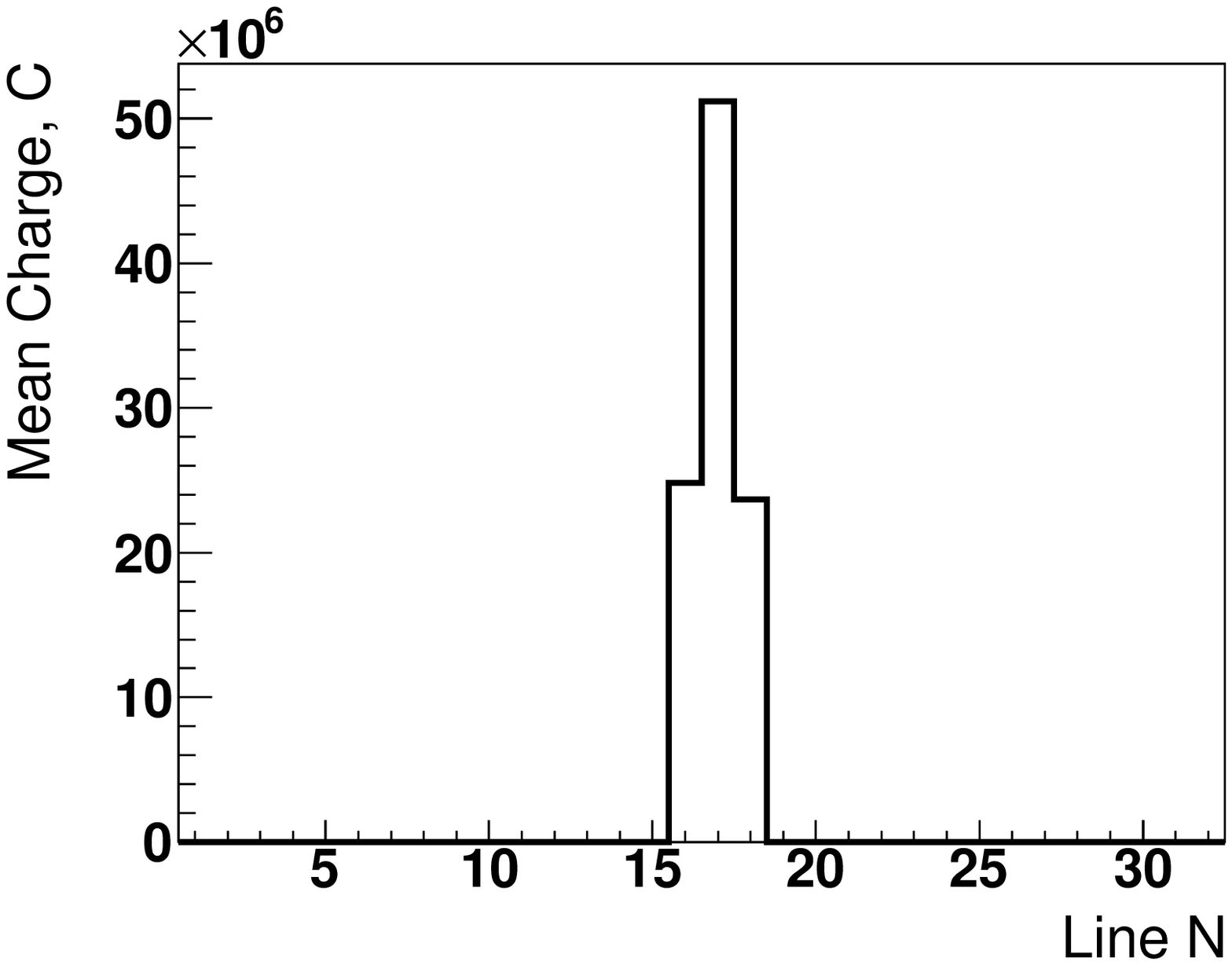}
\caption{Simulated average charge on TLs at photon position (X,Y) = (0.0 mm, 0.0 mm). } \label{Coverline_sim}
\end{subfigure}	
\hspace*{\fill}
\caption{Charge and amplitude on TLs in the one-photon regime.} \label{CAoverline}
\end{figure*}

The electron cloud induces a signal on the anode plane when it drifts from the MCP output to the anode. The typical surface size on which this signal is induced is comparable to the distance between the electron cloud and the anode plane \cite{Fong1967} and makes several mm, hence involving 2-3 lines. The charge density profile, $\sigma_c$ for this distance is modeled by the Gaussian distribution:
\begin{equation} \label{}
\sigma_c = {e^{-\frac{d^2}{2\sigma^2}}}\;,
\end{equation}
where $d$ is the distance in the XY-plane between the position of the initial photoelectron from the photocathode and the closest anode pad center, and $\sigma$ is the standard deviation of the distribution. We tuned $\sigma$ by observing the charge and the amplitude on the TLs in the one-photon regime. Figs. \labelcref{Coverline_mea575,Coverline_mea986} show the measured charge on each TL at different laser positions. We observe that the centerline \#27 has 2 to 3 times higher charge than the neighboring lines \#26 and \#28. The charge ratio between the centerline and the neighboring lines varies due to the different illumination positions depending on the contact between the MCP-PMT and the TL printed circuit board. We chose a value of 0.875 mm for $\sigma$. Fig. \labelcref{Coverline_sim} show the results of the implementation. The centerline has two times higher charge and amplitude than the neighboring lines. We will optimize this value when we will get a better detector performance.

Comparing the simulation against the measurements, we notice that there are tails in the measurements only. We think of two possible causes for this: first, the dark count of the MCP-PMT, and second, the non-Gaussian behavior for the charge sharing that was not simulated with our assumption.

\subsubsection{Signal readout}\label{subsubsection:RO}
\begin{figure*}[hbt!]
\centering
\begin{subfigure}[t]{0.31\textwidth}
\includegraphics[width=\linewidth]{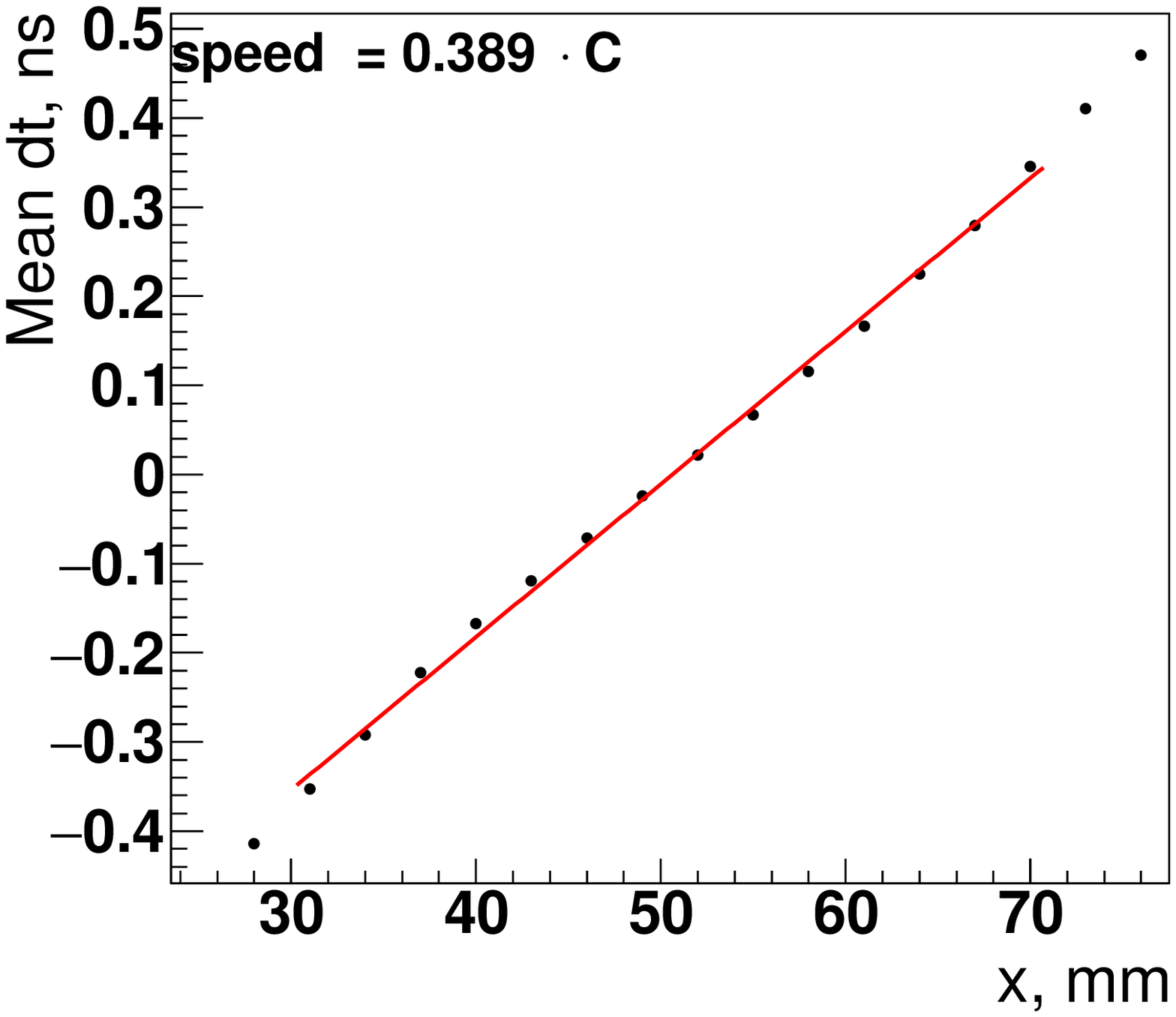}
\caption{Linear fit of the laser position versus the time difference between both ends of TL \#16.}\label{fit16}
\end{subfigure}
\hspace*{\fill} 
\begin{subfigure}[t]{0.31\textwidth}
\includegraphics[width=\linewidth]{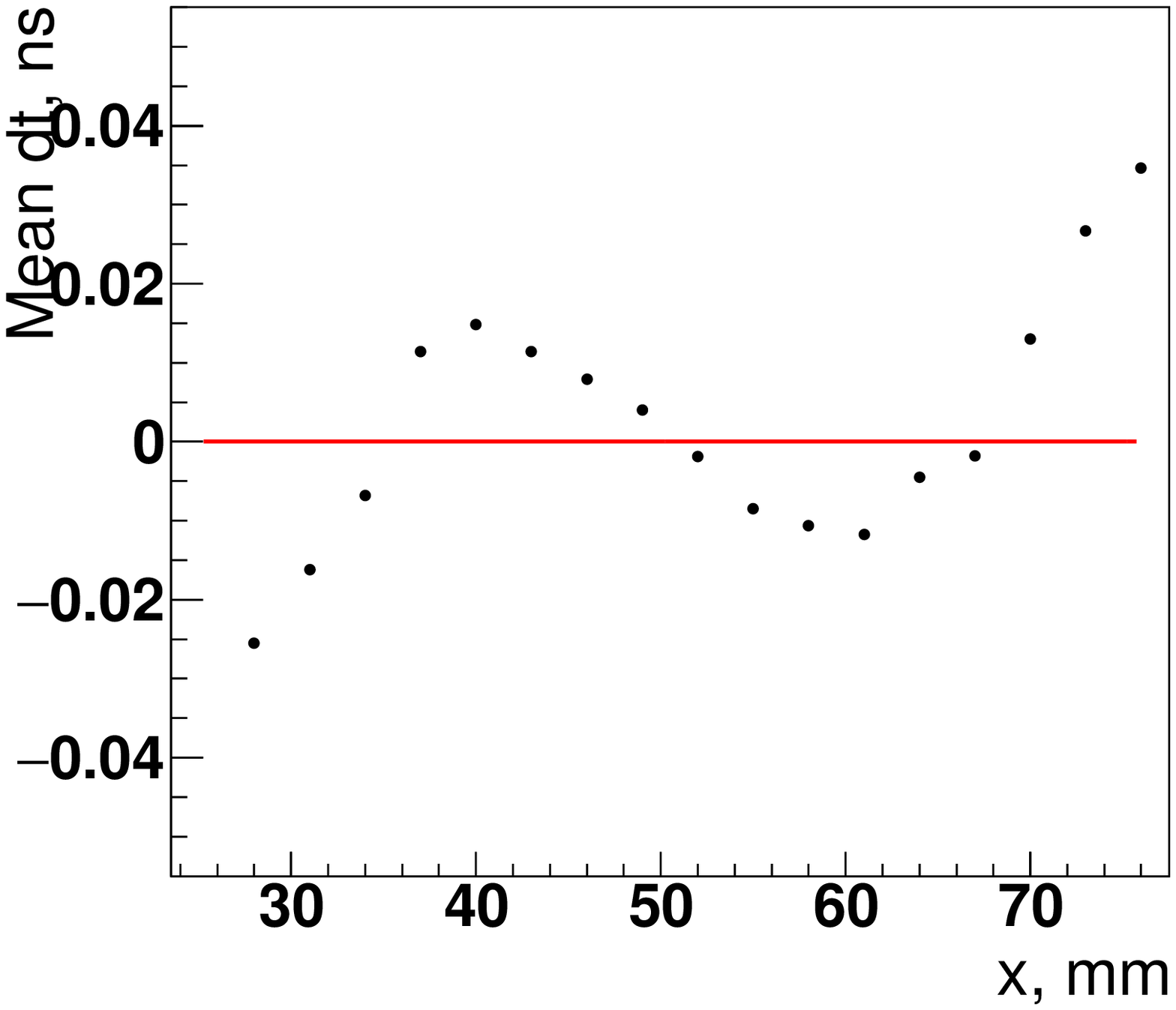}
\caption{Residual of the fit in Fig. \ref{fit16}.}\label{res16}
\end{subfigure}
\hspace*{\fill}
\begin{subfigure}[t]{0.31\textwidth}
\includegraphics[width=\linewidth]{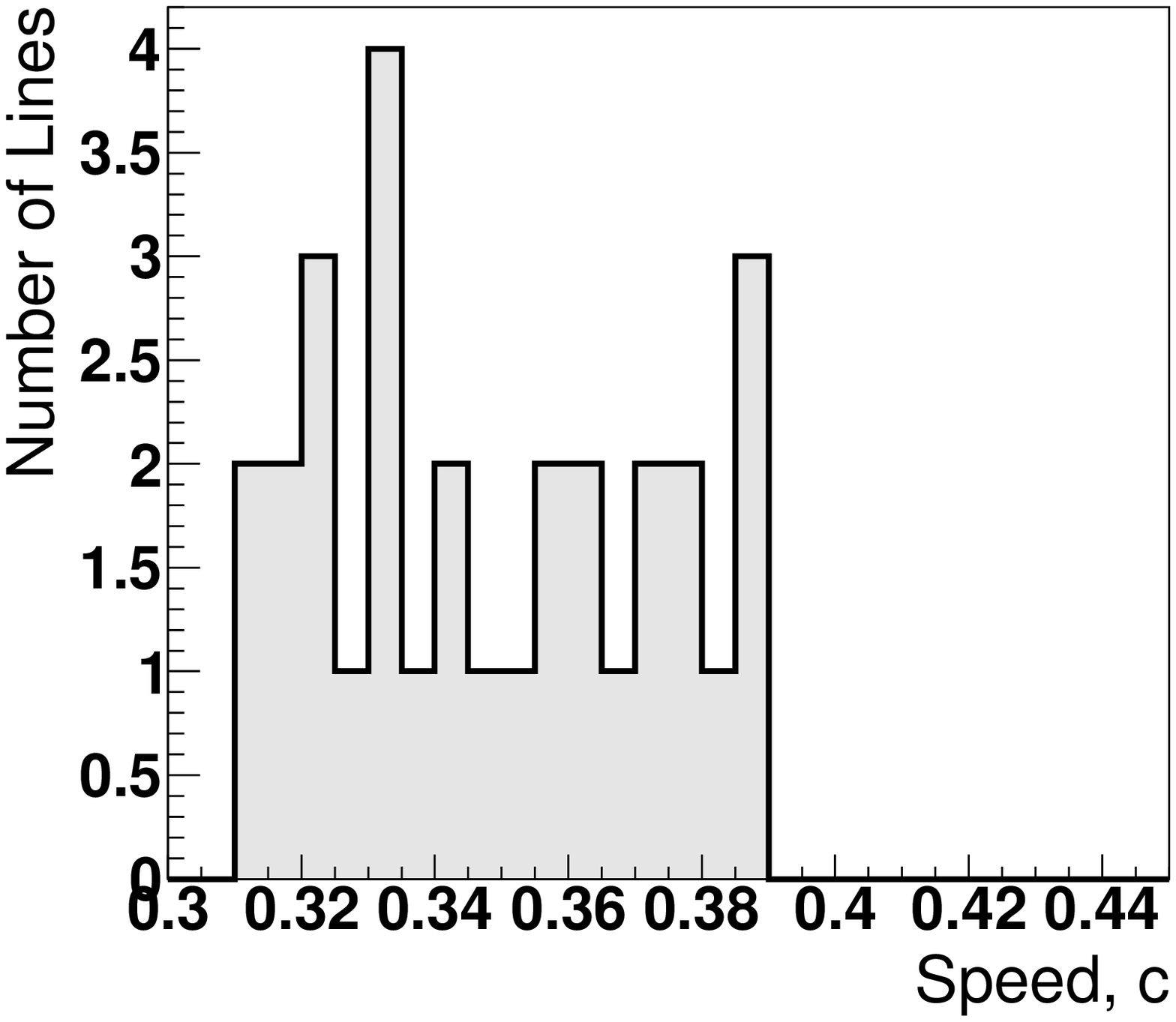}
\caption{Signal propagation speed on TLs in a unit of c (speed of light).}\label{speedall}
\end{subfigure}
\hspace*{\fill} 
\caption{Signal propagation calibration.} \label{}
\end{figure*}

The PMT has 4096 individual anode pads arranged in a 64 $\times$ 64 pattern with 0.828 mm pitch, resulting in a 53 $\times$ 53 mm$^2$ readout area. The signals were split into two equal parts that and propagated to the left and the right end of the TLs. Fig. \labelcref{fit16} shows an example of the correlation between the laser position and the time difference between both ends of TL \#16. We made a linear fit of such a dependence and calculated the signal propagation speed on each line. Fig. \ref{speedall} shows the speed on all the lines of the MAPMT253 detector. In the model, we thus assume 35\% of the speed of light as the signal propagation speed.

\subsubsection{Signal shape}\label{subsubsection:signal}
Signal shapes are important in this study because the event reconstruction algorithm is based on either the information extracted from the signal or the signal shape itself. In addition, only a realistic signal shape can simulate the superposition of the photoelectrons accurately. We noticed the dependence of the laser positions on the signal shape. Fig. \ref{sig_c} shows the registered signals at the left and right ends of TL \#27 when the laser is positioned at the center of the detector. The difference in the arrival times is small because the distances from the laser position to the ends of the TL are similar. There are two main peaks on both channels, one has a higher amplitude and the other one comes later with a lower amplitude. When the laser is on the left (Fig. \ref{sig_l}), the left channel receives the signal first. The registered signal shape still has two main peaks. However, the signal on the right channel has only one main peak. The opposite behavior appears when the laser is on the right side of the detector. We consider that the second peak results from the reflection that happens from the connectors due to the impedance mismatch. In order to simulate a realistic signal shape, including the dependence of the detector position, we first fitted the measured signals with the following models.
We simulated the first peak, so called main peak, with function $f(t)$:
\begin{equation} \label{}
f(t)=-e^{\frac{\;-t^2\;}{2\sigma^2}}+ a \cdot [b+ \tanh(c t)] \cdot e^{\frac{-t}{\tau}} \;,
\end{equation}
where $\sigma$ is the standard deviation for the main signal, $a$, $b$, and $c$ are the coefficients to adjust the hyperbolic tangent model of the signal rebound, and $\tau$ is the time constant for the relaxation. Next, we assumed that the reflection peak, defined by the function $g(t,dt)$, has a similar shape, including the rebound and the relaxation, but with a different standard deviation than the main peak:
\begin{equation} \label{}
g(t,dt)=-e^{-0.5 (t+dt)^2/ \;\sigma'^2}+ a \cdot [b+ \tanh(c (t+dt))] \cdot e^{-(t+dt)/\tau} \;,
\end{equation}
where $\sigma'$ is the standard deviation of the reflection peak, $a$, $b$, $c$, and $\tau$ are the same with the main peak, and $dt$ is a delay time depending on the signal induced position and the signal propagation speed. Finally, we combined the two functions and fit the measured signals with the function $F(t,dt)$:
\begin{equation} \label{}
F(t,dt)=\alpha[f(t)+ \beta \cdot g(t,dt) ] \;,
\end{equation}
where $\alpha$ is the signal amplitude, and $\beta$ is the ratio of the amplitude between the reflection peak and the main peak.

We fitted the signal shape for different laser positions (Table \ref{fitsig}), and we came to the conclusion that (i) the amplitude of the reflection peak is 25\% of the main peak, and (ii) the width of the reflection peak is slightly wider than for the main peak. 
\begin{table*}[!b]
\centering
\caption{Signal shape fit results.} \label{fitsig}
\begin{tabular}{C{1cm}C{1.5cm}C{1.5cm}C{2cm}C{1cm}C{1cm}C{1cm}C{1cm}C{1cm}}
$\beta$& $\sigma$ & $\sigma'$ & $dt$ & $a$ & $b$ & $c$ & $\tau$\\
\hline
25\% & 0.23 ns & 0.3 ns & 0.5 - 1.5 ns & 5\% & 1 & 10 & 10 ns\\
\hline
\end{tabular}
\end{table*}
The time delay $dt$ is between 0.5 ns to 1.5 ns, i.e., when the photoelectron is located on the edge of the detector, the closer end of the TL would get the signal with a time delay 1.5 ns between the main peak and the reflection peak. On the other channel, it would get a signal with a 0.5 ns delay time. It means that the main peak and the reflection peak would merge into one single peak like the signal from the right end of the TL displayed in Fig. \ref{sig_l} or the left end of the TL represented in Fig. \ref{sig_r}.

\begin{figure*}[hbt!]
\centering
\begin{subfigure}[t]{0.31\textwidth}
\includegraphics[width=\linewidth]{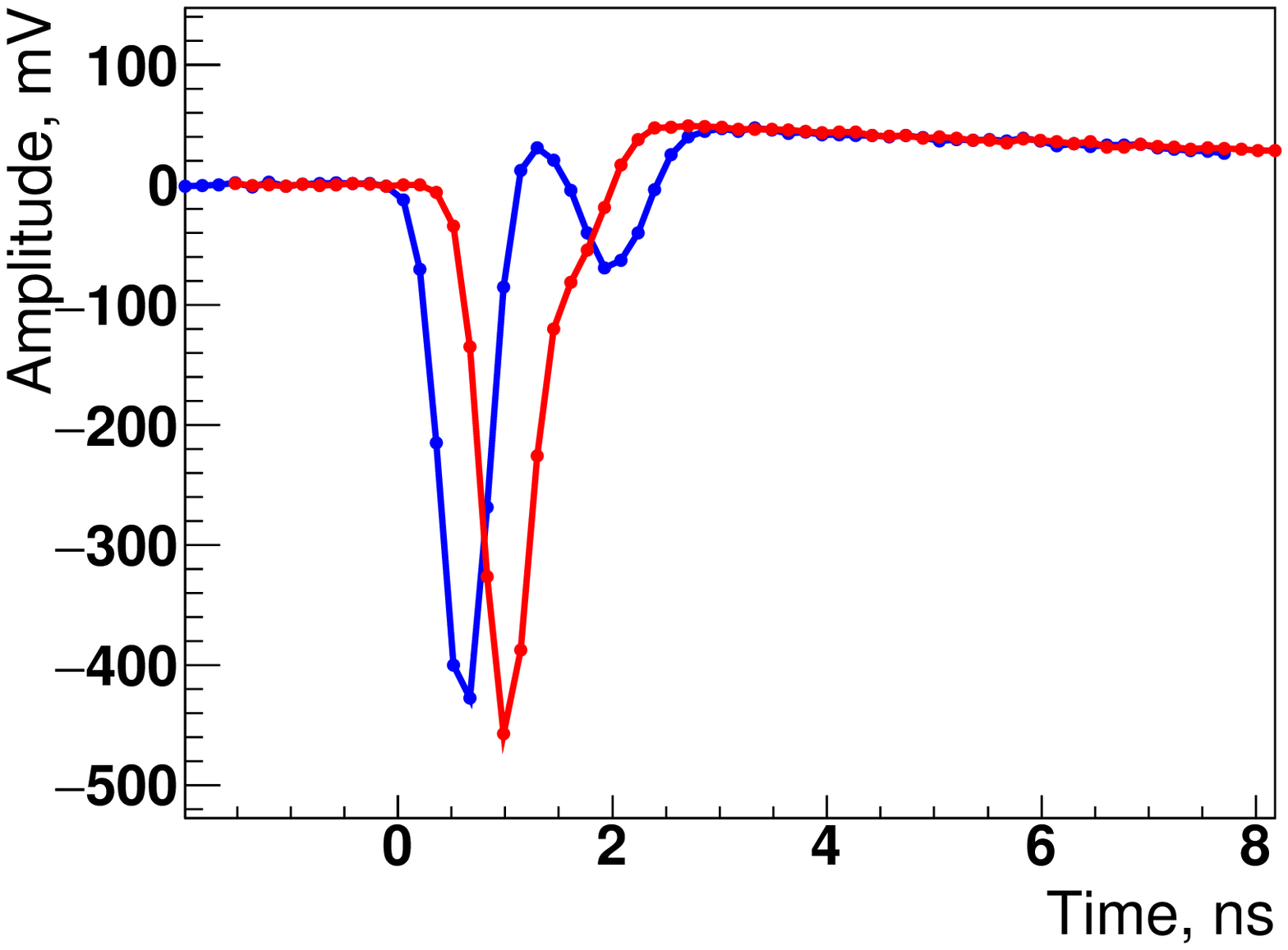}
\caption{Laser positioned $-20$ mm from the center of the detector.}\label{signal_l}
\end{subfigure}
\hspace*{\fill} 
\begin{subfigure}[t]{0.31\textwidth}
\includegraphics[width=\linewidth]{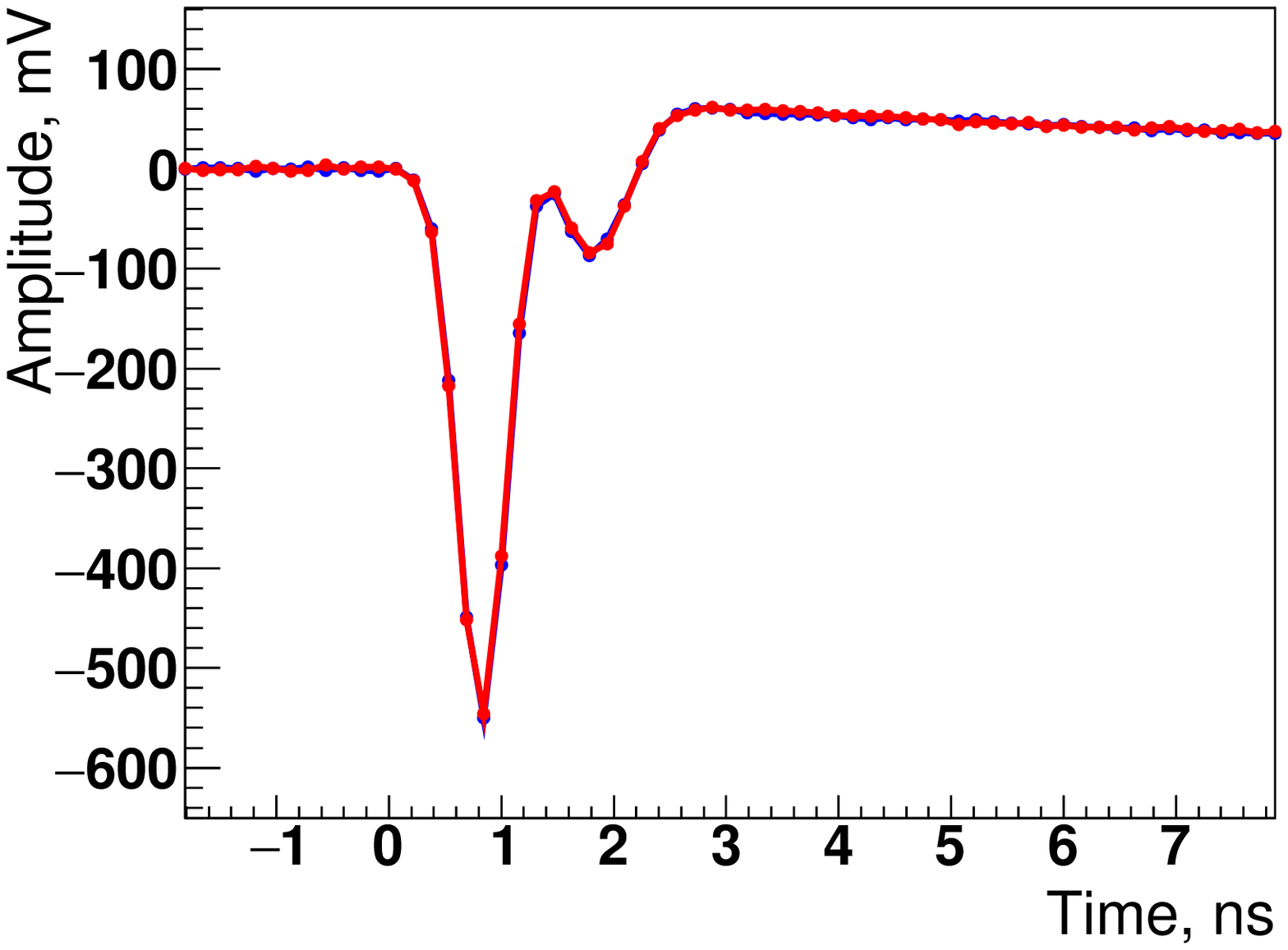}
\caption{Laser positioned at the center of the detector.}\label{signal_c}
\end{subfigure}
\hspace*{\fill} 
\begin{subfigure}[t]{0.31\textwidth}
\includegraphics[width=\linewidth]{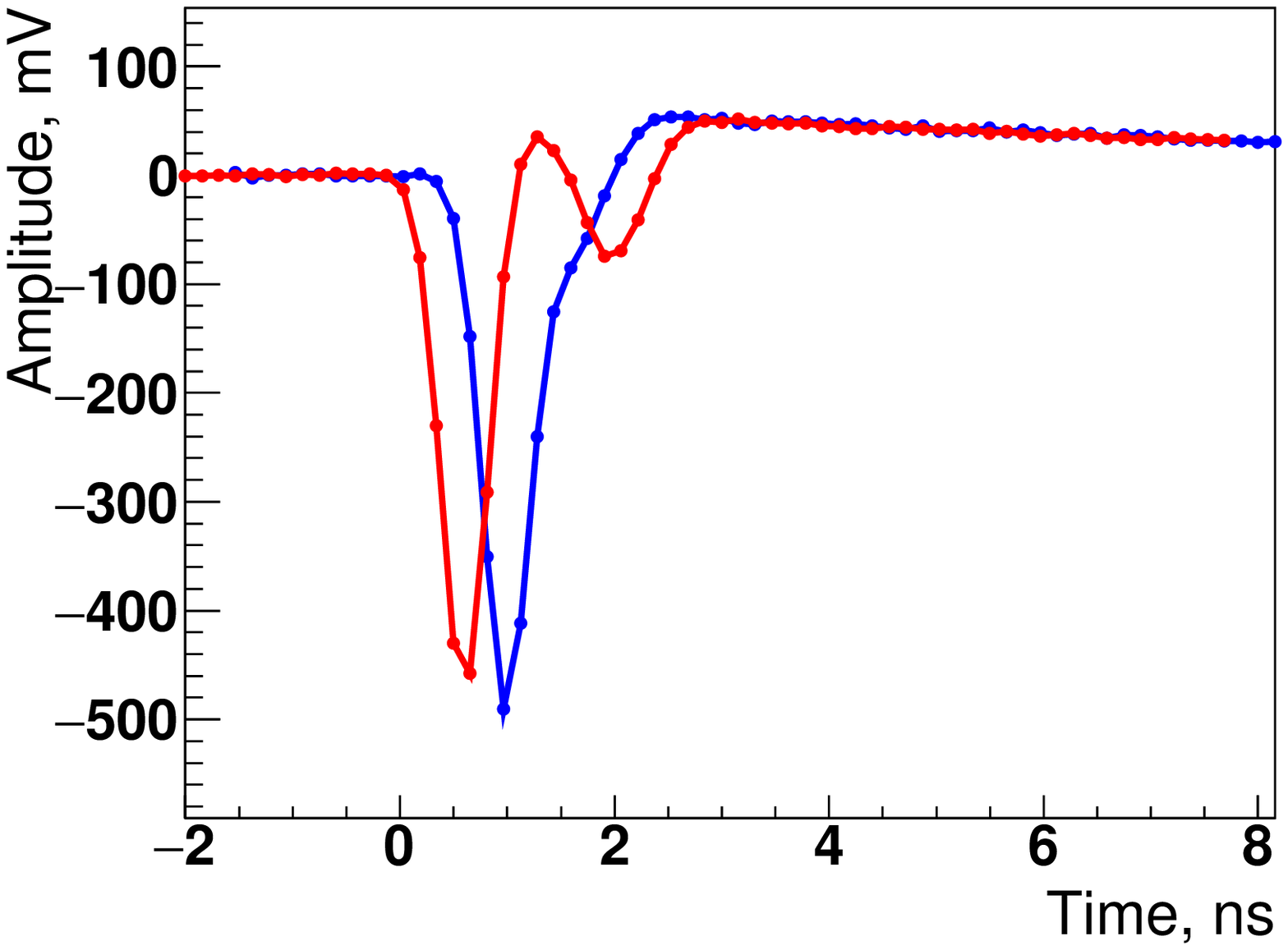}
\caption{Laser positioned at $+20$ mm from the center of the detector.} \label{signal_r}
\end{subfigure}
\hspace*{\fill} 
\caption{Simulated signals at the left (in blue) and right (in red) ends of a TL.} \label{sig_sim}
\end{figure*}

Fig. \ref{sig_sim} shows the implementation of the signal shape in the simulation. The simulation can generate a signal shape similar to the measured one and adjust the time delay for the reflection as a function of the position of the photoelectron along the line. At this phase of simulation, we did not try to simulate the amplifier ringing as observed in Fig. \ref{sig} to limit the complexity. The effect on the event reconstruction will be evaluated with the measurement data.

\subsubsection{Signal digitization}\label{subsubsection:sampic}
\begin{figure}[hbt!]
\centering
\includegraphics[width=0.7\linewidth]{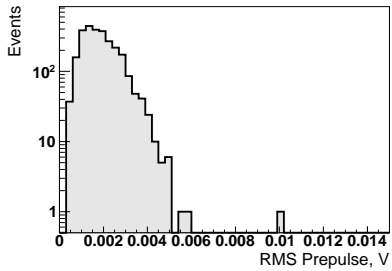}
\caption{Standard deviation calculated using 8 first samples of the signals at the laser position (X,Y)=(49.0,63.2).} \label{noise}
\end{figure}

The SAMPIC module digitizes the signals collected from both ends of each TL using 63 samples with 0.15625 ns steps. The threshold that triggers the data acquisition was set at 50 mV for each channel. The noise in the experiment was measured using the signal fluctuation around the baseline and included all the electronics contributions, such as amplification, propagation in the cables, and digitization by the SAMPIC module. Fig. \ref{noise} shows the standard deviation of all signals acquired at a fixed laser position.
Consequesntly, we used the peak value, 1.2 mV, as the noise value for our simulation. For this, the noise in the simulation was added to the signal readout process, by adding to each digitization sample a random value normally distributed with a standard deviation of 1.2 mV. 64 amplifiers amplified the signals from both ends with a gain of 70. This gain was also taken into account in the simulation. Finally, the signals were recorded using a sampling time in ns and a sampling amplitude in V.

\section{Simulated performances of the CM detection module prototype}\label{section:CMresults}

\subsection{Single photon spatial resolution}
\begin{figure}[hbt!]
\centering
\begin{subfigure}[t]{0.235\textwidth}
\includegraphics[width=\linewidth]{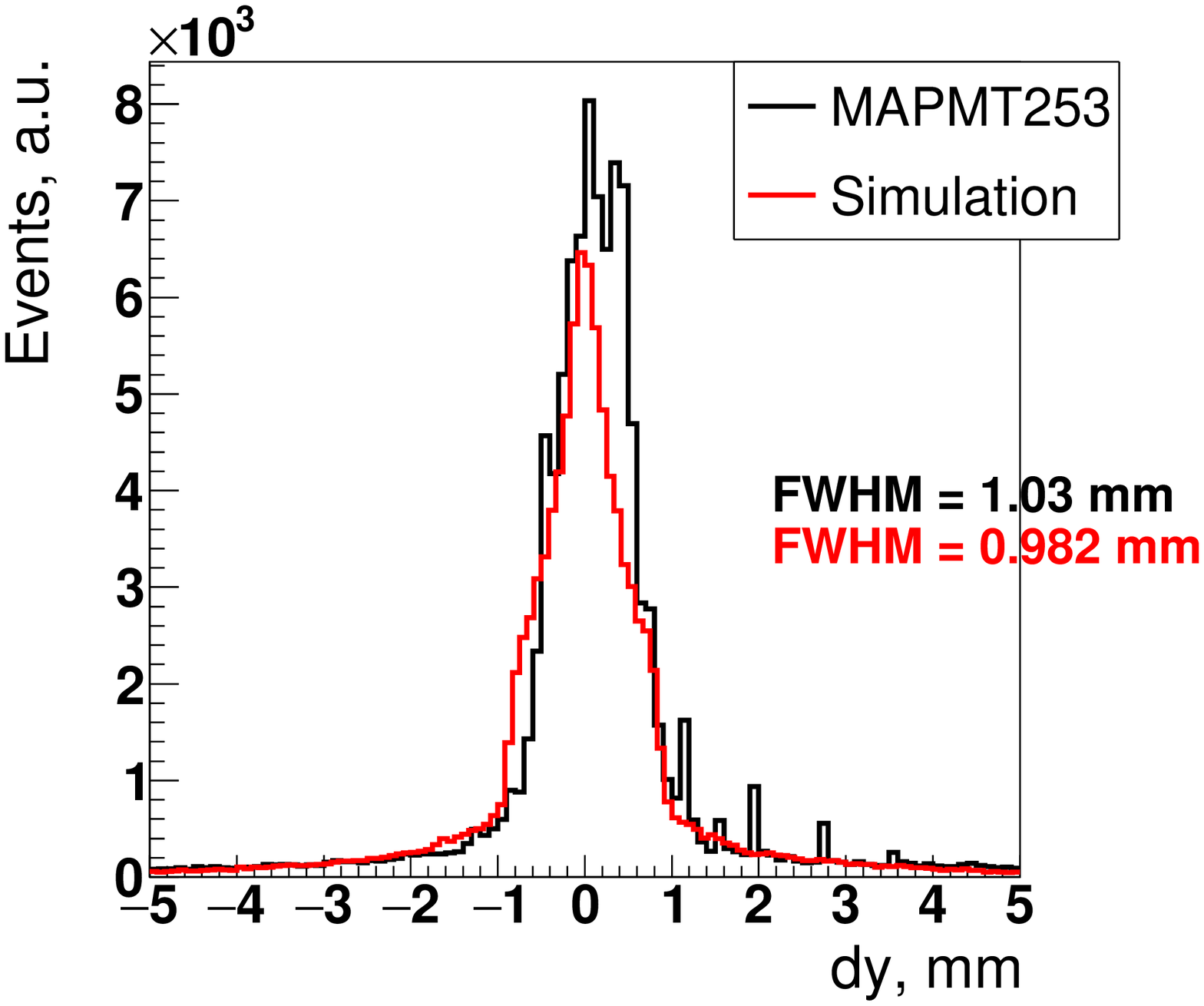}
\caption{Spatial resolution across the TLs. MAPMT253 measurement at the entire detector.}\label{dyboth}
\end{subfigure}
\hspace*{\fill} 
\begin{subfigure}[t]{0.235\textwidth}
\includegraphics[width=\linewidth]{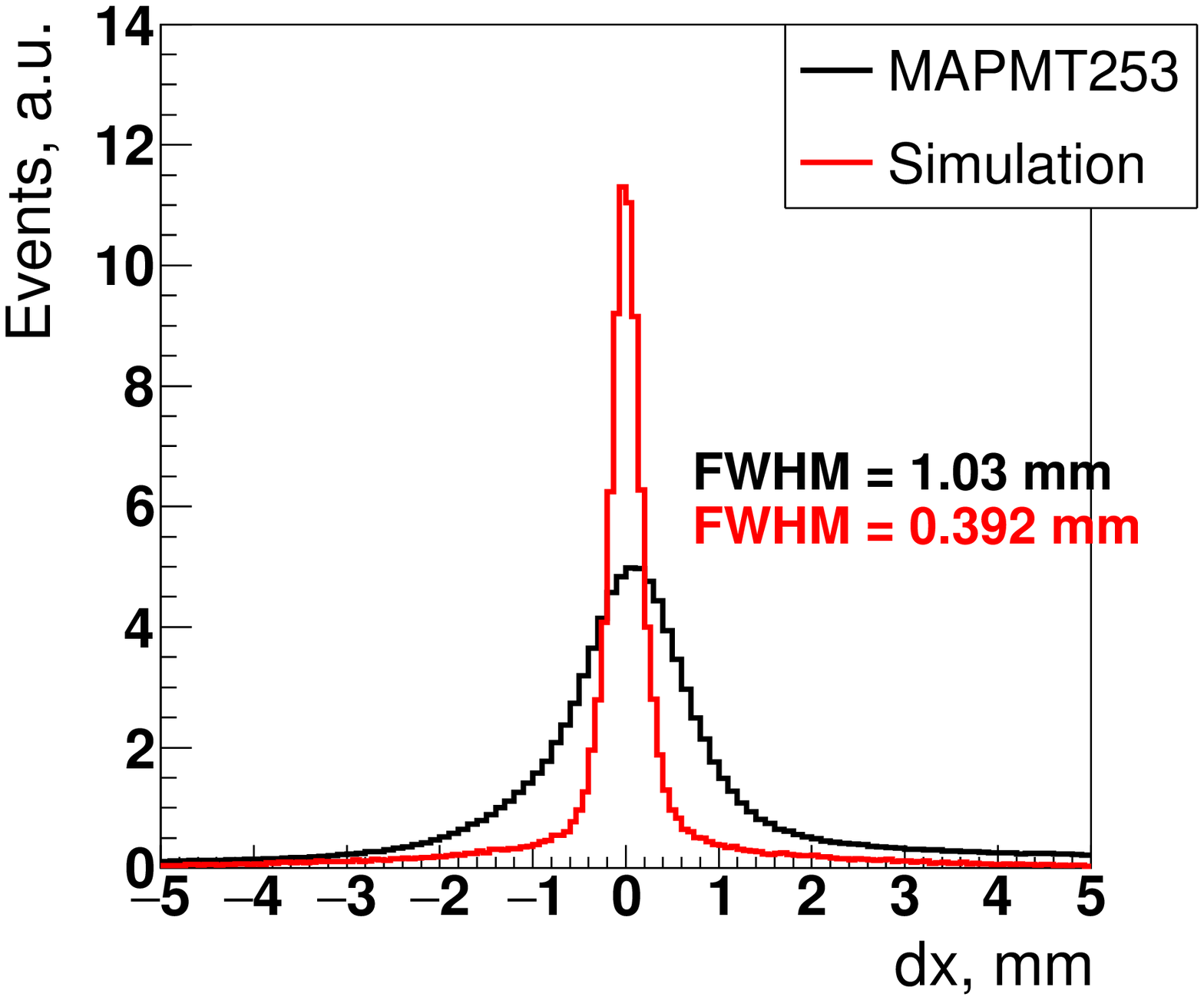}
\caption{Spatial resolution along the TLs. MAPMT253 measurement at (X,Y)=(49.0 mm, 63.2 mm).} \label{dxboth}
\end{subfigure}
\caption{Comparison between the spatial resolutions for the one-photon simulation and for data acquired with an MAPMT253 MCP-PMT.} \label{dxdyboth}
\end{figure}

The 2D position of the gamma interaction into the PWO crystal is determined by the statistical method described in Section \ref{subsection:stat}. Fig. \ref{dyboth} shows the spatial resolutions obtained across the TLs for the one-photon simulation. We obtained resolutions of $\sim$1 mm FWHM for both the simulation and data acquired with an MAPMT253 MCP-PMT.

Fig. \ref{dxboth} shows the spatial resolutions obtained along the TLs for the one-photon simulation and data acquisition at (X,Y)=(49.0 mm, 63.2 mm).
We obtained resolutions of 0.4 mm FWHM and 1 mm FWHM for the simulation and data acquired with an MAPMT253 MCP-PMT, respectively. The width of this distribution depends on the signal propagation speed. We observe that this speed is not constant along the line (Fig. \ref{res16}) and varies also between lines (Fig. \ref{speedall}). In the simulation, for now, we implemented a simplified model with a constant signal speed, the same for all lines. In addition, in the current version, the time jitter of the SAMPIC module (3 ps SD) is not considered. This leads to the too good resolution in the simulation. These effects will be taken into account in future implementations.

\subsection{Time resolution in single detection module}
\begin{figure}[hbt!]
\centering
\includegraphics[width=0.7\linewidth]{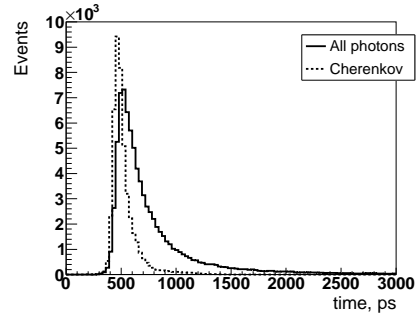}
\caption{Time resolution simulated for the CM detection module considering all gamma interactions. FWHM = $\sim$230 ps for all photons and FWHM = $\sim$117 ps for only Cherenkov photons.} \label{sigtime}
\end{figure}

Fig. \ref{sigtime} shows the simulated time difference between the gamma-ray emission time and the signal time simulated for the single CM detection module prototype (solid line) and the contribution of Cherenkov photons only (dashed line).
The time resolution degrades from $\sim$117 ps to $\sim$230 ps FWHM when scintillation photons are included in the simulation. It happens because the number of detected Cherenkov photons is small, $\sim$1 in average (Fig. \ref{Pe}), and some events do not have any Cherenkov photons at all. It proves that a high detection efficiency of Cherenkov photons is absolutely necessary to achieve good time resolution. The quantum efficiency of the CM detection module prototype is $\leq20$\% at 400 nm. The later phase of the CM detection module development will focus on improving Cherenkov photon detection efficiency and adding a second photoelectric layer (e.g., an SiPM array) on the other side of the crystal (i.e., on the entrance window of the detection module). Thanks to the increase of the detected photons and the decorrelation between DOI and time of gamma interaction, we expect to achieve a better time resolution \cite{Yvon2020}.

\begin{figure}[hbt!]
\centering
\begin{minipage}[t]{0.235\textwidth}
\includegraphics[width=\linewidth]{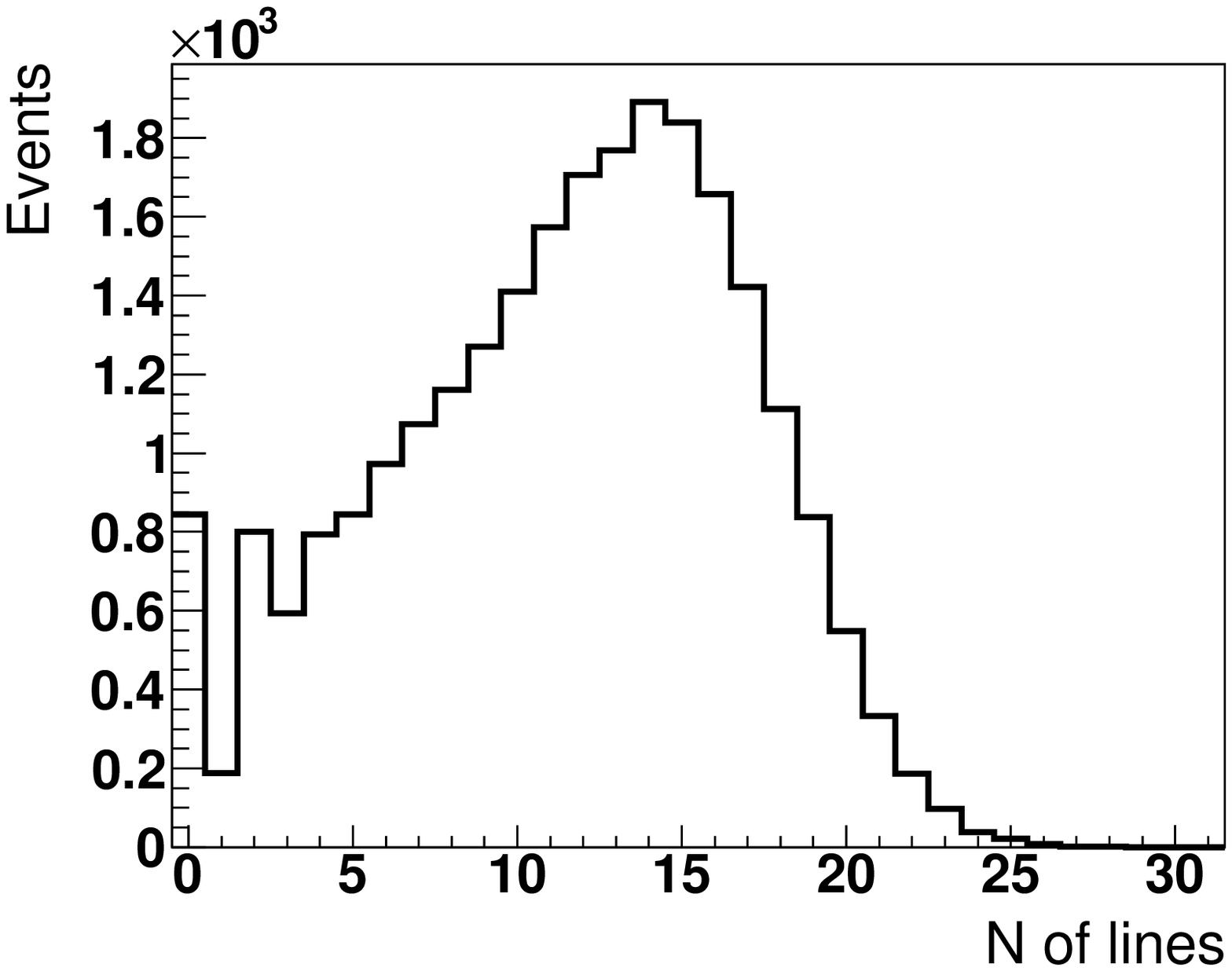}
\caption{Number of triggered lines.} \label{nlines}
\end{minipage}
\hspace*{\fill}
\begin{minipage}[t]{0.235\textwidth}
\includegraphics[width=\linewidth]{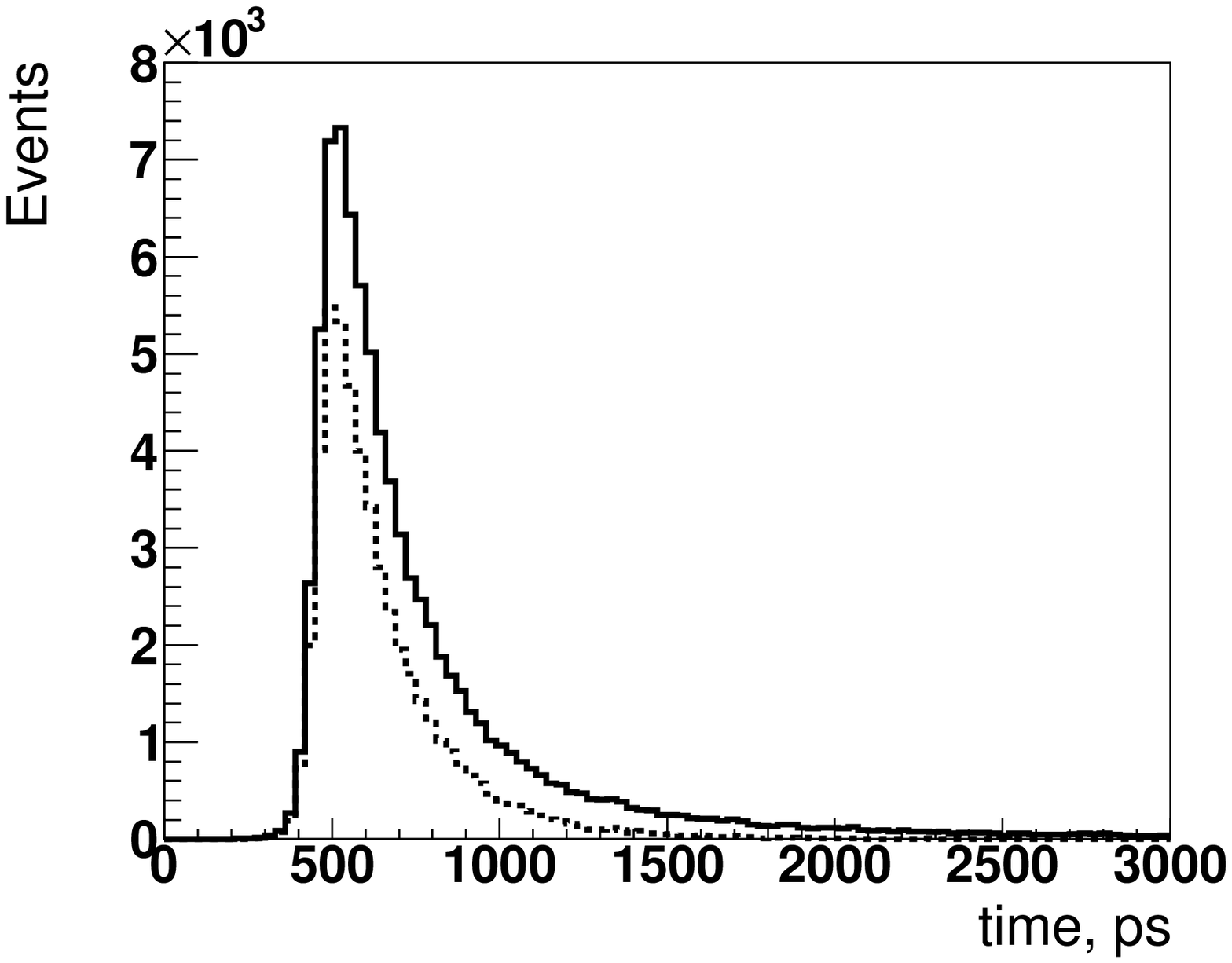}
\caption{Time resolution simulated for the CM detection module considering all gamma interactions and all events (solid line, $\sim$230 ps FWHM) and events with at least 13 triggered lines (dashed line, $\sim$209 ps FWHM). The histograms are normalized to the same number of events.} \label{sigtimecut}
\end{minipage}
\end{figure}

Fig. \ref{nlines} shows the number of lines triggered by the gamma-ray detection. It can be used to estimate the number of photoelectron produced by the photocathode. On average, approximately 30 photoelectrons will result in 14 triggered lines at peak. By applying a selection of the number of triggered lines, we can filter some of the low deposited energy events caused by Compton scattering. Fig. \ref{sigtimecut} shows the time resolution improvement by a selection of the events triggering at least 13 lines.

The time resolution is improved from $\sim$230 ps to $\sim$209 ps FWHM. Moreover, the selection of triggered lines reduces the tail of the time distribution, i.e., more fast photons are detected. The time performance verification is undergoing and will be the goal for our upcoming article.

\section{Event reconstruction}\label{section:Reco}
The reconstruction of the first gamma-conversion vertex in the crystal is crucial to get good detector performance. 

There are different approaches to estimating the interaction position for a monolithic crystal. The position estimation is mostly based on the pixelized SiPM or PMT array signals. Refs \cite{Tao2020,Kawula2021} applied convolutional neural network (CNN) to the LYSO or LaBr$_3$:Ce- and CeBr$_3$-based detectors and achieved spatial resolutions $<1$ mm FWHM in 2D. Ref. \cite{Babiano2019} applied analytical models and developed a neural network (NN) algorithm to the LaCl$_3$:Ce-based detector with different thicknesses. The authors achieved spatial resolutions $\sim$1 mm FWHM in 2D from the analytical models and $\sim$3 mm FWHM from the NN algorithm. Refs. \cite{Stockhoff2019,Stockhoff2021,Decuyper2021} used a $k$-nearest neighbor ($k$-NN) algorithm, a mean nearest neighbour (MNN) algorithm, and an NN algorithm to achieve spatial resolution below 1 mm FWHM in 2D with the LYSO-based detector. To estimate the depth-of-interaction (DOI), unlike Ref. \cite{Yang2008}, which is using dual-ended readout for the pixelized crystal, Refs. \cite{Babiano2019,Stockhoff2019,Stockhoff2021,Decuyper2021} used SiPM signals as input to train the algorithms and achieved a DOI resolution of a few mm FWHM. Ref. \cite{Jaliparthi2021} estimated the gamma interaction position analyzing light sharing with an analytical model and images obtained from the detector with a deep residual-CNN algorithm. They achieved reconstructed spatial resolutions of 0.6 mm FWHM in 3D.




With the CM detection module, we can reconstruct the 3D coordinates of the vertex, i.e., x, y, and DOI, using the signals registered on all the TLs. The following section investigates the possibility of such a reconstruction using a simplified configuration for the detection module and for the CM detection module prototype. For the simplified configuration, we assume that all the crystal faces, except the one with the deposited photocathode, are painted in black, i.e., all visible photons are absorbed by impinging the surfaces. This is the only difference with the configuration employed for the CM detection module prototype that we have described in Section \ref{section:CM}. We also consider only events where 511 keV gamma-rays are interacting in the PWO crystal by photoelectric effect.

In this study we investigated three approaches: a simple statistical reconstruction (Section \ref{subsection:stat}) and machine learning approaches (Sections \ref{subsection:bdt} and \ref{subsection:dnn}). As input variables, we did not use the signal's shape directly, but only signal parameters (e.g., CFD time, charge, etc.). The former is a subject of separate developments. We considered two different machine learning techniques: the gradient boosted decision tree (GBDT) and deep neural network (DNN) as implemented in the ROOT TMVA v.6.18/04 package \cite{TMVA}. Section \ref{subsection:resultsrecon} discusses the performance comparision of different methods and the ongoing work in the collaboration.

\subsection{Statistical method}\label{subsection:stat}
We selected the TL with maximum charge. To reconstruct the coordinate across the TLs (y-coordinate, $y_R$), we calculate the weighted average of coordinates for this line and the two neighboring lines:
\begin{equation} \label{}
  y_R=\frac{\sum_{k=i-1}^{i+1} y_kC_k}{\sum_{k=i-1}^{i+1} C_k} \;,
\end{equation}
where $y_k$ is a y-coordinate of the line center, $C_k$ is a charge of line $k$ (only the negative signal part is used for the charge calculation), $i$ is the line number that has the maximum charge.

The coordinate along lines (x-coordinate, $x_R$) is reconstructed as
\begin{equation} \label{xr}
  x_R = \frac{(t_R - t_L)}{2} \times s \;,
\end{equation}
where $t_R$ and $t_L$ are a time measured at the right and left ends of line $i$, respectively, and $s$ is a signal propagation speed, measured to be about 35\% of the speed of light.

\begin{figure}[hbt!]
  \centering
    \includegraphics[width=0.48\linewidth]{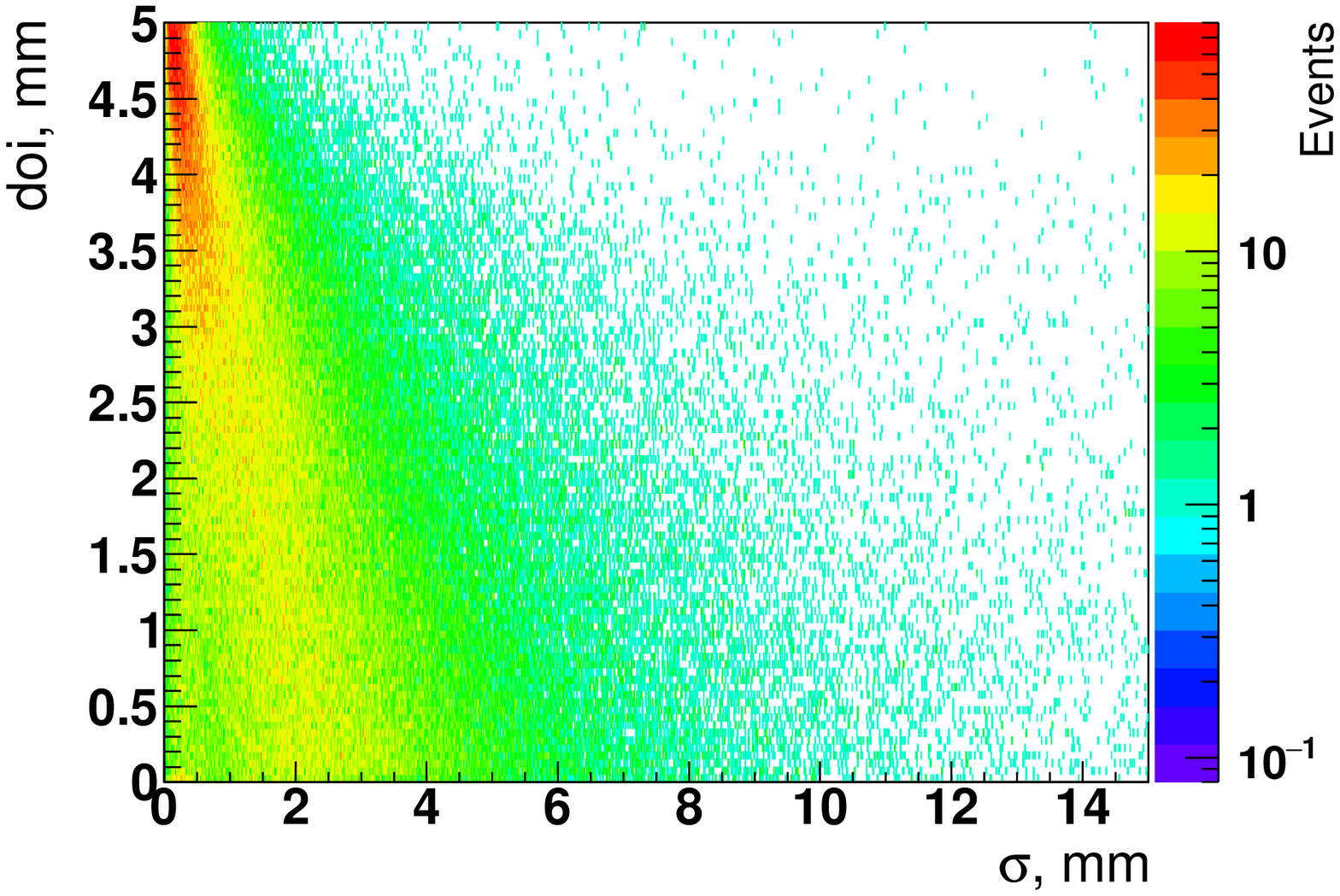}
  \hspace*{\fill} 
    \includegraphics[width=0.48\linewidth]{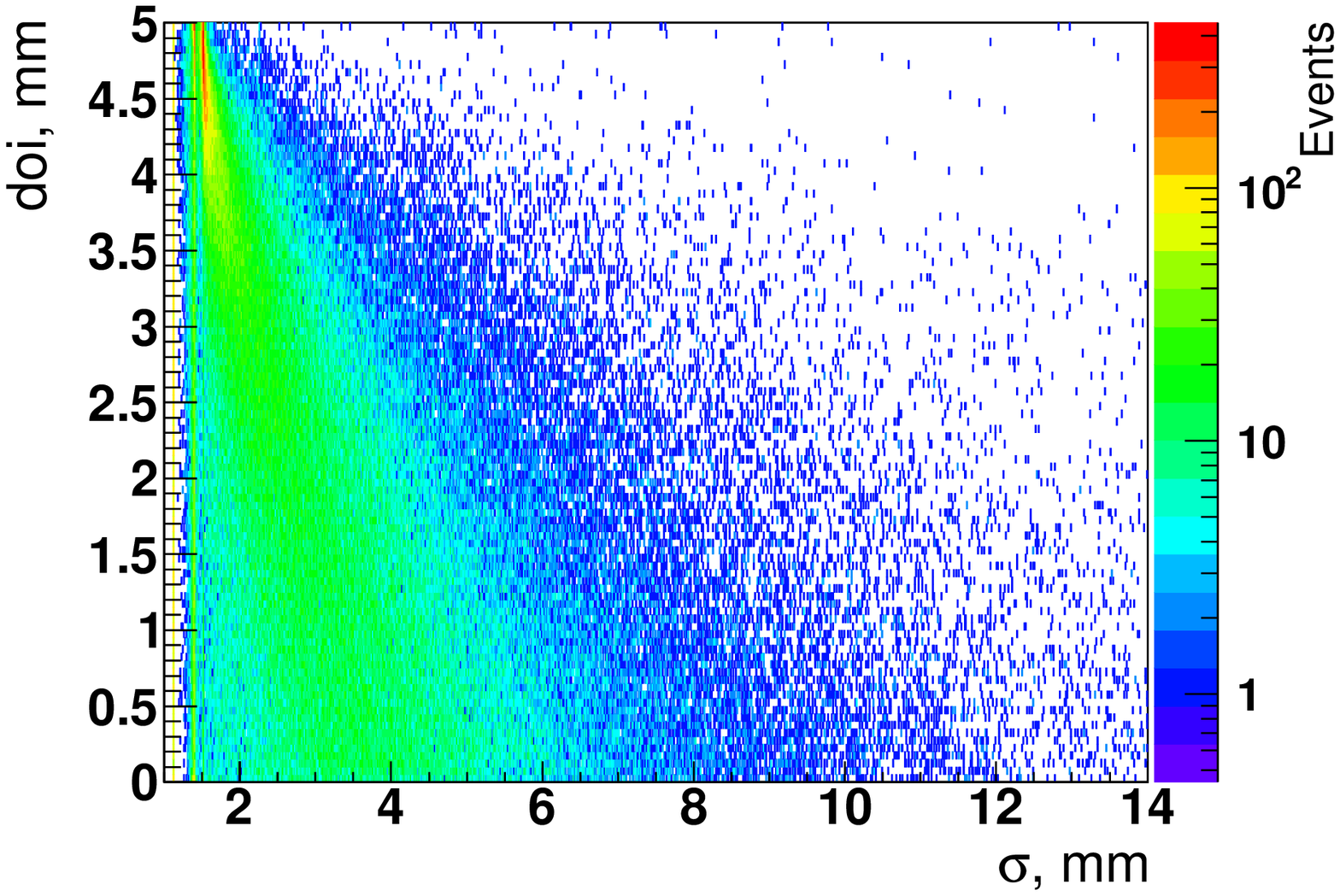}
  \caption{Scatter plots of (left) $\sigma_x$ and (right) $\sigma_y$ versus DOI for the simplified detector configuration.}\label{rms5black}
\end{figure}

\begin{figure}[hbt!]
  \centering
    \includegraphics[width=0.48\linewidth]{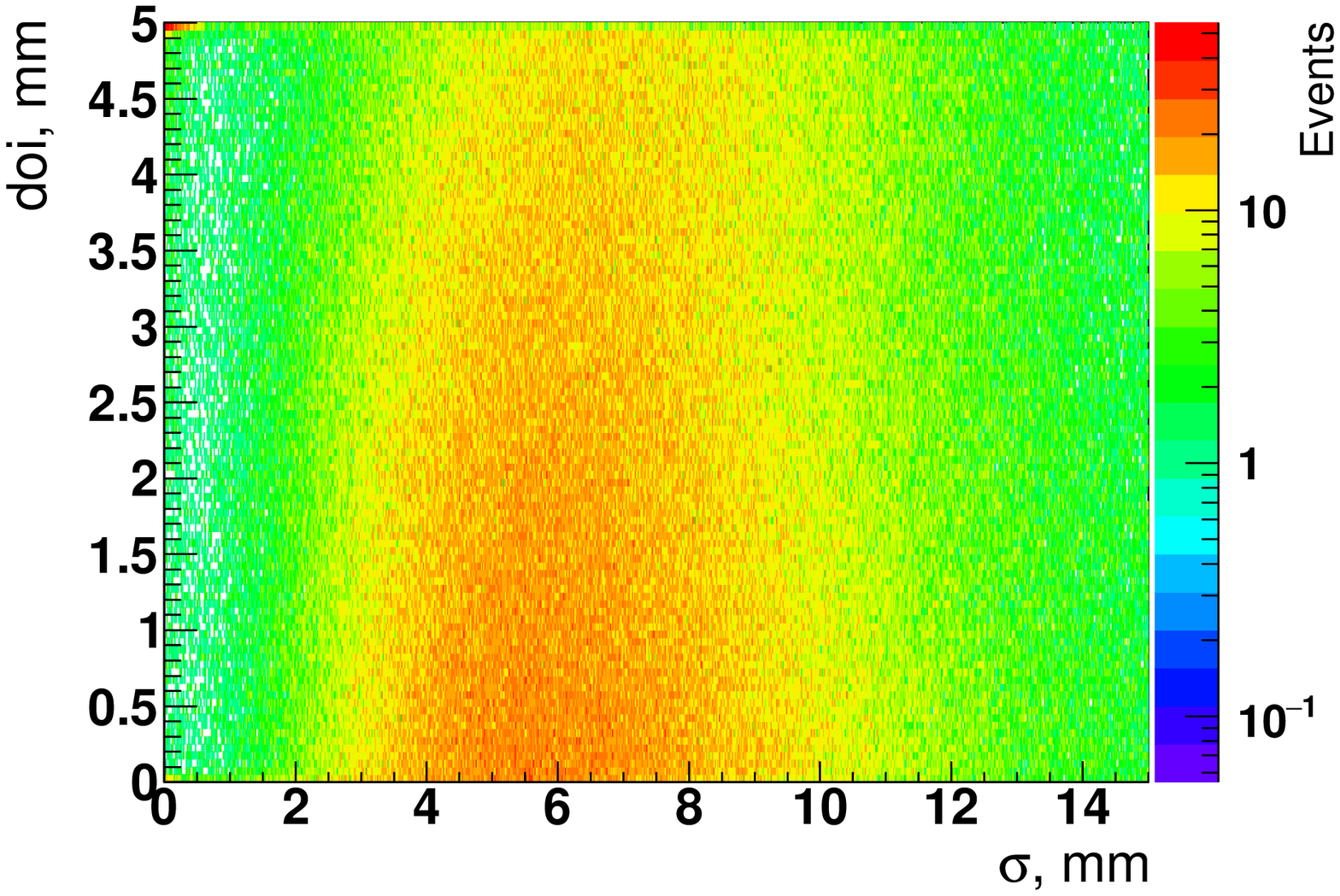}
  \hspace*{\fill} 
    \includegraphics[width=0.48\linewidth]{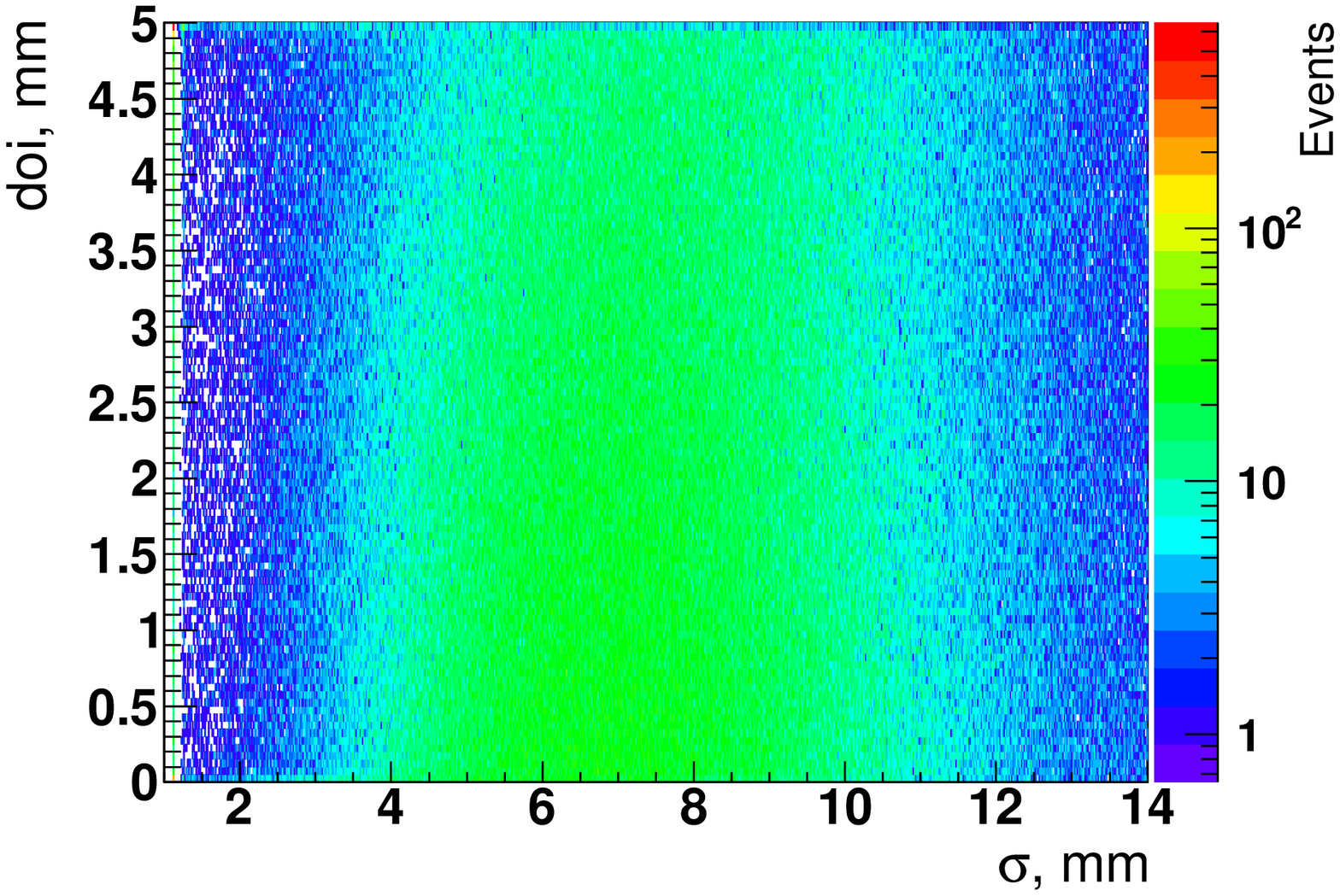}
  \caption{Scatter plots of (left) $\sigma_x$ and (right) $\sigma_y$ versus DOI for the CM prototype configuration.}\label{rms4black}
\end{figure}

To reconstruct the DOI, we used the correlation between estimators ($\sigma_x$ and $\sigma_y$) calculated by the weighted standard deviation (SD) of x- and y-coordinates, and the DOI. These estimators represent the spread of the detected photons in two directions, across the lines and along the lines:
\begin{equation} \label{}
  \sigma_{y}=\sqrt{\frac{\sum_{i=1}^{32}(y_{i}-\bar{y})^2\cdot C_{i}}{\alpha\cdot\sum_{i=1}^{32}C_{i}}} \;,
\end{equation} 
\begin{equation} \label{}
  \sigma_{x}=\sqrt{\frac{\sum_{i=1}^{32}(x_{i}-\bar{x})^2\cdot C_{i}}{\alpha\cdot\sum_{i=1}^{32}C_{i}}} \;,
\end{equation} 
\begin{equation} \label{}
\alpha = 1-\frac{\sum_{i=1}^{32} C_i^2}{(\sum_{i=1}^{32} C_i)^2} \;,
\end{equation} 
where $i$ is the line number, $y_i$ is a y-coordinate of the center of line $i$, $x_i$ is a x-coordinate calculated by the eq. \ref{xr} of line $i$, $\bar{y}$ and $\bar{x}$ are the weighted average of $y_i$ and $x_i$, $C_i$ is a charge at the line $i$, and $\alpha$ is a correction factor. We considered only the triggered TLs among all the 32 TLs. Fig. \ref{rms5black} shows the correlation of DOI and the estimators obtained from the simplified detector. The correlation determined across the lines is $DOI_y = 15.25 - 6.84 \ \sigma_{y} \; (1.5 < \sigma_y < 2.23)$, along the lines is $DOI_x = 5.14 - 1.06 \: \sigma_{x} \; (0.13 < \sigma_x < 4.83)$. To calculate the final DOI, we used meta-analysis to combine the results of $DOI_x$ and $DOI_y$.
Fig. \ref{rms4black} shows the correlation of DOI and the estimators calculated from the CM detection module prototype configuration. Unlike the simplified detector configuration, we could not see any obvious correlation between the estimators and the DOIs. Thus, the reconstruction for the CM detection module prototype will mainly focus on x- and y-coordinates.

\subsection{Gradient boosted decision tree}\label{subsection:bdt}
The GBDT regression \cite{BDT} is applied for the event reconstruction. To train the algorithm on 3D reconstruction, the hyper-parameters and the variables are shown in Table \ref{bdtg}.
\begin{table*}[]
\scriptsize
  \centering 
    \caption{Training parameters of GBDT model. For more details, see Ref. \cite{TMVA}.} \label{bdtg}
    \begin{tabular}{lcccccc}
    \hline\\[-1em]
    Configuration & \multicolumn{3}{c}{Simplified Detector}& \multicolumn{3}{c}{CM Prototype} \\
    Parameters& X & Y & DOI & X & Y & DOI\\
    \hline\\[-1em]
    Train samples (events) & 592k & 50k  & 75k  & 100k & 100k & 97k\\
    Test samples (events)  & 592k & 50k  & 75k  & 100k & 100k & 97k\\
    Maximum trees          & 2000 & 2000 & 500  & 3000 & 2500 & 2000\\
    Maximum tree depth     & 100   & 30   & 10   & 10  & 1000 & 100\\
    Seperation type & \multicolumn{2}{c}{RegressionVariance}  & GiniIndex& \multicolumn{3}{c}{RegressionVariance}\\
    Shrinkage factor       & 0.01 & 0.01 & 0.1  & 0.005 & 0.01 & 0.01\\
    Tree pruning method & \multicolumn{6}{c}{CostComplexity} \\
    Pruning strength       & 50   & 30   & 20   & 80   & 500  & 300 \\
    Variable transform &  Gaussian &\multicolumn{2}{c}{Normalization}&  Gaussian &\multicolumn{2}{c}{Normalization}\\
    Variables$^*$ & Charge$_i$ & Charge$_i$ & Charge$_i$ & Charge$_i$ & Charge$_i$ & Charge$_i$ \\
    {}        & $x_i$  &        & $\sigma_x$&$x_i$ & $y_R$ & $\sigma_x$\\
    {}        & {}     & {}     & $\sigma_y$&$x_R$ & {} & $\sigma_y$\\
    \hline\\[-1em]
    \multicolumn{6}{l}{$^*$ $i$ indicates the all TL numbers}
  \end{tabular}
\end{table*}
Selecting the events with one single 511 keV gamma-ray photoelectric conversion resulted in more than a hundred thousand training samples. We trained the algorithm with the same variables used in the statistical method for the simplified detector. For the y-coordinate, we used only the charges on all the lines. For the x-coordinate, we used the reconstructed x-coordinates and the charges on all the lines. For the DOI, we also used the charges on all lines and the pre-calculated estimators, $\sigma_x$ and $\sigma_y$. For the CM prototype configuration, we added the statistical reconstructed results, $x_R$ and $y_R$ to train the algorithms of the x- and y-coordinates, respectively.

\subsection{Deep neural network}\label{subsection:dnn}
To train the algorithm on 3D reconstruction with DNN regression, the hyper-parameters and the variables are shown in Table \ref{dnn} with a selection of the events with a single 511 keV gamma-ray photoelectric conversion. The variables are the same as those used in the GBDT algorithm training.
  
\begin{table*}[bht!]
\scriptsize
  \centering
  \caption{Training parameters of DNN model. For more details, see Ref. \cite{TMVA}.} \label{dnn}
  \begin{tabular}{m{3cm}C{1cm}C{1cm}C{1cm}C{1cm}C{1cm}C{1cm}}
    \hline\\[-1em]
    Configuration & \multicolumn{3}{c}{Simplified Detector}& \multicolumn{3}{c}{CM Prototype} \\
    Coordinates & X & Y & DOI & X & Y & DOI\\
    \hline\\[-1em]
    \textbf{Parameters} &&&&&&\\
    Train samples (events) & 592k & 50k  & 119k & 100k & 100k & 97k\\
    Test samples (events)  & 592k & 50k  & 119k & 100k & 100k & 97k\\
    Hidden layers          & 6    & 5    & 4    & 6    & 4    & 4\\
    Neurons per layer      & 300  & 500  & 300  & 100  & 300  & 100\\
    Activation function    & \multicolumn{6}{c}{RELU}\\
    Batch size             & 10   & 64   & 32   & 10   & 10   & 10\\[2pt]
    Variable transform &  Gaussian &\multicolumn{2}{c}{Normalization}&  Gaussian &\multicolumn{2}{c}{Normalization}\\
    Variables              & Charge$_i$ & Charge$_i$ & Charge$_i$ & Charge$_i$ & Charge$_i$ & Charge$_i$ \\
    {}                     & $x_i$      &            & $\sigma_x$ & $x_i$     & $y_R$      &$\sigma_x$\\
    {}                     & {}         & {}         & $\sigma_y$ & $x_R$     & {}        &$\sigma_y$\\
    Strategy I &&&&&&\\
    \hspace{2ex}Learning rate     & 5.e-4 & 1.e-3 & 1.e-3 & 5.e-4 & 5.e-4 & 5.e-4\\
    \hspace{2ex}Convergence steps & 34    & 15    & 15    & 34    & 9     & 9\\
    \hspace{2ex}Regularization    & L2    & None  & None  & L2    & None  & None\\
    \hspace{2ex}Weight decay      & 5$\times 10^{-6}$&0&0&5$\times 10^{-6}$&1$\times 10^{-6}$&1$\times 10^{-6}$\\
    \hspace{2ex}Momentum          & 0.5   & 0     & 0     & 0.5   & 0     & 0\\
    \hspace{2ex}Dropout fraction  & 10\%  & 0     & 0     & 10\%  & 0     & 0\\
    Strategy II &&&&&&\\
    \hspace{2ex}Learning rate     & 2.e-5 & 1.e-4 & 1.e-4 & 2.e-5 & 2.e-5 &2.e-5\\
    \hspace{2ex}Convergence steps & 34    & 20    & 20    & 34    & 14    &9\\
    \hspace{2ex}Weight decay      & 1$\times 10^{-6}$&0&0&1$\times 10^{-6}$&1$\times 10^{-6}$&1$\times 10^{-6}$\\
    \hspace{2ex}Dropout fraction  & 10\%  & 0     & 0     & 10\%  & 1\%   & 1\%\\
    Strategy III &&&&&&\\
    \hspace{2ex}Learning rate     & 9.e-6 & 1.e-5 & 1.e-5 & 9.e-6 & 1.e-6 &1.e-6\\
    \hspace{2ex}Convergence steps & 24    & 35    & 40    & 24    & 19    &14\\
    \hspace{2ex}Dropout fraction  & 2\%   & 0     & 0     & 2\%   & 2\%   & 2\%\\
    Strategy IV &&&&&&\\
    \hspace{2ex}Learning rate     & 1.e-6               &&& 1.e-6 & 5.e-7 &\\
    \hspace{2ex}Convergence steps & 24                  &&& 24    & 49    &\\
    \hspace{2ex}Dropout fraction  & 2\%                 &&& 2\%   & 0     &\\
    Strategy V &&&&&&\\
    \hspace{2ex}Learning rate                          &&&&       &1.e-7  &\\
    \hspace{2ex}Convergence steps                      &&&&       &49     &\\
    \hspace{2ex}Dropout fraction                       &&&&       &0      &\\
    \hline
  \end{tabular}
\end{table*}
We trained the algorithm with several consecutive strategies from a higher learning rate to a lower one and different convergence steps to let the algorithms converge faster in the beginning and be more precise in the minimum determination at the end.

\subsection{Results and discussion}\label{subsection:resultsrecon}
We evaluated the reconstruction performance in several ways. First, the FWHM of the distribution of the difference between the reconstructed results and the simulated (true) results shows the accuracy of the reconstruction resolution. Second, the SD of the same distribution shows the spread of the distribution, i.e., the ability to reconstruct the data with a reasonable agreement. Third, we observed the tails of the distributions and calculated the fraction of the center (within $\pm5$ mm for x- and y-coordinates and $\pm1.5$ mm for DOI). A smaller SD together with a larger central fraction represents a better ability to reconstruct the events.

\begin{figure}[t!] \centering
  \begin{subfigure}[t]{0.235\textwidth}
    \includegraphics[width=\linewidth]{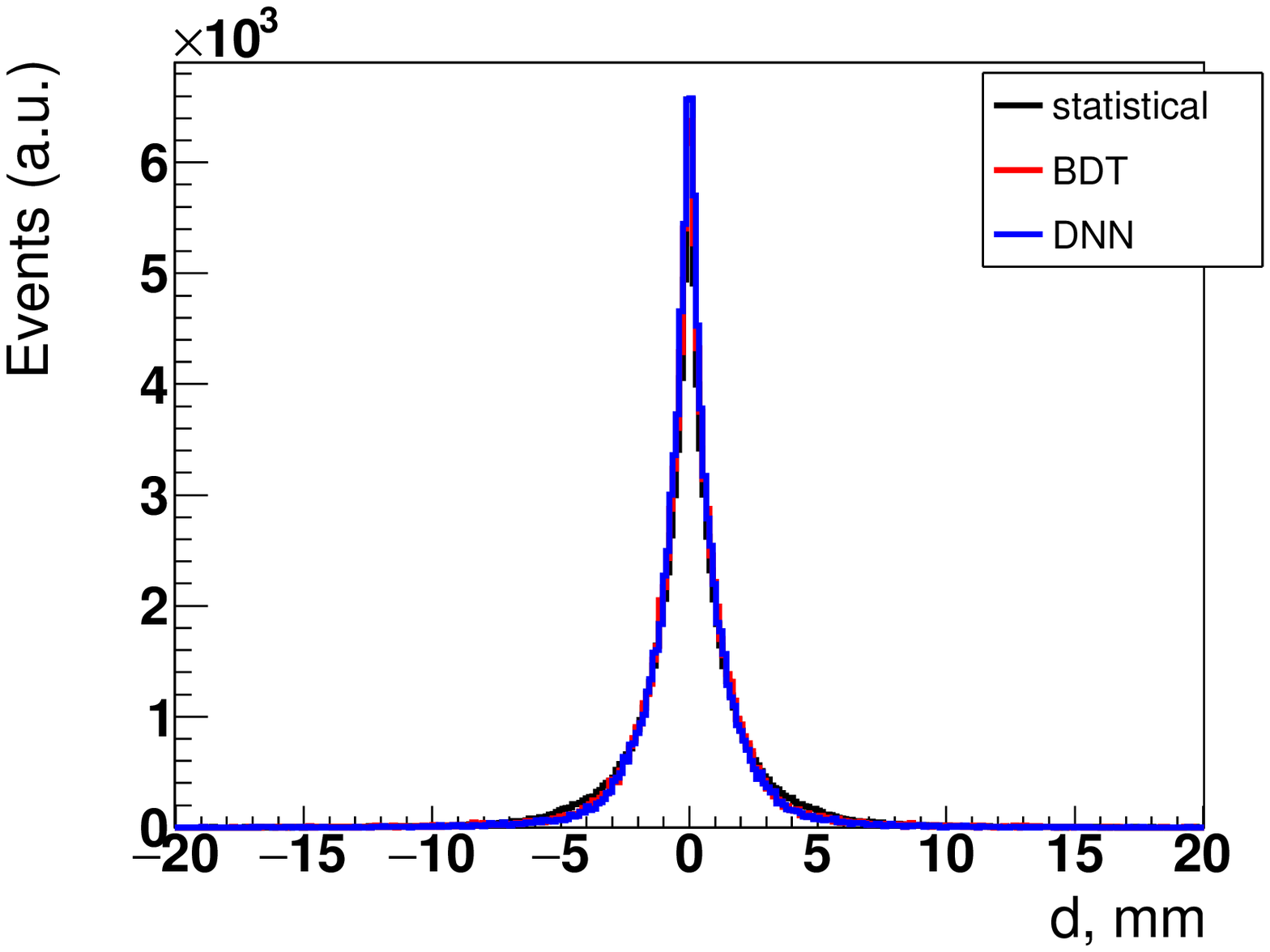}
    \caption{Difference between reconstructed and true y-coordinate.}\label{dy}
  \end{subfigure}
  \hspace*{\fill} 
  \begin{subfigure}[t]{0.235\textwidth}
    \includegraphics[width=\linewidth]{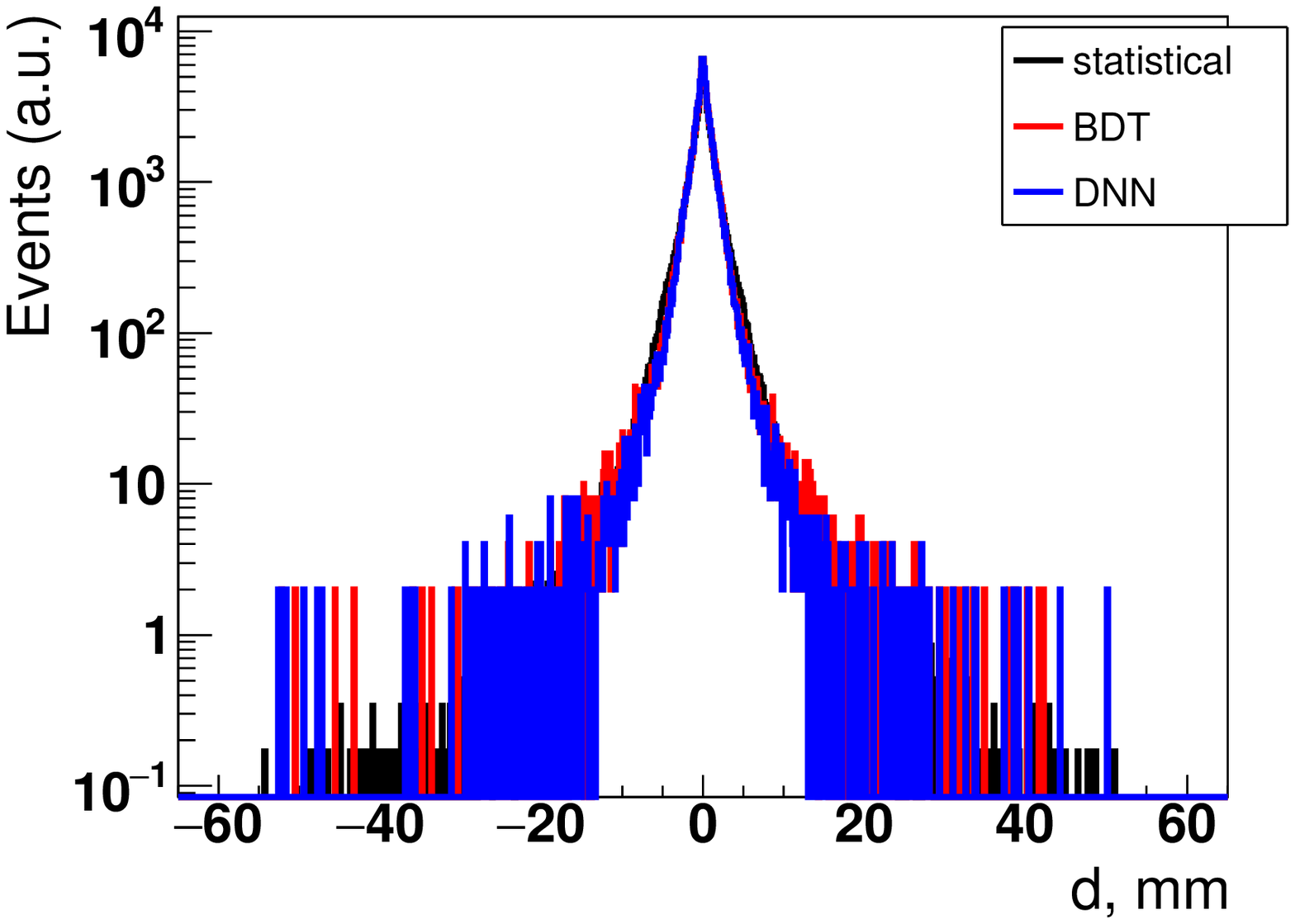}
    \caption{Difference between reconstructed and true y-coordinate in log scale.}\label{dylog}
  \end{subfigure}\\
  \begin{subfigure}[t]{0.235\textwidth}
    \includegraphics[width=\linewidth]{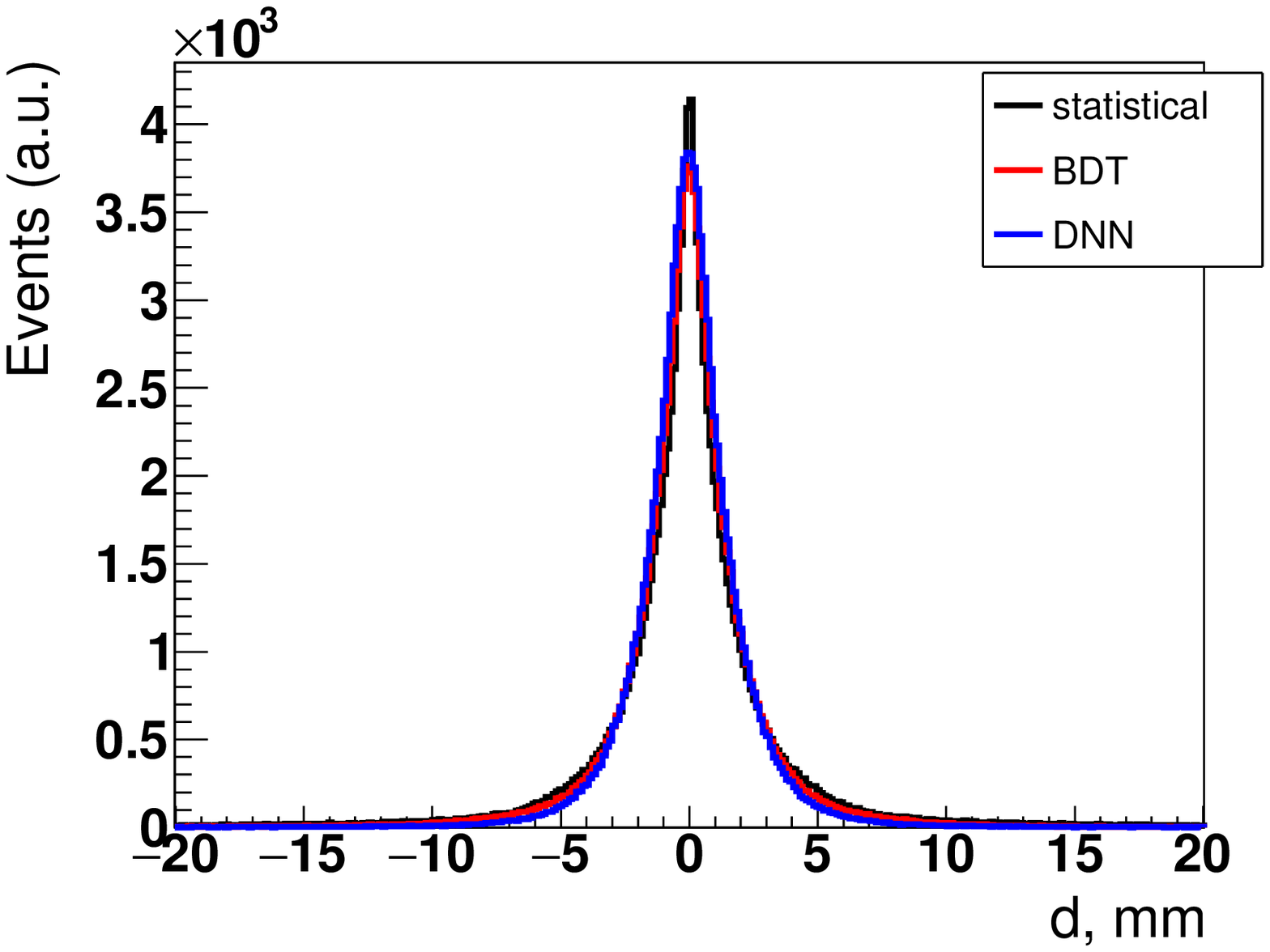}
    \caption{Difference between reconstructed and true x-coordinate.}\label{dx}
  \end{subfigure}
  \hspace*{\fill} 
  \begin{subfigure}[t]{0.235\textwidth}
    \includegraphics[width=\linewidth]{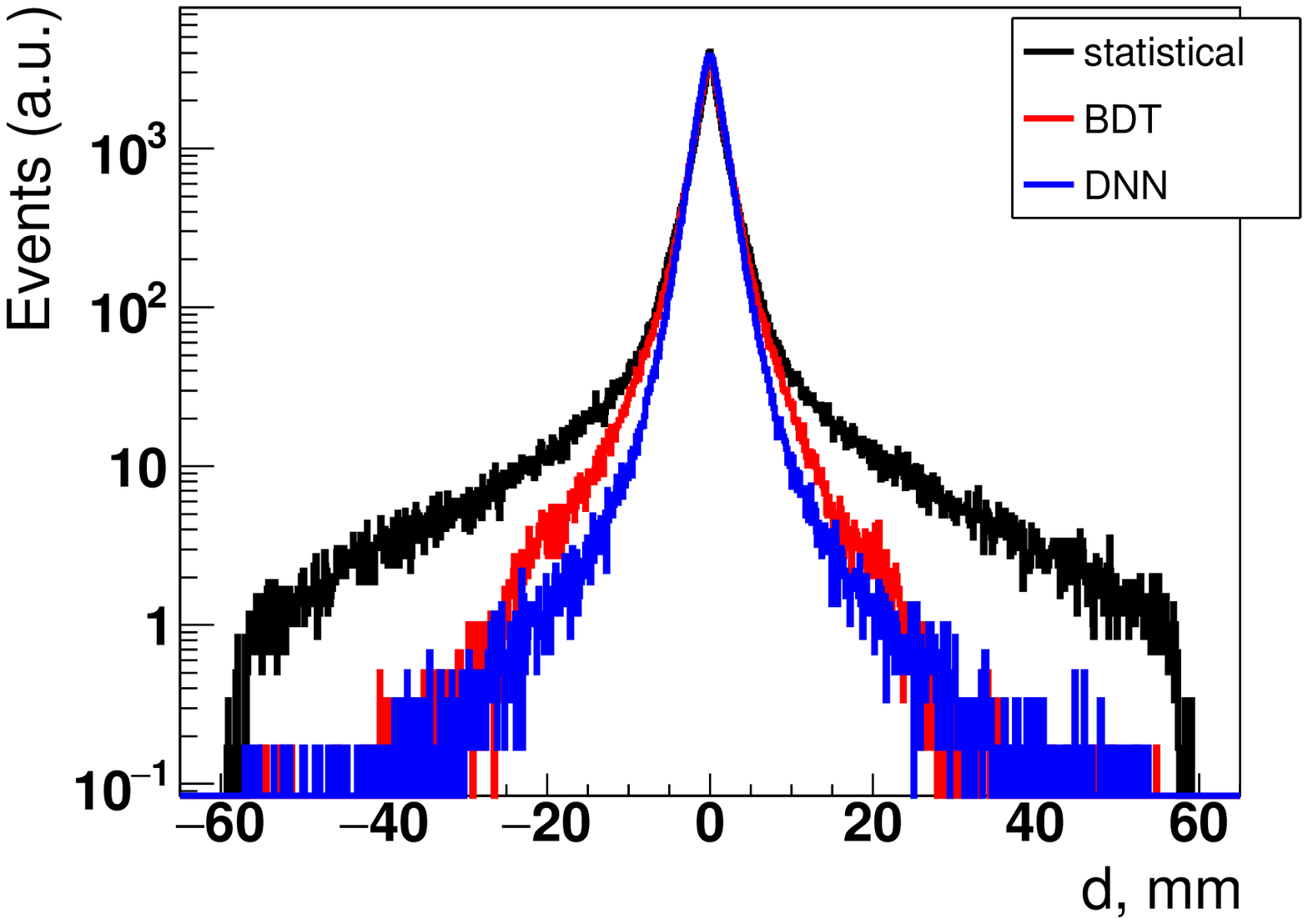}
    \caption{Difference between reconstructed and true x-coordinate in log scale.}\label{dxlog}
  \end{subfigure}\\
  \begin{subfigure}[t]{0.235\textwidth}
    \includegraphics[width=\linewidth]{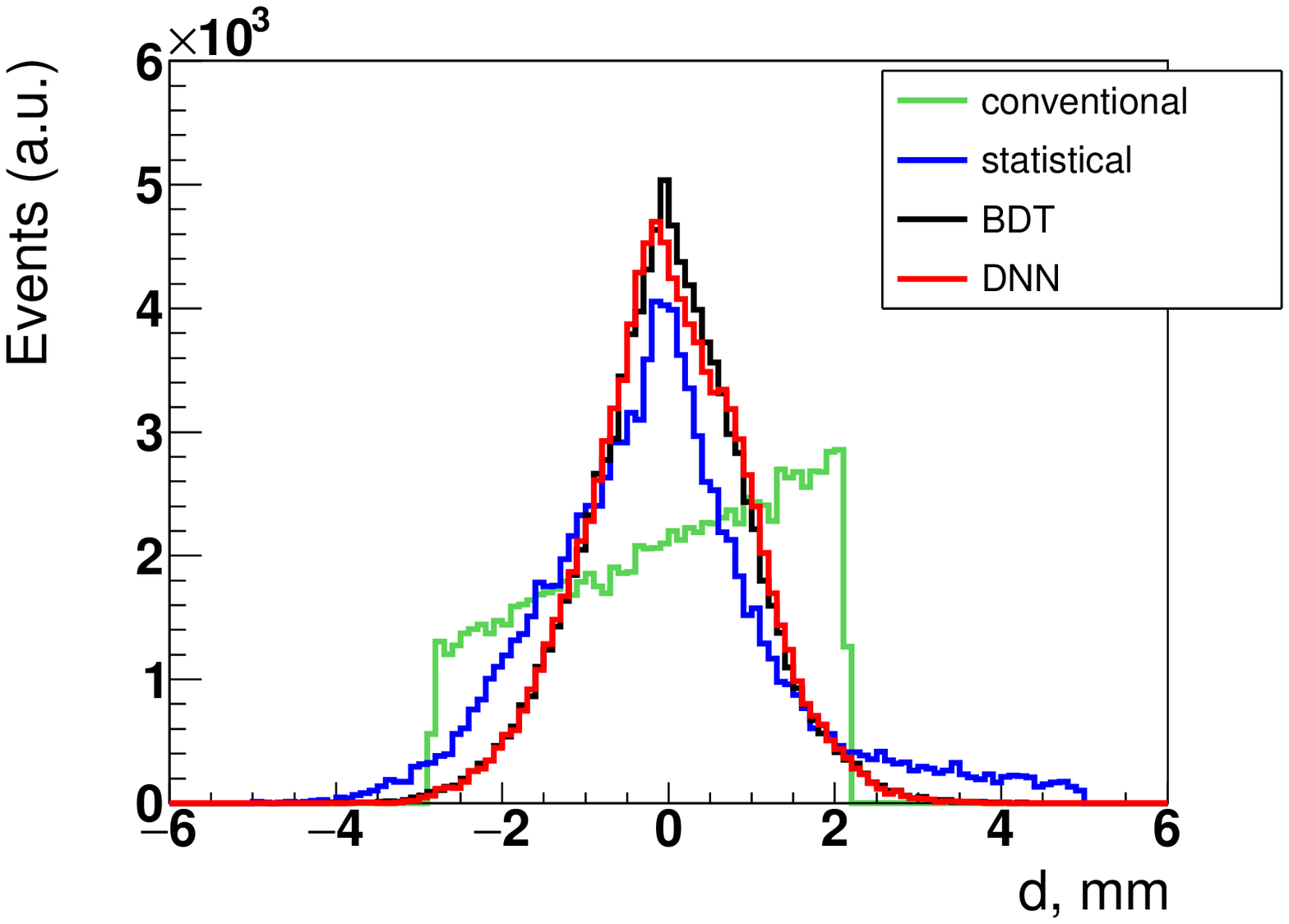}
    \caption{Difference between reconstructed and true DOI-coordinate.}\label{dz}
  \end{subfigure}
  \hspace*{\fill} 
  \caption{Reconstruction results of the simplified detector configuration.} \label{d}
\end{figure}

Fig. \ref{d} shows the resolution for the best achieved reconstruction for three coordinates in the simplified detector configuration.
All histograms are normalized to the same number of events. The resolution for the y-coordinate is $\sim$1 mm FWHM (Fig. \ref{dy}), and $\sim$2 mm FWHM for the x-coordinate (Fig. \ref{dy}). These three methods show similar resolutions. However, the machine learning methods have better results in the tails, i.e., the error $>5$ mm or $<-5$ mm, which are shown in Fig. \ref{dylog} and Fig. \ref{dxlog}. All the performances are summarized in Table \ref{R}. 
\begin{table*}[hbt!]
\scriptsize
\centering
\caption{Reconstruction performances.} \label{R}
\begin{tabular}{m{2cm}m{2cm}C{1cm}C{1cm}C{1cm}C{1cm}C{1cm}C{1cm}}
\hline\\[-1em]
\multicolumn{2}{l}{}& \multicolumn{3}{c}{Simplified Detector}& \multicolumn{3}{c}{CM Prototype} \\
\multicolumn{2}{l}{} & X & Y & DOI & X & Y & DOI\\
\hline\\[-1em]
FWHM (mm) &Statistical & 1.8 & 1.0 & 1.9 & 2.9 & 2.5 &     \\
{}   &GBDT        & 2.5 & 1.2 & 1.8 & 5.8 & 2.7 & 3.4\\
{}   &DNN         & 2.6 & 1.2 & 2.0 & 5.5 & 2.0 & 3.4\\
{}   &Conventional&     &     & 4.5 &     &     & 4.9 \\
SD$^{i}$ (mm)   &Statistical  &6.3&2.5&1.5&5.1&4.4&    \\
{}        &GBDT         &3.1&2.4&1.0&4.0&2.5&1.2\\
{}        &DNN          &2.5&2.3&1.0&3.3&2.0&1.2\\
{}        &Conventional &   &   &1.4&   &   &1.4\\
Fraction$^{ii}$ &Statistical    &87.8\%&95.5\%&65.6\%&77.0\%&82.9\%&      \\
{}               &GBDT         &93.3\%&96.4\%&89.5\%&85.8\%&95.6\%&79.1\%\\
{}               &DNN          &96.4\%&97.2\%&89.7\%&89.6\%&97.3\%&77.7\%\\
{}               &Conventional &   &         &65.6\%&      &      &64.3\%\\
\hline\\[-1em]
\multicolumn{8}{l}{$^{i}$ Standard deviation}\\
\multicolumn{8}{l}{$^{ii}$ Fraction of results within 5 mm for X and Y reconstruction, within 1.5 mm for the DOI}\\
\multicolumn{8}{l}{\hspace{2ex} reconstruction}\\
\end{tabular}
\end{table*}
An obvious improvement can be seen in the SD and the fraction of machine-learning-based results. Regarding the SD and the fraction of the center versus the others, the DNN algorithm performs the best among the three methods.

Fig. \ref{dz} shows that all the three methods have a resolution of $\sim$2 mm FWHM for the DOI reconstruction. The distribution of the statistical method looks more asymmetrical than the others. This can be related to the range limit of the correlation. We compared the DOI reconstruction performance with the conventional PET scanner. The conventional machine does not have DOI information so we fixed the reconstructed DOI value to $\sim$2.2 (to center the distribution) for the resolution comparison as shown in Fig. \ref{dz}. The DOI reconstruction we established improves the DOI resolution from $\sim$4.5 mm FWHM to $\sim$2 mm FWHM in a 5-mm thick crystal. The conventional and statistical methods only correctly reconstruct $\sim$65\% of events within an error of 1.5 mm. On the contrary, almost 90\% of the events using the GBDT and DNN methods are reconstructed within 1.5 mm.

\begin{figure}[t!]
  \centering
  \begin{subfigure}[t]{0.235\textwidth}
    \includegraphics[width=\linewidth]{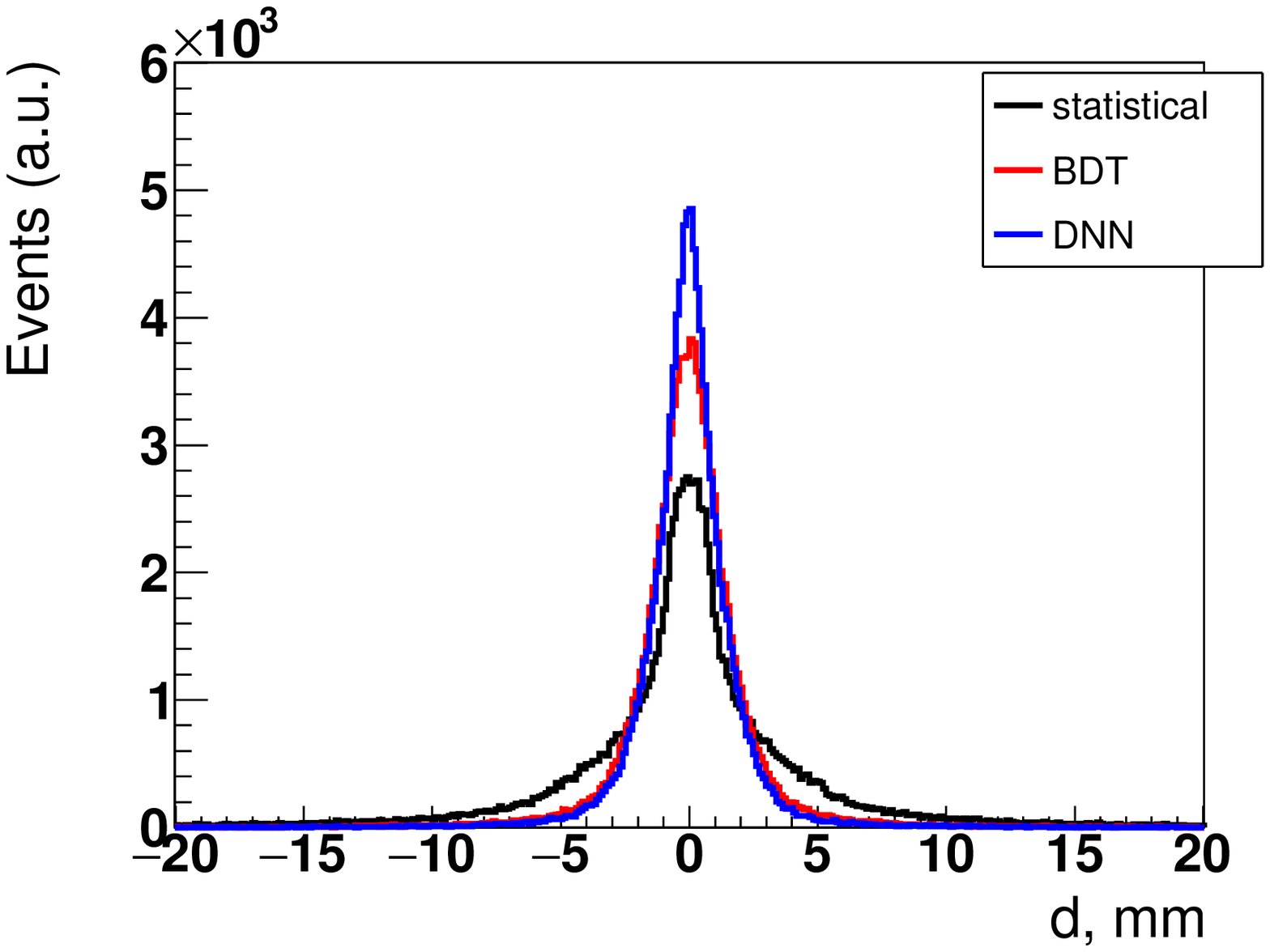}
    \caption{Difference between reconstructed and true y-coordinate.}\label{dycm}
  \end{subfigure}
  \hspace*{\fill} 
  \begin{subfigure}[t]{0.235\textwidth}
    \includegraphics[width=\linewidth]{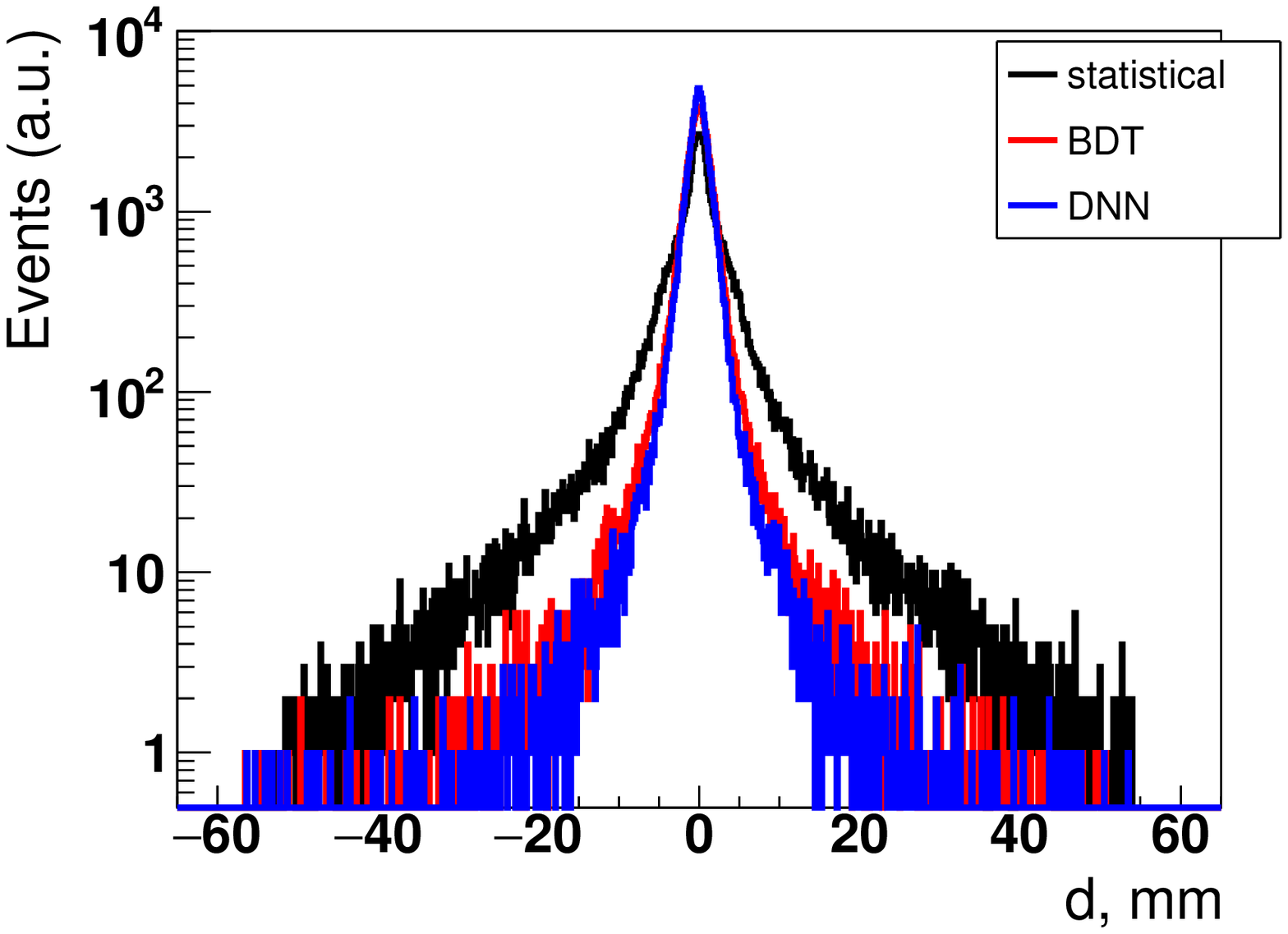}
    \caption{Difference between reconstructed and true y-coordinate in log scale.}\label{dycmlog}
  \end{subfigure}
  \begin{subfigure}[t]{0.235\textwidth}
    \includegraphics[width=\linewidth]{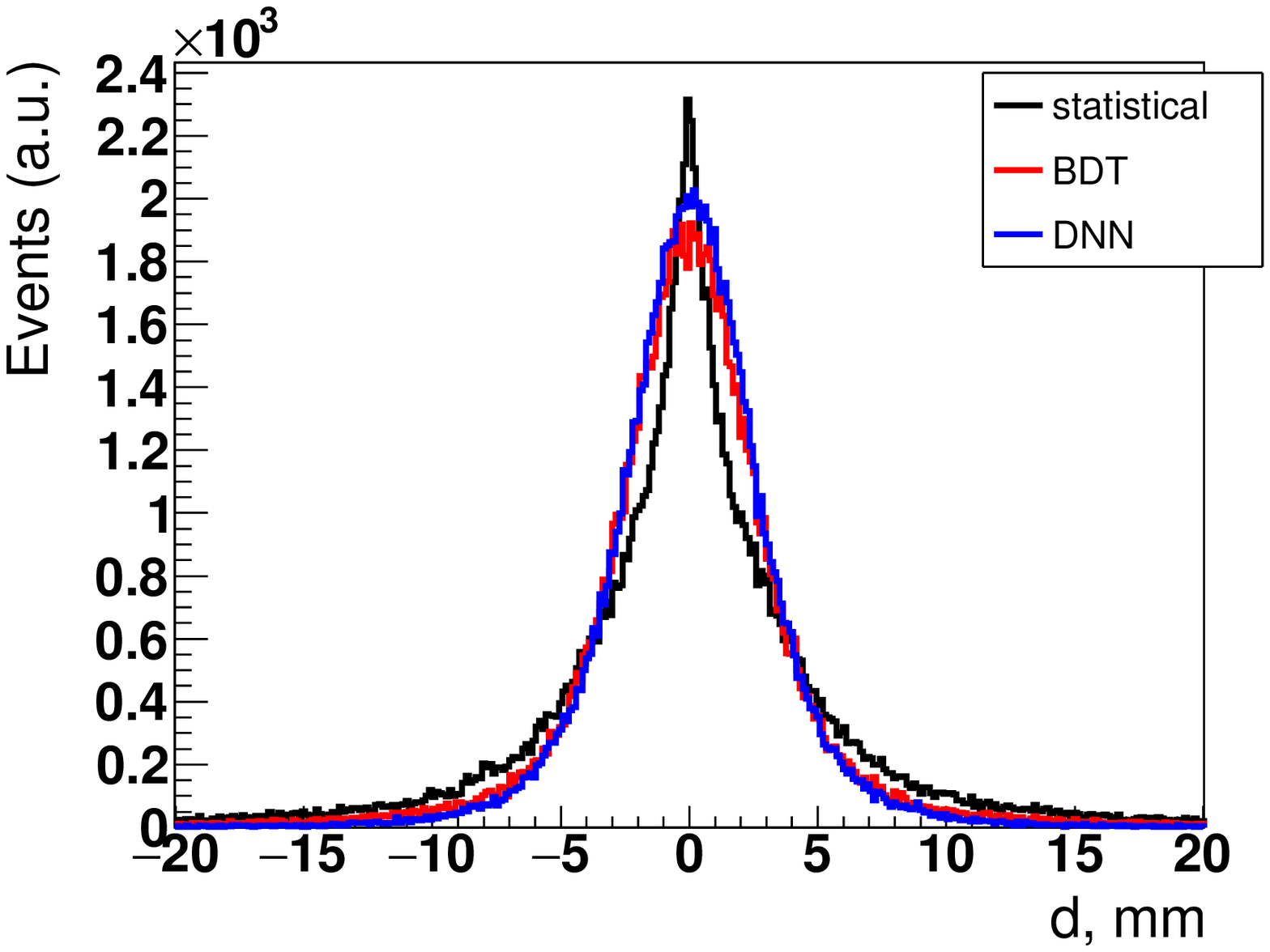}
    \caption{Difference between reconstructed and true x-coordinate.}\label{dxcm}
  \end{subfigure}
  \hspace*{\fill} 
  \begin{subfigure}[t]{0.235\textwidth}
    \includegraphics[width=\linewidth]{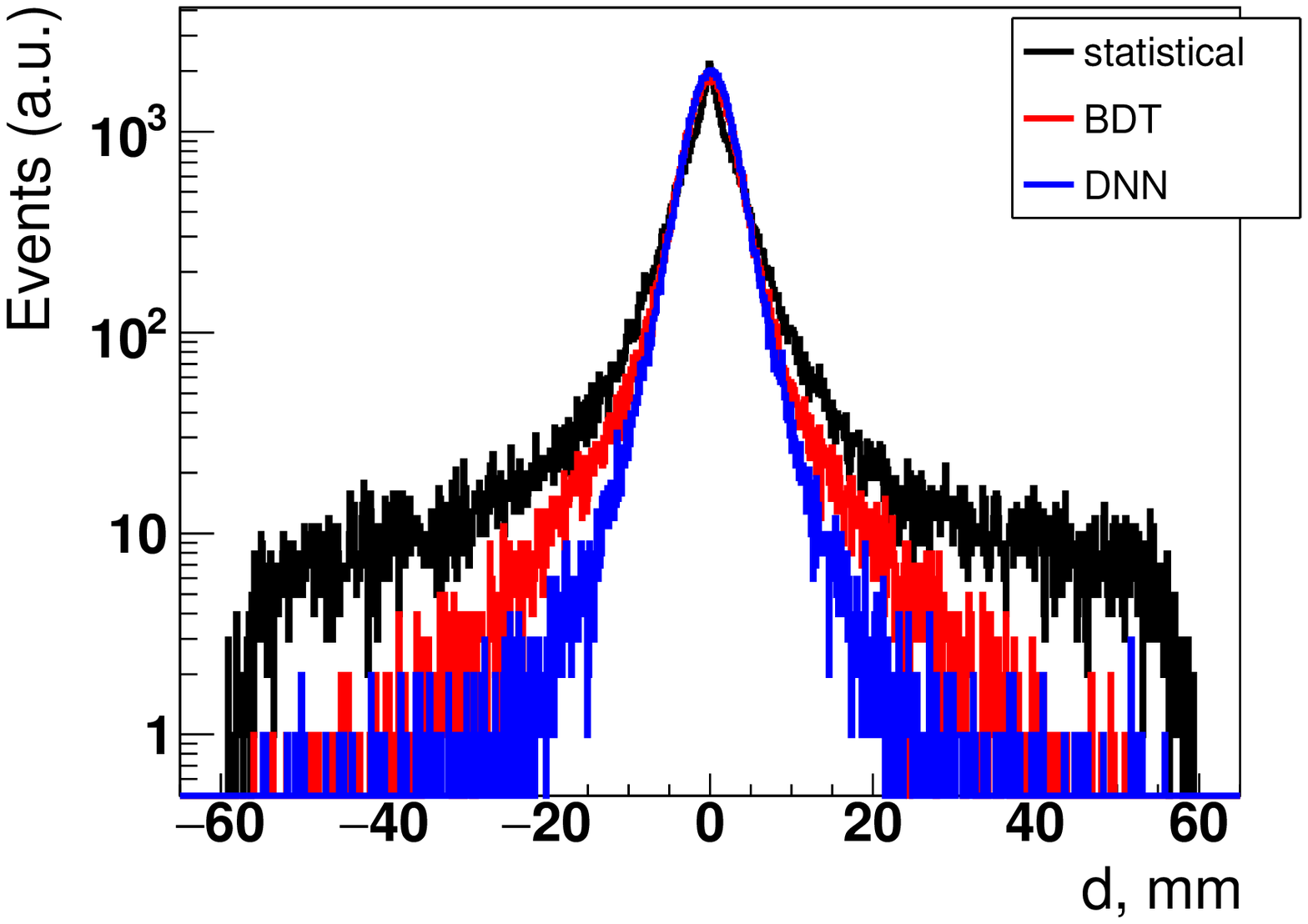}
    \caption{Difference between reconstructed and true x-coordinate in log scale.}\label{dxcmlog}
  \end{subfigure}\\
  \begin{subfigure}[t]{0.235\textwidth}
    \includegraphics[width=\linewidth]{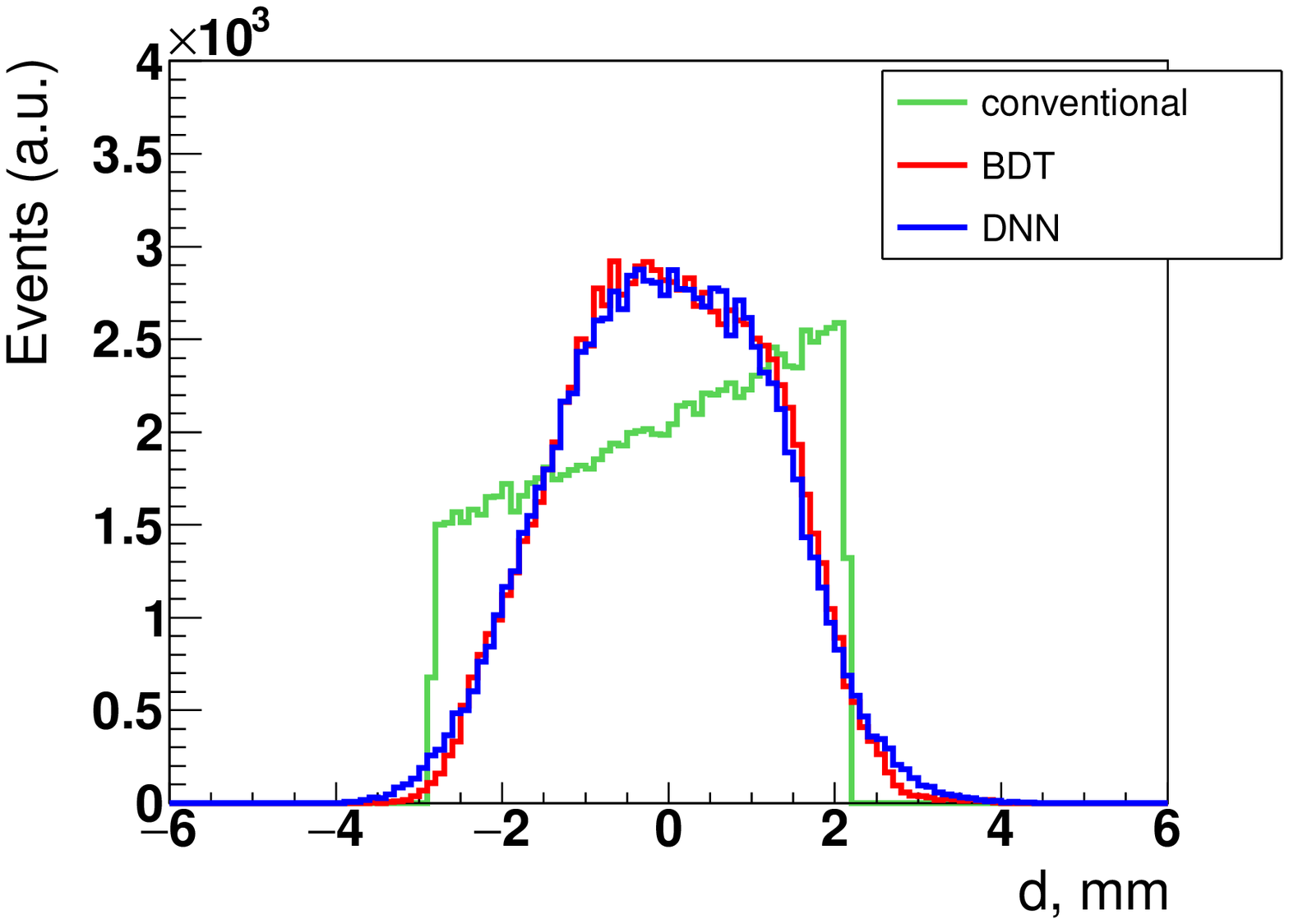}
    \caption{Difference between reconstructed and true DOI-coordinate.}\label{dzcm}
  \end{subfigure}
  \hspace*{\fill}
  \caption{Reconstruction results of the CM prototype.} \label{dcm}
\end{figure}

The reconstruction of the CM detection module prototype configuration is more challenging than the simplified detector one due to the reflection from the front side of the crystal. Fig. \ref{dcm} shows the resolution of the CM prototype configuration. The results from the machine learning methods also have smaller tails according to Fig. \ref{dycmlog} and Fig. \ref{dxcmlog}. 

From Table \ref{R}, we conclude that the machine learning methods have better performances since they reduce a lot the tails of the distribution, which represent worse reconstructed events. Fig. \ref{dzcm} shows the DOI reconstruction in the CM prototype configuration comparing with a conventional scanner. Despite the lack of the correlation between the estimators and the DOI, machine learning methods show a slightly better performance than a conventional scanner (no DOI information) in all the three aspects. GBDT algorithm allows the best resolution in FWHM, the smallest SD, and a high fraction of the center for the DOI reconstruction.

We tried using more variables to reconstruct the event, e.g., the rise time and threshold-to-threshold value of the signals. It did not show an improvement in the reconstruction results. Thus, we presented the simplest algorithm structures in Table \ref{bdtg} and Table \ref{dnn}. Our group also develops the reconstruction algorithms directly using the full signal shapes as inputs. It will bring more information to the algorithm and be expected to have better results especially for the events with the overlapped signals due to the photoelectron superpositions.

The DOI reconstruction will be more important for thicker crystals. The future stage of the CM detection module will have a 10-mm thickness crystal with a second photosensor on the other side of the crystal corresponding to the entrance face of the CM detection module. It will results in a better performance in DOI reconstruction due to the photon detection from both sides of the crystal. It will have less reflection from the crystal's front side and increase the photon detection efficiency.

The time resolution is a crucial point for our detector. Only the configuration with both sides instrumented and optimal photocathode performance will allow us to have a decent time resolution. In that case, the additional improvement due to the machine learning correction for the DOI-related bias will be helpful. 

\section{Perspective and conclusion}\label{section:conclusion}

In this study, we presented a detailed simulation of the CM detection module prototype, including the interaction in the crystal and the mechanics of each component. The reconstruction algorithms using only signal parameters showed a potential to obtain a 3D spatial resolution of a few mm FWHM for x- and y-coordinates.

As shown in this work, the Cherenkov photon efficiency is crucial for obtaining good timing performances for the ClearMind technology. The dedicated efforts are undergoing to increase it in the upcoming prototypes. Furthermore, a second photosensor will be used on the opposite face of the crystal. Such a configuration will increase photon detection efficiency and improve the reconstruction precision, especially for the DOI-coordinate. Finally, we plan to increase the thickness of the PWO crystal to 10 mm, thus increasing the overall gamma-ray detection efficiency, which is an important parameter for using this technology in PET.


\section*{Acknowledgements}
Chi-Hsun Sung, Ph.D. is supported by the CEA NUMERICS program, which has received funding from the European Union's Horizon 2020 research and innovation program under the Marie Sklodowska-Curie grant agreement No. 800945 $-$ (NUMERICS $-$ H2020-MSCA-COFUND-2017). We are grateful for the support and seed funding from the CEA, Programme Exploratoire Bottom-Up, under grant No. 17P103-CLEAR-MIND, and the French National Research Agency under grant No. ANR-19-CE19-0009-01. This work is conducted in the scope of the IDEATE International Associated Laboratory (LIA).






\end{document}